\documentclass[lettersize,journal]{IEEEtran}

\usepackage{vwcol}
\usepackage{url}
\usepackage{graphicx}
\usepackage{psfrag}
\usepackage{amsmath}
\usepackage{amsthm}
\usepackage{amsfonts}
\usepackage{amssymb}
\usepackage{bm}

\usepackage[ruled,linesnumbered]{algorithm2e}

\usepackage{algpseudocode}
\usepackage{xcolor}
\usepackage{float}
\usepackage{bbm}
\usepackage{mathrsfs}
\usepackage{ulem}
\usepackage{bbold}
\usepackage[justification=centering]{caption}
\usepackage{subfigure}
\usepackage{cite}
\usepackage{hyperref} 
\usepackage{multicol}
\usepackage{tikz}
\usetikzlibrary{arrows.meta}
\usepackage{setspace}
\usepackage{color}
\usepackage[numbers,sort&compress]{natbib}
\usepackage{latexsym}
\usepackage{extarrows}
\usepackage{cancel}
\usepackage{lipsum}

\usepackage{pifont}
\usepackage{multirow}
\usepackage{indentfirst}
\usepackage{color}
\usepackage{ulem}

\usepackage{cuted}

\usepackage{tcolorbox}

\usepackage{makecell}

\usepackage{colortbl}
\usepackage{booktabs}
\usepackage{arydshln}

\usepackage{textcomp}
\usepackage{manyfoot}
\usepackage{listings}
\usepackage{pifont}

\newcommand{\bs}{\boldsymbol}
\newcommand{\Sf}{\mathsf}
\newcommand{\TB}{\textbf}

\newtheorem{assumption}{Assumption}
\newtheorem{lemma}{Lemma}

\newtheorem{proposition}{Proposition}

\newcommand{\Eye}{\bs{I}}

\newcommand{\Trace}{\Sf{Tr}}
\newcommand{\Ts}{\Sf{T}}
\newcommand{\Cts}{\Sf{H}}

\newcommand{\Normal}{\Sf{N}}

\newcommand{\dd}{\Sf{d}}

\newcommand{\Proj}{\Sf{Proj}}
\newcommand{\Diag}{\Sf{D}}
\newcommand{\diag}{\Sf{d}}

\newcommand{\Mean}{\mathbb{E}}
\newcommand{\Var}{\mathbb{V}}

\newcommand{\Real}{\mathbb{R}}

\newcommand{\EXP}{\Sf{e}}

\newcommand{\imagUnit}{\mathbb{j}}

\newcommand{\Extr}{\mathop{\operatorname{extr}}}

\newcommand{\ArgMin}{\mathop{\operatorname{argmin}}}
\newcommand{\ArgMax}{\mathop{\operatorname{argmax}}}

\newcounter{magicrownumbers}
\newcommand\rownumber{\stepcounter{magicrownumbers}\arabic{magicrownumbers}}

\allowdisplaybreaks[4]

\ifCLASSINFOpdf

\else

\fi

\begin{document}

\title{
Vector Approximate Survey Propagation
}

\author{
	Qun Chen, Haochuan Zhang*, and Huimin Zhu
	\thanks{
		Q. Chen and H. Zhang are with Guangdong University of Technology, Guangzhou 510006, China (Emails:
		c.prey.q@gmail.com;
		haochuan.zhang@gdut.edu.cn).
		H. Zhu is with Guangzhou University of Chinese Medicine, Guangzhou 510006, China (email:
		hm\_zhu@gzucm.edu.cn).
		This work was supported by Guangdong Basic and Applied Basic Research Foundation under Grants 2022A1515010196 and 2023A1515110853.
	}
	\thanks{
		*Corresponding author:
		Haochuan Zhang.
	}
}

\maketitle

\begin{abstract}
Approximate Message Passing (AMP), originally designed to solve high-dimensional linear inverse problems, has found broad applications in signal processing and statistical inference. Among its key variants, Vector Approximate Message Passing (VAMP) and Generalized Approximate Survey Propagation (GASP) have demonstrated effectiveness even in scenarios where the assumed generative models differ from the true models. However, the maximum a posteriori (MAP) versions of VAMP and GASP have limitations: VAMP is restricted to differentiable priors and likelihoods, while GASP requires the measurement matrix to have independent identically distributed (i.i.d.) elements. To overcome these limitations, this paper introduces a new algorithm, Vector Approximate Survey Propagation (VASP). VASP utilizes survey propagation to handle non-differentiable priors and likelihoods, along with employs vector-form messages to account for correlations in the measurement matrix. Simulations reveal that VASP significantly surpasses VAMP and GASP in estimation accuracy, particularly when the assumed prior is discrete-supported and the measurement matrix is non-i.i.d.. Additionally, the state evolution (SE) of VASP, derived heuristically, accurately reflects the per-iteration mean squared error (MSE). A comparison between the SE and the free energy computed by Takahashi and Kabashima under the one-step replica symmetry breaking (1RSB) ansatz shows that the SE's fixed-point equations align with the free energy's saddle point equations. This suggests that VASP efficiently implements the postulated MAP estimator (which is NP-hard in the worst case) with cubic computational complexity, assuming the 1RSB ansatz is valid.
\end{abstract}


\IEEEpeerreviewmaketitle

\section{Introduction}

Approximate Message Passing (AMP) \cite{kabashima2003cdma,donoho2009message} has significantly influenced high-dimensional signal processing and statistical inference.
Originally developed to efficiently tackle large-scale linear inverse problems, AMP is renowned for achieving (near-)optimal performance while maintaining low computational complexity.
A key factor in AMP's success is its strong theoretical foundation, particularly its connection to rigorous asymptotic analysis via the state evolution (SE) framework\cite{bayati2011dynamics,javanmard2013state}.
This framework not only provides precise performance predictions for AMP but also sheds light on the fundamental limits of high-dimensional estimation problems.
Over time, AMP has been extended and refined to address various data types and models, resulting in numerous variants.
These include Generalized AMP (GAMP) \cite{rangan2011generalized} for non-linear measurements, Vector AMP (VAMP) \cite{rangan2019vector,schniter2016vector,fletcher2018inference,he2018bayesian} and Orthogonal AMP (OAMP) \cite{ma2017orthogonal} for structured measurement matrices, and Approximate Survey Propagation (ASP) \cite{antenucci2019approximate} and Generalized ASP (GASP) \cite{lucibello2019generalized} for mismatched models.
These advancements have expanded AMP's applicability across various domains, including compressed sensing, image reconstruction, and machine learning.

Takahashi and Kabashima \cite{takahashi2022macroscopic} recently demonstrated that for VAMP \cite{rangan2019vector}, even in the presence of a mismatch between the assumed and true generative models, the algorithm can still approximate the postulated posterior estimator (PPE), which is NP-hard in complexity, in the large system limit (LSL).
This holds as long as the algorithm converges to the SE fixed point and the replica symmetry (RS) ansatz remains valid for the PPE.
However, they also highlighted that VAMP is unsuitable for inference in mismatched models where the log-posterior is non-convex.
In particular, the maximum a posteriori (MAP) version of VAMP requires second-order differentiability of both the prior and the likelihood \cite[Tab. 1]{rangan2011generalized} \cite[Eq. (13)]{rangan2019vector}.
For example, in cases involving the estimation of digitally modulated signals, such as QPSK symbols, MAP-VAMP is inapplicable because the prior is supported on discrete values.
To address such challenges, the GASP algorithm \cite{lucibello2019generalized} effectively handles discrete-supported priors in mismatched models.
GASP integrates Gaussian smoothing into the survey definition, creating a differentiable free entropy regardless of the assumed prior or likelihood.
However, GASP has a fundamental limitation: it requires the measurement matrix to be i.i.d. Gaussian, which constrains its applicability.

In this paper, we aim to develop an algorithm that addresses the aforementioned challenges.
We focus on the estimation of a high-dimensional signal based on observation generated from a generalized linear model (GLM) \cite{takahashi2022macroscopic,lucibello2019generalized}.
Specifically, we consider an
$ N \times 1 $
random signal
$
\bs{x}_{0}
\sim
p(\bs{x}_{0})
$,
which is first linearly transformed into
$
\bs{z}_{0}
\triangleq
\bs{H} \bs{x}_{0}
$
using an
$ M \times N $
deterministic measurement matrix
$ \bs{H} $.
The resulting
$ M \times 1 $
vector is then randomly mapped to produce the observation
$
\bs{y}
\sim
p(\bs{y} | \bs{z}_{0})
$,
i.e.,
\begin{align*}
p( \bs{x}_{0} )
& \triangleq
\prod_{i = 1}^{N}
p( x_{0, i} )
,
\quad
p( \bs{y} | \bs{z}_{0} )
\triangleq
\prod_{a = 1}^{M}
p( y_{a} | z_{0, a} )
.
\end{align*}
Similar to \cite{takahashi2022macroscopic,lucibello2019generalized},
we consider a model-mismatched setting for estimating the input signal
$ \bs{x}_0 $.
In this scenario, we assume that the prior and likelihood information used by the estimator differ from the true ones in the GLM.
Let the postulated prior and likelihood be denoted as
$ q( \bs{x}_{0} ) $
and
$ q( \bs{y} | \bs{z}_{0} ) $,
where
\begin{align*}
q( \bs{x}_{0} )
& \triangleq
\prod_{i = 1}^{N}
q( x_{0, i} )
, \quad
q( \bs{y} | \bs{z}_{0} )
\triangleq
\prod_{a = 1}^{M}
q( y_{a} | z_{0, a} )
.
\end{align*}
We can express the (postulated) ``optimal'' estimate as \cite{takahashi2022macroscopic,rangan2012asymptotic}:
\begin{align}
\hat{\bs{x}}
& =
\int{
	\dd \bs{z}_{0}
	\dd \bs{x}_{0}
} \,
q( \bs{z}_{0}, \bs{x}_{0} )
\bs{x}_{0}
, \label{Eq:Postulated_Posterior_Estimator} \\
q( \bs{z}_{0}, \bs{x}_{0} )
& =
\frac{1}{ Z(\bs{y}) }
q^{\beta}( \bs{y} | \bs{z}_{0} )
\delta(
	\bs{z}_{0}
	-
	\bs{H} \bs{x}_{0}
)
q^{\beta}(\bs{x}_{0})
, \label{Eq:Postulated_Posterior} \\
Z( \bs{y} )
& =
\int{
	\dd \bs{z}_{0}
	\dd \bs{x}_{0}
} \,
q^{\beta}( \bs{y} | \bs{z}_{0} )
\delta(
	\bs{z}_{0}
	-
	\bs{H} \bs{x}_{0}
)
q^{\beta}( \bs{x}_{0} )
. \label{Eq:Normalization_Factor}
\end{align}
Here, the estimate
$ \hat{\bs{x}} $
is referred to as the postulated MAP (PMAP) estimate \cite{rangan2012asymptotic} when
$ \beta \to + \infty $,
and
the postulated minimum mean squared error (PMMSE) estimate \cite{rangan2012asymptotic} when
$ \beta = 1 $.
In the LSL, where
$ M \rightarrow \infty $
and
$ N \rightarrow \infty $
with a fixed ratio
$ \alpha = M / N $,
solving the PMAP estimation problem becomes highly non-trivial.
In fact, this estimation is non-convex and NP-hard in the worst case \cite{antenucci2019approximate}.

To develop a computationally efficient algorithm, we adopt a message-passing framework, following the approach of VAMP and GASP.
However, a key question arises:
what type of message should we use?
VAMP approximates the PMAP estimator \eqref{Eq:Postulated_Posterior_Estimator} using belief propagation (BP), while GASP employs survey propagation (SP)\footnote{
	Survey propagation \cite{mezard1987spin,braunstein2005survey,mezard2001bethe,mezard2003cavity} is a message passing procedure parallel to belief propagation \cite{pearl1988probabilistic}.
	Generally speaking, a survey is a weighted sum of beliefs. For more discussion, see the conclusion section of this paper.
}.
SP is more general than BP but less efficient, as it encompasses BP as a special case but requires additional integration or summation.
To address this question, we consider the prediction for our target problem \eqref{Eq:Postulated_Posterior_Estimator} presented by Takahashi and Kabashima \cite{takahashi2022macroscopic}, assuming the validity of the one-step replica symmetry breaking (1RSB) ansatz.
This prediction also characterizes the performance of VAMP and GASP.
However, the original 1RSB expressions just provide little insight into a practical algorithm.

To overcome this, we reformulate the 1RSB results using the Gaussian convolution trick\cite{bromiley2003products}, derive a new set of saddle point equations that are mathematically equivalent to those in \cite[Eqs. (137)-(142), (168)-(174)]{takahashi2022macroscopic}, and conclude that vectorized SP is the optimal choice for message passing in this context.
We then define the vector survey, design the update rules, schedule the message propagation, and ultimately propose a new algorithm called vector ASP (VASP).
Our contributions are twofold:

(i)
VASP significantly outperforms VAMP and GASP in estimation accuracy for both matched and mismatched models, particularly when the postulated prior or likelihood has discrete support and/or the measurement matrix
$ \bs{H} $
is correlated.

(ii)
The dynamics of VASP are well characterized by the SE we present.
Interestingly, the SE fixed point equations align with the saddle points of the 1RSB free energy derived by Takahashi and Kabashima \cite{takahashi2022macroscopic}.
This correspondence suggests that, once the 1RSB ansatz is valid and the SE fixed point is reached, VASP can accurately compute estimator \eqref{Eq:Postulated_Posterior_Estimator} in LSL.

This paper is organized as follows:
Section \ref{sec:Reformulation} reformulates the 1RSB prediction of Takahashi and Kabashima, leveraging the Gaussian convolution formula. 
Section \ref{sec:VASP_algo} analyzes the reformulated result and designs accordingly the VASP algorithm. 
Section \ref{sec:SE} presents an SE that characterizes the MSE performance of the algorithm.
Section \ref{sec:simulation} validates the effectiveness of the algorithm and its SE via Monte Carlo simulations,
and the last section concludes the paper.
Throughout the paper, we use the following notation conventions:
Plain letters (e.g.,
$ m $
and
$ v $)
represent scalars.
Boldface lowercase letters (e.g.,
$ \bs{m} $
and
$ \bs{v} $)
represent column vectors.
Boldface lowercase letters with a vector arrow (e.g.,
$ \vec{\bs{m}} $
and
$ \vec{\bs{v}} $)
represent row vectors.
Boldface uppercase letters (e.g.,
$ \bs{V} $
and
$ \bs{C} $)
represent matrices.
$ \EXP(\cdot) $
denotes the exponential operator.
$ \delta(\cdot) $
represents a Dirac delta function.
$ \Normal[ \bs{x} | \bs{m}, \bs{C} ] $
denotes a Gaussian distribution with mean
$ \bs{m} $
and covariance
$ \bs{C} $,
defined as
$
\Normal[ \bs{x} | \bs{m}, \bs{C} ]
\triangleq
| 2 \pi \bs{C} |^{ - \frac{1}{2} }
\EXP[
	-
	\frac{1}{2}
	( \bs{x} - \bs{m} )^{\Ts}
	\bs{C}^{-1}
	( \bs{x} - \bs{m} )
]
$.
For any matrix
$ \bs{A} $,
$ A^{ ( i, j ) } $
represents the element at the
$ i $-th row and
$ j $-th column,
$ \bs{A}^{\Ts} $
denotes the transpose of matrix
$ \bs{A} $,
$ \Trace(\bs{A}) $
represents the trace of matrix
$ \bs{A} $.
$ \bs{e}_{N} $
is a
$ N $-dimensional column vector with all zeros except for a
$ 1 $
in the first element.
$ \bs{1}_{N} $
is a
$ N $-dimensional column vector consisting of all ones.
$\Eye_{N}$
represents the identity matrix of size
$ N $.
$ \mathbb{1}_{N} $
is a matrix of size
$ N $
with all entries equal to one.
$ \Diag(\bs{v}) $
is a diagonal matrix with diagonal elements equal to the entries of vector
$ \bs{v} $.
$ \diag(\bs{C}) $
is a diagonal operator that returns an
$ N $-dimensional column vector containing the diagonal elements of the matrix
$ \bs{C} $.

\section{Reformulation of Kabashima's Result}
\label{sec:Reformulation}
\subsection{Revisit to Kabashima's Replica Prediction}

In \cite{takahashi2022macroscopic}, Takahashi and Kabashima established a connection between the VAMP algorithm and the theoretical estimator \eqref{Eq:Postulated_Posterior_Estimator}, which is NP-hard, using the replica method.
They began by calculating the average free energy:
\begin{align*}
f
& \triangleq
-
\lim_{ N \rightarrow \infty }
\frac{1}{ N \beta }
\Mean_{
	\bs{y}, \bs{H}, \bs{x}_{0}
}[
	\log Z( \bs{y} )
]
.
\end{align*}
This average free energy serves as the cumulant generating function \cite{mezard2009information}, capturing all the cumulants of the Boltzmann distribution \eqref{Eq:Postulated_Posterior}.
Using the replica method, they expressed the average free energy as follows \cite{takahashi2022macroscopic}:
\begin{align*}
f
& =
-
\lim_{ \tau \rightarrow 0 }
\frac{1}{\tau}
\Extr_{
	\bs{Q}_{\Sf{z}},
	\bs{Q}_{\Sf{x}}
}[
	g_{\Sf{z}}
	+
	g_{\Sf{h}}
	+
	g_{\Sf{x}}
	-
	g_{\Sf{e}}
]
, \\
g_{\Sf{z}}
& \triangleq
\Extr_{
	\tilde{\bs{Q}}_{\Sf{z}}
}[
	\frac{\alpha}{\beta}
	\log \int{
		\dd y
		\dd \vec{\bs{z}}
	} \,
	p(
		y \vec{\bs{1}} | \vec{\bs{z}}
	)
	\EXP(
		-
		\frac{
			\vec{\bs{z}}
			\tilde{\bs{Q}}_{\Sf{z}}
			\vec{\bs{z}}^{\Ts}
		}{2}
	)
	+
	\frac{
		\alpha
		\Trace(
			\bs{Q}_{\Sf{z}}
			\tilde{\bs{Q}}_{\Sf{z}}
		)
	}{2 \beta}
]
, \\
g_{\Sf{h}}
& \triangleq
\Extr_{
	\bs{\Lambda}_{\Sf{z}},
	\bs{\Lambda}_{\Sf{x}}
}[
	\frac{
		\alpha
		\Trace(
			\bs{Q}_{\Sf{z}}
			\bs{\Lambda}_{\Sf{z}}
		)
	}{2 \beta}
	-
	\frac{
		\Mean_{\lambda}[
			\log
			|
				\bs{\Lambda}_{\Sf{x}}
				+
				\lambda \bs{\Lambda}_{\Sf{z}}
			|
		]
	}{2 \beta}
	+
	\frac{
		\Trace(
			\bs{Q}_{\Sf{x}}
			\bs{\Lambda}_{\Sf{x}}
		)
	}{2 \beta}
]
, \\
g_{\Sf{x}}
& \triangleq
\Extr_{
	\tilde{\bs{Q}}_{\Sf{x}}
}[
	\frac{1}{\beta}
	\log
	\int{
		\dd \vec{\bs{x}}
	} \,
	\EXP(
		-
		\frac{
			\vec{\bs{x}}
			\tilde{\bs{Q}}_{\Sf{x}}
			\vec{\bs{x}}^{\Ts}
		}{2}
	)
	p(
		\vec{\bs{x}}
	)
	+
	\frac{
		\Trace(
			\bs{Q}_{\Sf{x}}
			\tilde{\bs{Q}}_{\Sf{x}}
		)
	}{2 \beta}
]
, \\
g_{\Sf{e}}
& \triangleq
\frac{\alpha}{2 \beta}
\log
| \bs{Q}_{\Sf{z}} |
+
\frac{1}{2 \beta}
\log
| \bs{Q}_{\Sf{x}} |
.
\end{align*}
Here,
$ \bs{Q}_{\Sf{z}} $,
$ \bs{Q}_{\Sf{x}} $,
$ \bs{\Lambda}_{\Sf{z}} $,
$ \bs{\Lambda}_{\Sf{x}} $,
$ \tilde{\bs{Q}}_{\Sf{z}} $,
$ \tilde{\bs{Q}}_{\Sf{x}} $
are
$ (\tau + 1) \times (\tau + 1) $
real symmetric matrices,
$ \Extr[ \cdot ] $
denotes an extremum operator,
$ \lambda $
follows the limiting eigenvalue distribution of
$ \bs{H}^{\Ts} \bs{H} $,
and the integration measures
$ \dd \vec{\bs{z}} $
and
$ \dd \vec{\bs{x}} $
are defined as:
$
\dd \vec{\bs{z}}
\triangleq
\dd z_{0}
\dd \tilde{z}_{1}
\cdots
\dd \tilde{z}_{\tau}
$,
$
\dd \vec{\bs{x}}
\triangleq
\dd x_{0}
\dd \tilde{x}_{1}
\cdots
\dd \tilde{x}_{\tau}
$
with
$
\vec{\bs{z}}
\triangleq
[
	z_{0},
	\tilde{z}_{1},
	\cdots,
	\tilde{z}_{\tau}
]
$,
$
\vec{\bs{x}}
\triangleq
[
	x_{0},
	\tilde{x}_{1},
	\cdots,
	\tilde{x}_{\tau}
]
$,
$
p(
	y \vec{\bs{1}}_{\tau + 1} |
	\vec{\bs{z}}
)
\triangleq
p( y | z_{0} )
\prod_{i = 1}^{\tau}
q^{\beta}( y | \tilde{z}_{i} )
$,
and
$
p( \vec{\bs{x}} )
\triangleq
p( x_{0} )
\prod_{i = 1}^{\tau}
q^{\beta}( \tilde{x}_{i} )
$.
The extremum conditions were then formulated as in \cite[Eqs. (124)-(126)]{takahashi2022macroscopic}:
\begin{subequations}
\begin{align}
\bs{0}
& =
\bs{\Lambda}_{\Sf{z}}
-
\bs{Q}_{\Sf{z}}^{-1}
+
\tilde{\bs{Q}}_{\Sf{z}}
, \label{Eq:Extr_Cond:Z} \\
\bs{0}
& =
\bs{\Lambda}_{\Sf{x}}
-
\bs{Q}_{\Sf{x}}^{-1}
+
\tilde{\bs{Q}}_{\Sf{x}}
, \label{Eq:Extr_Cond:X} \\
\bs{Q}_{\Sf{z}}
& =
\frac{1}{\alpha}
\Mean_{\lambda}[
	\lambda (
		\bs{\Lambda}_{\Sf{x}}
		+
		\lambda
		\bs{\Lambda}_{\Sf{z}}
	)^{-1}
]
, \label{Eq:Extr_Cond:Exp_Z} \\
\bs{Q}_{\Sf{x}}
& =
\Mean_{\lambda}[
	(
		\bs{\Lambda}_{\Sf{x}}
		+
		\lambda
		\bs{\Lambda}_{\Sf{z}}
	)^{-1}
]
, \label{Eq:Extr_Cond:Exp_X} \\
\bs{Q}_{\Sf{z}}
& =
\frac{
	\int{
		\dd y
		\dd \vec{\bs{z}}
	} \,
	p(
		y \vec{\bs{1}} | \vec{\bs{z}}
	)
	\EXP(
		-
		\frac{1}{2}
		\vec{\bs{z}}
		\tilde{\bs{Q}}_{\Sf{z}}
		\vec{\bs{z}}^{\Ts}
	)
	\vec{\bs{z}}^{\Ts} \vec{\bs{z}}
}{
	\int{
		\dd y
		\dd \vec{\bs{z}}
	} \,
	p(
		y \vec{\bs{1}} | \vec{\bs{z}}
	)
	\EXP(
		-
		\frac{1}{2}
		\vec{\bs{z}}
		\tilde{\bs{Q}}_{\Sf{z}}
		\vec{\bs{z}}^{\Ts}
	)
}
, \label{Eq:Extr_Cond:Int_Z} \\
\bs{Q}_{\Sf{x}}
& =
\frac{
	\int{
		\dd \vec{\bs{x}}
	} \,
	\EXP(
		-
		\frac{1}{2}
		\vec{\bs{x}}
		\tilde{\bs{Q}}_{\Sf{x}}
		\vec{\bs{x}}^{\Ts}
	)
	p(
		\vec{\bs{x}}
	)
	\vec{\bs{x}}^{\Ts} \vec{\bs{x}}
}{
	\int{
		\dd \vec{\bs{x}}
	} \,
	\EXP(
		-
		\frac{1}{2}
		\vec{\bs{x}}
		\tilde{\bs{Q}}_{\Sf{x}}
		\vec{\bs{x}}^{\Ts}
	)
	p(
		\vec{\bs{x}}
	)
}
. \label{Eq:Extr_Cond:Int_X}
\end{align}
\label{Eq:Extreme_Condition}
\end{subequations}

Takahashi and Kabashima demonstrated the following conclusions \cite{takahashi2022macroscopic}:

(i)
VAMP-RS correspondence:
Under the RS ansatz, there is a correspondence between the SE fixed point equations of VAMP and the saddle point equations of the estimator \eqref{Eq:Postulated_Posterior_Estimator}.
VAMP implements \eqref{Eq:Postulated_Posterior_Estimator} in the LSL, provided that the RS ansatz holds and SE converges to a fixed point.

(ii)
VAMP Limitation:
VAMP is unsuitable for estimating mismatched problems where RSB emerges.

(iii)
1RSB Prediction:
Under the 1RSB ansatz, the saddle point equations for the average free energy were given in \cite[Eqs. (137)-(142), (168)-(174)]{takahashi2022macroscopic}, where the matrices follow a common structure, expressed as:
\begin{subequations}
\begin{align}
\bs{Q}_{\Sf{z}}
& =
\mathcal{Q}^{ \tau + 1 }_{ \tilde{L} }
(
	C_{\Sf{z}}, D_{\Sf{z}}, F_{\Sf{z}},
	\frac{ \chi_{\Sf{z}} }{\beta},
	H_{\Sf{z, 1}}
)
, \\
\bs{Q}_{\Sf{x}}
& =
\mathcal{Q}^{ \tau + 1 }_{ \tilde{L} }
(
	C_{\Sf{x}}, D_{\Sf{x}}, F_{\Sf{x}},
	\frac{ \chi_{\Sf{x}} }{\beta},
	H_{\Sf{x, 1}}
)
, \label{eq:def_Dx} \\
\bs{\Lambda}_{\Sf{z}}
& =
\mathcal{Q}^{ \tau + 1 }_{ \tilde{L} }
(
	\hat{C}_{\Sf{2z}},
	- \beta \hat{D}_{\Sf{2z}},
	- \beta^{2} \hat{F}_{\Sf{2z}},
	\beta \hat{\chi}_{\Sf{2z}},
	- \beta^{2} \hat{H}_{\Sf{2z, 1}}
)
, \\
\bs{\Lambda}_{\Sf{x}}
& =
\mathcal{Q}^{ \tau + 1 }_{ \tilde{L} }
(
	\hat{C}_{\Sf{2x}},
	- \beta \hat{D}_{\Sf{2x}},
	- \beta^{2} \hat{F}_{\Sf{2x}},
	\beta \hat{\chi}_{\Sf{2x}},
	- \beta^{2} \hat{H}_{\Sf{2x, 1}}
)
, \\
\tilde{\bs{Q}}_{\Sf{z}}
& =
\mathcal{Q}^{ \tau + 1 }_{ \tilde{L} }
(
	\hat{C}_{\Sf{1z}},
	- \beta \hat{D}_{\Sf{1z}},
	- \beta^{2} \hat{F}_{\Sf{1z}},
	\beta \hat{\chi}_{\Sf{1z}},
	- \beta^{2} \hat{H}_{\Sf{1z, 1}}
)
, \\
\tilde{\bs{Q}}_{\Sf{x}}
& =
\mathcal{Q}^{ \tau + 1 }_{ \tilde{L} }
(
	\hat{C}_{\Sf{1x}},
	- \beta \hat{D}_{\Sf{1x}},
	- \beta^{2} \hat{F}_{\Sf{1x}},
	\beta \hat{\chi}_{\Sf{1x}},
	- \beta^{2} \hat{H}_{\Sf{1x, 1}}
)
,
\end{align}
\end{subequations}
with
$
\tilde{L}
\triangleq
\frac{L}{\beta}
$,
$
\mathcal{A}^{\tau}_{ \tilde{L} }(
	F, \chi, H
)
\triangleq
F \mathbb{1}_{\tau}
+
\chi \Eye_{\tau}
+
H \Eye_{ \frac{ \tau }{ \tilde{L} } }
\otimes
\mathbb{1}_{ \tilde{L} }
$,
$
\mathcal{Q}^{ \tau + 1 }_{ \tilde{L} }(
	C, D, F, \chi, H
)
\triangleq
\left[
\begin{array}{cc}
	C
	&
	D \vec{\bs{1}}_{\tau}
	\\
	D \bs{1}_{\tau}
	&
	\mathcal{A}^{\tau}_{
		\tilde{L}
	}( F, \chi, H )
	\\
\end{array}
\right]
$.
Here, we highlight a limitation in Kabashima's 1RSB prediction.
Specifically, the quantity
$ D_{\Sf{x}} $
in \eqref{eq:def_Dx} does not directly correspond to any practical estimator.
In previous works \cite{zou2021multi,zou2023high},
$ D_{\Sf{x}} $
was referred to as the overlap \cite{mezard1987spin} between the ground truth
$x_{0}$
and its estimate, expressed as:
$
D_{\Sf{x}}
=
\Mean_{
	x_{0}, \mu_{\Sf{x}}
}[ x_{0} \hat{x}( \mu_{\Sf{x}} ) ]
$,
where the overlap serves as the link between the theoretical results and practical AMP variants
$ \hat{x} $.
However, in Kabashima's approach,
$ D_{\Sf{x}} $
is formulated differently, as shown in \cite[Eq. (171)]{takahashi2022macroscopic}:
\begin{align}
D_{\Sf{x}}
& =
\Mean_{
	x_{0}, \xi
}[ x_{0} \bar{x}( x_{0}, \xi ) ]
, \label{Eq:Overlap_D}
\end{align}
where
$ \bar{x}(x_{0}, \xi) $
explicitly depends on
$ x_{0} $.
Clearly,
$ \bar{x}(x_{0}, \xi) $
is not a practical estimator because no estimator can take the ground truth
$ x_{0} $
as an input prior to performing the estimation.

\subsection{Reformulation of the 1RSB Saddle Point Equations}

To address this limitation, we reformulate the results of \cite{takahashi2022macroscopic} by applying a Gaussian density convolution technique \cite{bromiley2003products} to the extremum conditions.
This yields an overlap
$ D_{\Sf{x}} $
that closely resembles the expressions found in \cite{zou2021multi,zou2023high}.
Below is an outline of the derivation.
For a more detailed explanation, readers are encouraged to refer to the supplementary material of this paper.
For convenience, all relevant definitions are summarized in Tab. \ref{Tab:notations} in Appendix \ref{Appendix:Summary}.
The reformulation process involves five key steps:

(S1)
Under the 1RSB ansatz, equations \eqref{Eq:Extr_Cond:Z}–\eqref{Eq:Extr_Cond:X} are reformulated as shown below, in conjunction with Lines 7–14 of Tab. \ref{Tab:Saddle_Point}.
\begin{align}
0
& =
\hat{C}_{\Sf{1z}}
+
\hat{C}_{\Sf{2z}}
-
\frac{1}{C_{\Sf{z}}}
, \label{Eq:HCz_Equal} \\
0
& =
\hat{C}_{\Sf{1x}}
+
\hat{C}_{\Sf{2x}}
-
\frac{1}{C_{\Sf{x}}}
. \label{Eq:HCx_Equal}
\end{align}

\begin{figure*}[!t]
\small
\begin{align}
&
\EXP(
	-
	\frac{1}{2}
	\vec{\bs{x}}
	\tilde{\bs{Q}}_{\Sf{x}}
	\vec{\bs{x}}^{\Ts}
)
\overset{ \text{(a)} }{=}
\sqrt{
	\frac{\eta_{\Sf{x}}}{\pi}
}
\int{ \dd \mu_{\Sf{x}} } \,
\EXP(
	-
	\eta_{\Sf{x}} \mu_{\Sf{x}}^{2}
)
b_{\Sf{x}}
\prod_{i = 1}^{
	\frac{\tau}{ \tilde{L} }
}
\EXP
\left[
	-
	\frac{1}{2}
	\beta \hat{\chi}_{\Sf{1x}}
	\sum_{j = 1}^{ \tilde{L} }
	\tilde{x}_{
		\tilde{L} (i - 1) + j
	}^{2}
	+
	\imagUnit
	\sqrt{
		2 \eta_{\Sf{x}}
		(
			-
			\beta^{2} \hat{F}_{\Sf{1x}}
		)
	}
	\mu_{\Sf{x}}
	\sum_{j = 1}^{ \tilde{L} }
	\tilde{x}_{
		\tilde{L} (i - 1) + j
	}
	-
	\frac{1}{2}
	(
		-
		\beta^{2} \hat{H}_{\Sf{1x, 1}}
	)
	(
		\sum_{j = 1}^{ \tilde{L} }
		\tilde{x}_{
			\tilde{L} (i - 1) + j
		}
	)^{2}
\right]
\nonumber \\
&
\overset{ \text{(b)} }{=}
\sqrt{
	\frac{\eta_{\Sf{x}}}{\pi}
}
\int{\dd \mu_{\Sf{x}}} \,
\EXP(
	-
	\eta_{\Sf{x}} \mu_{\Sf{x}}^{2}
)
b_{\Sf{x}}
\prod_{i = 1}^{
	\frac{\tau}{ \tilde{L} }
}
\underbrace{
	(2 \pi)^{
		\frac{ \tilde{L} }{2}
	}
	[
		(
			\beta \hat{\chi}_{\Sf{1x}}
		)^{
			\tilde{L} - 1
		}
		(
			\beta \hat{\chi}_{\Sf{1x}}
			-
			\beta^{2} \tilde{L}
			\hat{H}_{\Sf{1x, 1}}
		)
	]^{
		-
		\frac{1}{2}
	}
	\EXP\left[
		\frac{
			-
			\tilde{L} \eta_{\Sf{x}}
			(
				-
				\beta^{2} \hat{F}_{\Sf{1x}}
			)
			\mu_{\Sf{x}}^{2}
		}{
			\beta \hat{\chi}_{\Sf{1x}}
			-
			\beta^{2} \tilde{L}
			\hat{H}_{\Sf{1x, 1}}
		}
	\right]
}_{
	\triangleq C( \mu_{\Sf{x}} )
}
\int{\dd m} \,
\Normal[
	m | m_{\Sf{x}}, \Delta_{\Sf{x1}}
]
\prod_{j = 1}^{ \tilde{L} }{}
\Normal[
	\tilde{x}_{
		\tilde{L} (i - 1) + j
	}
	| m,
	\frac{1}{ \beta }
	\Delta_{\Sf{x0}}
]
\label{Eq:Saddle_Point:Eq_1}
\end{align}
\hrule
\end{figure*}

(S2)
Likewise, equations \eqref{Eq:Extr_Cond:Exp_Z}–\eqref{Eq:Extr_Cond:Exp_X} are transformed into Lines 17–18 and 21–22 in Tab. \ref{Tab:Saddle_Point}, with the definitions:
$
\theta_{0}
\triangleq
\hat{\chi}_{\Sf{2x}}
+
\lambda
\hat{\chi}_{\Sf{2z}}
$,
$
\theta
\triangleq
(
	\hat{\chi}_{\Sf{2x}}
	-
	L \hat{H}_{\Sf{2x, 1}}
)
+
\lambda
(
	\hat{\chi}_{\Sf{2z}}
	-
	L \hat{H}_{\Sf{2z, 1}}
)
$,
along with the following expressions:
\begin{align}
C_{\Sf{z}}
& =
\frac{1}{\alpha}
\Mean_{\lambda}[
	\frac{\lambda}{
		\hat{C}_{\Sf{2x}}
		+
		\lambda
		\hat{C}_{\Sf{2z}}
	}
]
, \label{Eq:Saddle_Point:Exp_Cz} \\
D_{\Sf{z}}
& =
\frac{1}{\alpha}
\Mean_{\lambda}[
	\frac{
		\lambda
		(
			\hat{D}_{\Sf{2x}}
			+
			\lambda
			\hat{D}_{\Sf{2z}}
		)
	}{
		(
			\hat{C}_{\Sf{2x}}
			+
			\lambda
			\hat{C}_{\Sf{2z}}
		)
		\theta
	}
]
, \label{Eq:Saddle_Point:Exp_Dz} \\
F_{\Sf{z}}
& =
\frac{1}{\alpha}
\Mean_{\lambda}[
	\frac{
		\lambda
		(
			\hat{D}_{\Sf{2x}}
			+
			\lambda
			\hat{D}_{\Sf{2z}}
		)^{2}
	}{
		(
			\hat{C}_{\Sf{2x}}
			+
			\lambda
			\hat{C}_{\Sf{2z}}
		)
		\theta^{2}
	}
]
+
\frac{1}{\alpha}
\Mean_{\lambda}[
	\frac{
		\lambda
		(
			\hat{F}_{\Sf{2x}}
			+
			\lambda
			\hat{F}_{\Sf{2z}}
		)
	}{
		\theta^{2}
	}
]
, \label{Eq:Saddle_Point:Exp_Fz} \\
C_{\Sf{x}}
& =
\Mean_{\lambda}[
	\frac{1}{
		\hat{C}_{\Sf{2x}}
		+
		\lambda
		\hat{C}_{\Sf{2z}}
	}
]
, \label{Eq:Saddle_Point:Exp_Cx}\\
D_{\Sf{x}}
& =
\Mean_{\lambda}[
	\frac{
		\hat{D}_{\Sf{2x}}
		+
		\lambda
		\hat{D}_{\Sf{2z}}
	}{
		(
			\hat{C}_{\Sf{2x}}
			+
			\lambda
			\hat{C}_{\Sf{2z}}
		)
		\theta
	}
]
, \label{Eq:Saddle_Point:Exp_Dx} \\
F_{\Sf{x}}
& =
\Mean_{\lambda}[
	\frac{
		(
			\hat{D}_{\Sf{2x}}
			+
			\lambda
			\hat{D}_{\Sf{2z}}
		)^{2}
	}{
		(
			\hat{C}_{\Sf{2x}}
			+
			\lambda
			\hat{C}_{\Sf{2z}}
		)
		\theta^{2}
	}
]
+
\Mean_{\lambda}[
	\frac{
		\hat{F}_{\Sf{2x}}
		+
		\lambda
		\hat{F}_{\Sf{2z}}
	}{
		\theta^{2}
	}
]
. \label{Eq:Saddle_Point:Exp_Fx}
\end{align}

(S3)
The reformulation of \eqref{Eq:Extr_Cond:Int_X} begins by expanding the term
$
\EXP(
	-
	\frac{1}{2}
	\vec{\bs{x}}
	\tilde{\bs{Q}}_{\Sf{x}}
	\vec{\bs{x}}^{\Ts}
)
$:
\begin{align*}
&
\EXP(
	-
	\frac{1}{2}
	\vec{\bs{x}}
	\tilde{\bs{Q}}_{\Sf{x}}
	\vec{\bs{x}}^{\Ts}
)
=
\EXP
\left\{
	-
	\frac{1}{2}
	\left[
		\hat{C}_{\Sf{1x}}
		-
		\frac{
			(
				-
				\beta \hat{D}_{\Sf{1x}}
			)^{2}
		}{
			-
			\beta^{2} \hat{F}_{\Sf{1x}}
		}
	\right]
	x_{0}^{2}
\right\}
\times
\nonumber \\
&
\quad \quad
\underbrace{
	\EXP
	\left[
		-
		\frac{1}{2}
		\left(
			\frac{
				-
				\beta \hat{D}_{\Sf{1x}}
			}{
				\sqrt{
					-
					\beta^{2} \hat{F}_{\Sf{1x}}
				}
			}
			x_{0}
			+
			\sum_{i = 1}^{\tau}
			\sqrt{
				-
				\beta^{2} \hat{F}_{\Sf{1x}}
			}
			\tilde{x}_{i}
		\right)^{2}
	\right]
}_{ E_{1} }
\times
\nonumber \\
&
\quad \quad
\EXP
(
	-
	\frac{1}{2}
	\beta \hat{\chi}_{\Sf{1x}}
	\sum_{i = 1}^{\tau}
	\tilde{x}_{i}^{2}
)
\underbrace{
	\EXP
	[
		-
		\frac{1}{2}
		(
			-
			\beta^{2} \hat{H}_{\Sf{1x, 1}}
		)
		\sum_{i = 1}^{
			\frac{\tau}{ \tilde{L} }
		}
		(
			\sum_{j = 1}^{ \tilde{L} }
			\tilde{x}_{
				\tilde{L} (i - 1) + j
			}
		)^{2}
	]
}_{ E_{2} }
,
\end{align*}
where the equality is derived using Lemma \ref{lemma:Decom_Formula}.
The primary technical challenge here arises from the integral in \eqref{Eq:Extr_Cond:Int_X}, involving the terms
$ E_{1} $
and
$ E_{2} $.
These terms contain multiple cross terms, making the integration process highly complex and seemingly intractable.

A standard approach to address this complexity is to apply the Hubbard-Stratonovich (H-S) transformation \cite{hubbard1959calculation,stratonovich1957method,tulino2013support}, which states that
$
\EXP( - x^{2} )
=
\sqrt{\frac{\eta}{\pi}}
\int{ \dd \mu } \,
\EXP(
	-
	\eta \mu^{2}
	+
	2 \imagUnit \sqrt{\eta} x \mu
)
$,
where
$ \eta $
is a positive real constant.
In \cite{takahashi2022macroscopic}, Takahashi and Kabashima applied the H-S transformation twice, once for
$ E_{1} $
and once for
$ E_{2} $.
In this paper, we apply the transformation only once, and additionally use a formula for the convolution of Gaussian densities \cite{bromiley2003products}.
Our result can then be expressed as \eqref{Eq:Saddle_Point:Eq_1}, where (a) follows from the H-S transformation, and (b) from the Gaussian density convolution \cite{bromiley2003products}.
The quantities
$ m_{\Sf{x}} $,
$ \Delta_{\Sf{x0}} $,
and
$ \Delta_{\Sf{x1}} $
are defined in Lines 1-3 of Tab. \ref{Tab:notations}.
At this stage, the denominator of \eqref{Eq:Extr_Cond:Int_X} can be rewritten as
\begin{align*}
&
\quad
\int{ \dd \vec{\bs{x}} } \,
\EXP(
	-
	\frac{1}{2}
	\vec{\bs{x}}
	\tilde{\bs{Q}}_{\Sf{x}}
	\vec{\bs{x}}^{\Ts}
)
p( \vec{\bs{x}} )
\\
& =
\sqrt{
	\frac{ \eta_{\Sf{x}} }{\pi}
}
\int{
	\dd \mu_{\Sf{x}}
} \,
\EXP(
	-
	\eta_{\Sf{x}} \mu_{\Sf{x}}^{2}
)
[
	\int{
		\dd x_{0}
	} \,
	b_{\Sf{x}}
	p( x_{0} )
]
\EXP\{
	\frac{\tau}{ \tilde{L} }
	[
		\log
		C( \mu_{\Sf{x}} )
		+
		g_{\Sf{x1}}
	]
\}
,
\end{align*}
with
$ b_{\Sf{x}} $,
$ \langle a \rangle $,
$ g_{\Sf{x0}} $,
$ g_{\Sf{x1}} $
defined in Lines 4-7 of Tab. \ref{Tab:notations}.

The extremum condition in \eqref{Eq:Extr_Cond:Int_X} is now transformed into
\begin{align}
C_{\Sf{x}}
& =
\frac{1}{C}
\int{
	\dd \mu_{\Sf{x}}
} \,
\EXP(
	-
	\eta_{\Sf{x}} \mu_{\Sf{x}}^{2}
)
[
	\int{
		\dd x_{0}
	} \,
	b_{\Sf{x}}
	p( x_{0} )
	x_{0}^{2}
]
, \label{Eq:Saddle_Point:Int_Cx} \\
D_{\Sf{x}}
& =
\frac{1}{C}
\int{
	\dd \mu_{\Sf{x}}
} \,
\EXP(
	-
	\eta_{\Sf{x}} \mu_{\Sf{x}}^{2}
)
[
	\int{
		\dd x_{0}
	} \,
	b_{\Sf{x}}
	p( x_{0} )
	x_{0}
]
\times
\nonumber \\
& \quad
\frac{
	\int{ \dd m } \,
	\Normal[
		m | m_{\Sf{x}}, \Delta_{\Sf{x1}}
	]
	\EXP[
		\tilde{L} g_{\Sf{x0}}
	]
	\langle \tilde{x} \rangle
}{
	\EXP[
		g_{\Sf{x1}}
	]
}
, \label{Eq:Saddle_Point:Int_Dx} \\
F_{\Sf{x}}
& =
\frac{1}{C}
\int{
	\dd \mu_{\Sf{x}}
} \,
\EXP(
	-
	\eta_{\Sf{x}} \mu_{\Sf{x}}^{2}
)
[
	\int{
		\dd x_{0}
	} \,
	b_{\Sf{x}}
	p( x_{0} )
]
\times
\nonumber \\
& \quad
\left(
	\frac{
		\int{\dd m} \,
		\Normal[
			m | m_{\Sf{x}}, \Delta_{\Sf{x1}}
		]
		\EXP[
			\tilde{L} g_{\Sf{x0}}
		]
		\langle \tilde{x} \rangle
	}{
		\EXP[
			g_{\Sf{x1}}
		]
	}
\right)^{2}
, \label{Eq:Saddle_Point:Int_Fx} \\
H_{\Sf{x, 1}}
& =
\frac{1}{C}
\int{
	\dd \mu_{\Sf{x}}
} \,
\EXP(
	-
	\eta_{\Sf{x}} \mu_{\Sf{x}}^{2}
)
[
	\int{
		\dd x_{0}
	} \,
	b_{\Sf{x}}
	p( x_{0} )
]
\times
\nonumber \\
& \quad
\frac{
	\int{\dd m} \,
	\Normal[
		m | m_{\Sf{x}}, \Delta_{\Sf{x1}}
	]
	\EXP[
		\tilde{L} g_{\Sf{x0}}
	]
	\langle \tilde{x} \rangle^{2}
}{
	\EXP[
		g_{\Sf{x1}}
	]
}
-
F_{\Sf{x}}
, \label{Eq:Saddle_Point:Int_Fx_Hx} \\
\frac{
	\chi_{\Sf{x}}
}{
	\beta
}
& =
\frac{1}{C}
\int{
	\dd \mu_{\Sf{x}}
} \,
\EXP(
	-
	\eta_{\Sf{x}} \mu_{\Sf{x}}^{2}
)
[
	\int{
		\dd x_{0}
	} \,
	b_{\Sf{x}}
	p( x_{0} )
]
\times
\nonumber \\
& \quad
\frac{
	\int{\dd m} \,
	\Normal[
		m | m_{\Sf{x}}, \Delta_{\Sf{x1}}
	]
	\EXP[
		\tilde{L} g_{\Sf{x0}}
	]
	\langle \tilde{x}^{2} \rangle
}{
	\EXP[
		g_{\Sf{x1}}
	]
}
-
F_{\Sf{x}}
-
H_{\Sf{x, 1}}
, \label{Eq:Saddle_Point:Int_chix_Fx_Hx}
\end{align}
with
$
C
\triangleq
\sqrt{
	\frac{
		\pi
	}{
		\eta_{\Sf{x}}
	}	
}
\int{ \dd x_{0} } \,
\EXP(
	-
	\frac{
		\hat{C}_{\Sf{1x}}
	}{2}
	x_{0}^{2}
)
p( x_{0} )
$.

(S4)
The condition in \eqref{Eq:Extr_Cond:Int_Z} can be similarly transformed into the expressions shown below and in Line 5 of Tab. \ref{Tab:Saddle_Point}
\begin{align}
D_{\Sf{z}}
& =
\sqrt{
	\frac{
		\eta_{\Sf{z}}
		\hat{C}_{\Sf{1z}}
	}{
		2 \pi^{2}
	}
}
\int{
	\dd y \dd \mu_{\Sf{z}}
} \,
\EXP(
	-
	\eta_{\Sf{z}} \mu_{\Sf{z}}^{2}
)
[
	\int{
		\dd z_{0}
	} \,
	b_{\Sf{z}}
	p( y | z_{0} )
	z_{0}
]
\times
\nonumber \\
& \quad
\frac{
	\int{\dd m} \,
	\Normal[
		m | m_{\Sf{z}}, \Delta_{\Sf{z1}}
	]
	\EXP[
		\tilde{L} g_{\Sf{z0}}
	]
	\langle \tilde{z} \rangle
}{
	\EXP[
		g_{\Sf{z1}}
	]
}
, \label{Eq:Saddle_Point:Int_Dz} \\
F_{\Sf{z}}
& =
\sqrt{
	\frac{
		\eta_{\Sf{z}}
		\hat{C}_{\Sf{1z}}
	}{
		2 \pi^{2}
	}
}
\int{
	\dd y \dd \mu_{{z}}
} \,
\EXP(
	-
	\eta_{\Sf{z}} \mu_{\Sf{z}}^{2}
)
[
	\int{
		\dd z_{0}
	} \,
	b_{\Sf{z}}
	p( y | z_{0} )
]
\times
\nonumber \\
& \quad
\left(
	\frac{
		\int{\dd m} \,
		\Normal[
			m | m_{\Sf{z}}, \Delta_{\Sf{z1}}
		]
		\EXP[
			\tilde{L} g_{\Sf{z0}}
		]
		\langle \tilde{z} \rangle
	}{
		\EXP[
			g_{\Sf{z1}}
		]
	}
\right)^{2}
, \label{Eq:Saddle_Point:Int_Fz} \\
H_{\Sf{z, 1}}
& =
\sqrt{
	\frac{
		\eta_{\Sf{z}}
		\hat{C}_{\Sf{1z}}
	}{
		2 \pi^{2}
	}
}
\int{
	\dd y \dd \mu_{\Sf{z}}
} \,
\EXP(
	-
	\eta_{\Sf{z}} \mu_{\Sf{z}}^{2}
)
[
	\int{
		\dd z_{0}
	} \,
	b_{\Sf{z}}
	p( y | z_{0} )
]
\times
\nonumber \\
& \quad
\frac{
	\int{\dd m} \,
	\Normal[
		m | m_{\Sf{z}}, \Delta_{\Sf{z1}}
	]
	\EXP[
		\tilde{L} g_{\Sf{z0}}
	]
	\langle \tilde{z} \rangle^{2}
}{
	\EXP[
		g_{\Sf{z1}}
	]
}
-
F_{\Sf{z}}
, \label{Eq:Saddle_Point:Int_Fz_Hz} \\
\frac{
	\chi_{\Sf{z}}
}{
	\beta
}
& =
\sqrt{
	\frac{
		\eta_{\Sf{z}}
		\hat{C}_{\Sf{1z}}
	}{
		2 \pi^{2}
	}
}
\int{
	\dd y \dd \mu_{\Sf{z}}
} \,
\EXP(
	-
	\eta_{\Sf{z}} \mu_{\Sf{z}}^{2}
)
[
	\int{
		\dd z_{0}
	} \,
	b_{\Sf{z}}
	p( y | z_{0} )
]
\times
\nonumber \\
& \quad
\frac{
	\int{\dd m} \,
	\Normal[
		m | m_{\Sf{z}}, \Delta_{\Sf{z1}}
	]
	\EXP[
		\tilde{L} g_{\Sf{z0}}
	]
	\langle \tilde{z}^{2} \rangle
}{
	\EXP[
		g_{\Sf{z1}}
	]
}
-
F_{\Sf{z}}
-
H_{\Sf{z, 1}}
, \label{Eq:Saddle_Point:Int_chiz_Fz_Hz}
\end{align}
with
$ m_{\Sf{z}} $,
$ \Delta_{\Sf{z0}} $,
$ \Delta_{\Sf{z1}} $,
$ b_{\Sf{z}} $,
$ \langle a \rangle $,
$ g_{\Sf{z0}} $,
$ g_{\Sf{z1}} $
from Lines 1-7 of Tab. \ref{Tab:notations}.

(S5)
So far, we have derived Lines 5, 7–14, 17–18, and 21–22 of Tab. \ref{Tab:Saddle_Point}.
To obtain the remaining lines, follow these steps:
(i)
Substitute Line 5 of Tab. \ref{Tab:Saddle_Point} into \eqref{Eq:HCz_Equal} to obtain Line 6;
(ii)
Substitute Line 6 into \eqref{Eq:Saddle_Point:Exp_Cx} to derive Line 3;
(iii)
Substitute Lines 3 and 6 into \eqref{Eq:HCx_Equal}-\eqref{Eq:Saddle_Point:Exp_Fz} and \eqref{Eq:Saddle_Point:Exp_Dx}-\eqref{Eq:Saddle_Point:Int_Cx} to derive Lines 1–2, 4, 15–16, and 19–20;
(iv)
Set the values of
$ \eta_{\Sf{z}} $
and
$ \eta_{\Sf{x}} $
as given in Line 8 of Tab. \ref{Tab:notations};
(v) 
Transform \eqref{Eq:Saddle_Point:Int_Dx}-\eqref{Eq:Saddle_Point:Int_chiz_Fz_Hz} into Lines 23–30 of Tab. \ref{Tab:Saddle_Point} using
$ a_{\Sf{z}} $
and
$ a_{\Sf{x}} $
defined in Line 9 of Tab. \ref{Tab:notations}.

\begin{proposition}
\label{Proposition:Saddle_Point}
The 1RSB saddle point equations derived by Takahashi and Kabashima \cite{takahashi2022macroscopic} are mathematically equivalent \footnote{
	We do not claim the rigorousness of Kabashima's results, as they were obtained from the replica method.
}
to the formulas provided in Tab. \ref{Tab:Saddle_Point} in Appendix \ref{Appendix:Summary}.
\end{proposition}

Tab. \ref{Tab:Saddle_Point} shows that the overlap
$ D_{\Sf{x}} $
is now expressed as
\begin{align*}
D_{\Sf{x}}
& =
\Mean[
	x_{0}
	\frac{
		\int{\dd m} \,
		\Normal[
			m | \mu_{\Sf{x}},
			\Delta_{\Sf{x1}}
		]
		\EXP[
			\tilde{L} g_{\Sf{x0}}
		]
		\langle \tilde{x} \rangle
	}{
		\EXP[
			g_{\Sf{x1}}
		]
	}
]
\triangleq
\Mean_{
	x_{0}, \mu_{\Sf{x}}
}[
	x_{0}
	\hat{x}( \mu_{\Sf{x}} )
]
,
\end{align*}
where
$ \hat{x}(\mu_{\Sf{x}}) $
is independent of
$ x_{0} $.
Compared to \eqref{Eq:Overlap_D}, this new expression is more suitable for practical implementation.

\section{The VASP Algorithm}
\label{sec:VASP_algo}

In this section, we develop a new algorithm using message passing on a factor graph.
Two key aspects of message passing are determined:
1)
The form of the message, which can either be a belief or a survey.
2)
The type of variable node in the factor graph, which can be either scalar or vector.
The reformulated results from the previous section are instrumental in this process.

\subsection{Analysis of Reformulated Result}

To determine the form of the message (belief or survey), we note that the log partition function (also known as the free entropy)
$ g_{\Sf{x1}} $,
as derived in Line 23 of Tab. \ref{Tab:Saddle_Point}, can be expressed by combining Lines 6-7 of Tab. \ref{Tab:notations}:
\begin{align*}
g_{\Sf{x1}}
& =
\log
\underbrace{
	\int{\dd m} \,
	\Normal[
		m | \mu_{\Sf{x}},
		\Delta_{\Sf{x1}}
	]
	\left[
		\int{
			\dd \tilde{x}
		} \,
		\Normal[
			\tilde{x}
			| m,
			\frac{
				\Delta_{\Sf{x0}}
			}{ \beta }
		]
		q^{\beta}(
			\tilde{x}
		)
	\right]^{ \tilde{L} }
}_{ \triangleq g }
.
\end{align*}

In the equation above,
$ g $
can be interpreted as a normalizing factor for the approximate posterior:
\begin{align}
\frac{1}{g}
\underbrace{
	\prod_{l = 1}^{ \tilde{L} }
	q^{\beta}(\tilde{x}_{l})
}_{ \text{postulated prior} }
\underbrace{
	\int{\dd m} \,
	\Normal[
		m | m_{\Sf{x}},
		\Delta_{\Sf{x1}}
	]
	\prod_{l = 1}^{ \tilde{L} }
	\Normal[
		\tilde{x}_{l} | m,
		\frac{
			\Delta_{\Sf{x0}}
		}{ \beta }
	]
}_{
	\text{
		approximate likelihood
		(i.e., a message)
	}
}
.
\label{eq:post}
\end{align}
Within the message passing framework \cite{bishop2006pattern}, an approximate posterior typically factors into a prior and a message, or alternatively, into a likelihood and a message.
The final term of \eqref{eq:post} can be seen as a message, aligning with the form of a survey as defined in \cite{lucibello2019generalized,antenucci2019approximate,gradenigo2020solving}.
Therefore, using a survey is more appropriate than using a belief in this context.
To determine the form of the variable nodes (scalar or vector), we observe that the measurement matrix
$ \bs{H} $
influences our saddle point equations through its limiting eigenvalue distribution, as indicated by
$ \Mean_{\lambda} [ \cdot ] $
in Lines 15-22 of Tab. \ref{Tab:Saddle_Point}, where
$ \lambda $
represents the unordered eigenvalues of
$ \bs{H}\bs{H}^{\Ts} $.
This differs from AMP and GASP, where the distribution of each matrix element directly affects the saddle point equations.
This scenario is similar to VAMP \cite{takahashi2022macroscopic}, where the limiting eigenvalue distribution results from message passing in a vector-valued factor graph \cite{rangan2019vector}, modeling the entire vector
$ \bs{x}_{0} $
as a single variable node.
Hence, we choose the vector form for the variable nodes.

In summary, we adopt survey propagation in the vector form as the backbone scheme for message passing.

\subsection{Definition of Vector Survey}
Let
$
\vec{\bs{x}}_{i}
\triangleq
[
	x_{i, 1},
	\cdots,
	x_{ i, \tilde{L} }
]
\in
\Real^{ 1 \times \tilde{L} }
$
be a replicated row vector of the scalar
$ x_{ 0, i } $,
which is the $ i $-th element of
$ \bs{x}_{0} $.
The scalar survey for
$
\vec{\bs{x}}_{i}
$
is defined by GASP \cite{lucibello2019generalized} as:
\begin{align}
p( \vec{\bs{x}}_{i} )
& \triangleq
\int{\dd m} \,
\Normal[
	m | \mu_{i}, v_{\Sf{1}, i}
]
\prod_{l = 1}^{ \tilde{L} }
\Normal[
	x_{i, l} | m,
	\frac{ v_{\Sf{0}, i} }{ \beta }
]
\label{eq:Gaussian_smoothing1} \\
& =
\Normal[
	\vec{\bs{x}}_{i}
	| \mu_{i}
	\vec{\bs{1}}_{ \tilde{L} },
	v_{\Sf{1}, i}
	\mathbb{1}_{ \tilde{L} }
	+
	\frac{ v_{\Sf{0}, i} }{ \beta }
	\Eye_{ \tilde{L} }
]
, \label{Eq:Def_VS:Gau_Form}
\end{align}
where
$ \tilde{L} $
is the number of copies.
In the context of the survey propagation algorithm, a survey is a probabilistic measure that describes the distribution of beliefs or messages across a large set of possible solutions to an inference problem.
Specifically, in problems with multiple (often exponentially many) solutions, a survey represents the probability distribution over the potential messages exchanged between nodes (or variables) in a factor graph.

Here, we generalize the concept of a survey for a scalar variable by defining a replicated matrix
$ \bs{X} $
for the vector
$ \bs{x}_{0} $
as follows:
\begin{align*}
\bs{X}
& \triangleq
[
	\vec{\bs{x}}_{1}^{\Ts},
	\cdots,
	\vec{\bs{x}}_{N}^{\Ts}
]^{\Ts}
\in
\Real^{ N \times \tilde{L} }
.
\end{align*}
The corresponding vector survey is then defined:
\begin{align*}
p(\bs{X})
& \triangleq
\prod_{i = 1}^{N}
p( \vec{\bs{x}}_{i} )
=
\Normal[
	\tilde{\bs{x}}
	|
	\bs{1}_{ \tilde{L} }
	\otimes
	\bs{\mu},
	\Delta(
		\frac{
			\bs{v}_{\Sf{0}}
		}{ \beta },
		\bs{v}_{\Sf{1}}
	)
]
,
\end{align*}
where
$
\tilde{\bs{x}}
\triangleq
\text{vec}(\bs{X})
$
and
$
\Delta(
	\frac{
		\bs{v}_{\Sf{0}}
	}{ \beta },
	\bs{v}_{\Sf{1}}
)
\triangleq
\mathbb{1}_{ \tilde{L} } \otimes
\Diag( \bs{v}_{\Sf{1}} )
+
\Eye_{ \tilde{L} } \otimes
\Diag(
	\frac{ \bs{v}_{\Sf{0}} }{ \beta }
)
\in
\Real^{
	N \tilde{L}
	\times
	N \tilde{L}
}
$
with
$ \otimes $
denoting the Hadamard product.
Similarly, we define the vector survey for
$
\vec{\bs{z}}_{m}
\triangleq
\vec{\bs{h}}_{m} \bs{X}
\in
\Real^{ 1 \times \tilde{L} }
$
with
$
\vec{\bs{h}}_{m}
$
given as
$ m $-th row vector of
$ \bs{H} $,
$
\bs{Z}
\triangleq
[
	\vec{\bs{z}}_{1}^{\Ts},
	\cdots,
	\vec{\bs{z}}_{M}^{\Ts}
]^{\Ts}
\in
\Real^{ M \times \tilde{L} }
$,
and vectorize
$\bs{Z}$
into to a column vector:
$
\tilde{\bs{z}}
\triangleq
\text{vec}(\bs{Z})
=
\tilde{\bs{H}}
\tilde{\bs{x}}
$,
where
$
\tilde{\bs{H}}
\triangleq
\Eye_{ \tilde{L} }
\otimes
\bs{H}
$.

\subsection{Propagation of Vector Survey}

\begin{figure}[!t]
\centering
\includegraphics[width=0.45\textwidth]{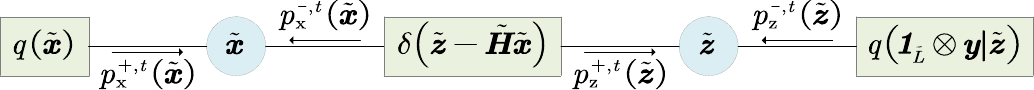}
\caption{Replicated factor graph for the inference problem}
\label{Fig:VASP}
\end{figure}

Given the vector surveys, we now introduce a replicated factor graph and propagate the surveys across it.
Fig. \ref{Fig:VASP} illustrates the replicated factor graph, where the surveys or messages sent from left to right are denoted with a superscript ``$+$'', and those sent in the reverse direction are denoted with a superscript ``$-$''.

Next, we iteratively update these surveys in a manner similar to VAMP and its variants \cite{zou2018concise,pandit2020inference}:
\begin{itemize}

\item
Backward propagation:
\begin{align}
p_{\Sf{z}}^{-, t}( \tilde{\bs{z}} )
& \propto
\frac{
	\Proj[
		q^{ \beta }(
			\bs{1}_{ \tilde{L} }
			\otimes
			\bs{y} | \tilde{\bs{z}}
		)
		p_{\Sf{z}}^{+, t}( \tilde{\bs{z}} )
	]
}{
	p_{\Sf{z}}^{+, t}( \tilde{\bs{z}} )
}
, \label{Eq:VSP:P_sz} \\
f_{\Sf{x}}^{-, t}( \tilde{\bs{x}} )
& \triangleq
\int{ \dd \tilde{\bs{z}} } \,
\delta(
	\tilde{\bs{z}}
	-
	\tilde{\bs{H}} \tilde{\bs{x}}
)
p_{\Sf{z}}^{-, t}( \tilde{\bs{z}} )
, \label{Eq:VSP:f_sx} \\
p_{\Sf{x}}^{-, t}( \tilde{\bs{x}} )
& \propto
\frac{
	\Proj[
		f_{\Sf{x}}^{-, t}( \tilde{\bs{x}} )
		p_{\Sf{x}}^{+, t}( \tilde{\bs{x}} )
	]
}{
	p_{\Sf{x}}^{+, t}( \tilde{\bs{x}} )
}
; \label{Eq:VSP:P_sx}
\end{align}

\item
Forward propagation:
\begin{align}
p_{\Sf{x}}^{+, t + 1}( \tilde{\bs{x}} )
& \propto
\frac{
	\Proj[
		q^{ \beta }( \tilde{\bs{x}} )
		p_{\Sf{x}}^{-, t}( \tilde{\bs{x}} )
	]
}{
	p_{\Sf{x}}^{-, t}( \tilde{\bs{x}} )
}
, \label{Eq:VSP:P_px} \\
f_{\Sf{z}}^{+, t + 1}( \tilde{\bs{z}} )
& \triangleq
\int{ \dd \tilde{\bs{x}} } \,
\delta(
	\tilde{\bs{z}}
	-
	\tilde{\bs{H}} \tilde{\bs{x}}
)
p_{\Sf{x}}^{+, t + 1}( \tilde{\bs{x}} )
, \label{Eq:VSP:f_pz} \\
p_{\Sf{z}}^{+, t + 1}( \tilde{\bs{z}} )
& \propto
\frac{
	\Proj[
		f_{\Sf{z}}^{+, t + 1}(
			\tilde{\bs{z}}
		)
		p_{\Sf{z}}^{-, t}( \tilde{\bs{z}} )
	]
}{
	p_{\Sf{z}}^{-, t}( \tilde{\bs{z}} )
}
; \label{Eq:VSP:P_pz}
\end{align}

\item
Iteration:
Then, back to the backward propagation and continue iterating until a stopping criterion is met.
\end{itemize}

The projection
$
\Proj[
	g( \tilde{\bs{x}} )
]
$
is crucial for the propagation process.
Its objective is to find the multivariate Gaussian distribution closest to any positive function
$
g( \tilde{\bs{x}} )
$
in terms of minimizing the Kullback-Leibler divergence (KLD).
Formally, it is defined as:
$
\Proj[
	g( \tilde{\bs{x}} )
]
\triangleq
\ArgMin\limits_{ h \in \Normal }
\mathrm{KL}[ g \| h ]	
$,
where KLD is given by:
$
\mathrm{KL}[ g \| h ]
\triangleq
\int{ \dd \tilde{\bs{x}} } \,
g( \tilde{\bs{x}} )
\log
\frac{
	g( \tilde{\bs{x}} )
}{
	h( \tilde{\bs{x}} )
}
$.
The projection is achieved by matching moments \cite{bishop2006pattern,opper2005expectation,minka2001family}, resulting in:
\begin{align*}
\Proj[g( \tilde{\bs{x}} )]
& =
\Normal[
	\tilde{\bs{x}} |
	\bs{1}_{\tilde{L}}
	\otimes
	\bs{m}_{ \Sf{g} },
	\Delta(
		\frac{
			\bs{v}_{ \Sf{g, 0} }
		}{ \beta },
		\bs{v}_{ \Sf{g, 1} }
	)
]	
, \\
\bs{1}_{\tilde{L}}
\otimes
\bs{m}_{ \Sf{g} }
& =
\frac{
	\int{ \dd \tilde{\bs{x}} } \,
	g( \tilde{\bs{x}} )
	\tilde{\bs{x}}
}{
	\int{ \dd \tilde{\bs{x}} } \,
	g( \tilde{\bs{x}} )
}
, \\
\Delta(
	\frac{
		\bs{v}_{ \Sf{g, 0} }
	}{ \beta },
	\bs{v}_{ \Sf{g, 1} }
)
& =
\mathbb{d}[
	\frac{
		\int{ \dd \tilde{\bs{x}} } \,
		g( \tilde{\bs{x}} )
		(
			\tilde{\bs{x}}
			-
			\bs{1}_{\tilde{L}}
			\otimes
			\bs{m}_{ \Sf{g} }
		)
		(
			\tilde{\bs{x}}
			-
			\bs{1}_{\tilde{L}}
			\otimes
			\bs{m}_{ \Sf{g} }
		)^{\Ts}
	}{
		\int{ \dd \tilde{\bs{x}} } \,
		g( \tilde{\bs{x}} )
	}
]
,
\end{align*}
where
$
\mathbb{d}(\bs{A})
\triangleq
[
	\diag(
		\bs{A}^{(i, j)}
	)
]
$
with
$
\bs{A}
\triangleq
[
	\bs{A}^{ ( i, j ) }
]
$,
$
\bs{A}^{ ( i, j ) }
\in
\Real^{ N \times N }
$,
$ i \in [ 1, M ] $,
$ j \in [ 1, M ] $.
Next, we explicitly compute \eqref{Eq:VSP:P_sz}-\eqref{Eq:VSP:P_pz}.

For \eqref{Eq:VSP:P_sz}, we begin by evaluating the numerator:
$
\Normal[
	\tilde{\bs{z}} |
	\bs{1}_{ \tilde{L} }
	\otimes
	\hat{\bs{\mu}}_{\Sf{z}}^{-, t},
	\Delta(
		\frac{
			\hat{\bs{v}}_{\Sf{z, 0}}^{+, t}
		}{ \beta },
		\hat{\bs{v}}_{\Sf{z, 1}}^{+, t}
	)
]
$
with
\begin{align*}
\hat{\mu}_{\Sf{z}, a}^{-, t}
& \triangleq
\frac{
	\int{\dd m} \,
	\Normal[
		m |
		\mu_{\Sf{z}, a}^{+, t},
		v_{\Sf{z, 1}, a}^{+, t}
	]
	\EXP[
		\tilde{L} g^{t}_{\Sf{z0}}
	]
	\langle z \rangle
}{
	\EXP[
		g^{t}_{\Sf{z1}}
	]
}
, \\
\hat{v}_{\Sf{z, 1}, a}^{-, t}
& \triangleq
\frac{
	\int{\dd m} \,
	\Normal[
		m |
		\mu_{\Sf{z}, a}^{+, t},
		v_{\Sf{z, 1}, a}^{+, t}
	]
	\EXP[
		\tilde{L} g^{t}_{\Sf{z0}}
	]
	\langle z \rangle^{2}
}{
	\EXP[
		g^{t}_{\Sf{z1}}
	]
}
-
(
	\hat{\mu}_{\Sf{z}, a}^{-, t}
)^{2}
, \\
\frac{
	\hat{v}_{\Sf{z, 0}, a}^{-, t}
}{ \beta }
& \triangleq
\frac{
	\int{\dd m} \,
	\Normal[
		m |
		\mu_{\Sf{z}, a}^{+, t},
		v_{\Sf{z, 1}, a}^{+, t}
	]
	\EXP[
		\tilde{L} g^{t}_{\Sf{z0}}
	]
	(
		\langle z^{2} \rangle
		-
		\langle z \rangle^{2}
	)
}{
	\EXP[
		g^{t}_{\Sf{z1}}
	]
}
,
\end{align*}
and the parameters
$ \langle a \rangle $,
$ g^{t}_{\Sf{z0}} $,
$ g^{t}_{\Sf{z1}} $
defined in Lines 10-12 of Tab. \ref{Tab:notations}.
Using the Gaussian reproduction property, we have:
\begin{align}
\eqref{Eq:VSP:P_sz}
& =
\Normal[
	\tilde{\bs{z}} |
	\bs{1}_{\tilde{L}}
	\otimes
	\bs{\mu}_{\Sf{z}}^{-, t},
	\Delta(
		\frac{
			\bs{v}_{\Sf{z, 0}}^{-, t}
		}{ \beta },
		\bs{v}_{\Sf{z, 1}}^{-, t}
	)
]
,
\end{align}
where
\begin{align*}
\Delta(
	\frac{
		\bs{v}_{\Sf{z, 0}}^{-, t}
	}{ \beta },
	\bs{v}_{\Sf{z, 1}}^{-, t}
)
& =
[
	\Lambda(
		\frac{
			\hat{\bs{v}}_{\Sf{z, 0}}^{-, t}
		}{ \beta },
		\hat{\bs{v}}_{\Sf{z, 1}}^{-, t}
	)
	-
	\Lambda(
		\frac{
			\bs{v}_{\Sf{z, 0}}^{+, t}
		}{ \beta },
		\bs{v}_{\Sf{z, 1}}^{+, t}
	)
]^{- 1}
, \\
\bs{1}_{\tilde{L}}
\otimes
\bs{\mu}_{\Sf{z}}^{-, t}
& =
\Delta(
	\frac{
		\bs{v}_{\Sf{z, 0}}^{-, t}
	}{ \beta },
	\bs{v}_{\Sf{z, 1}}^{-, t}
)
[
	\Lambda(
		\frac{
			\hat{\bs{v}}_{\Sf{z, 0}}^{-, t}
		}{ \beta },
		\hat{\bs{v}}_{\Sf{z, 1}}^{-, t}
	)
	(
		\bs{1}_{\tilde{L}}
		\otimes
		\hat{\bs{\mu}}_{\Sf{z, 0}}^{-, t}
	)
	-
\nonumber \\
& \quad
	\Lambda(
		\frac{
			\bs{v}_{\Sf{z, 0}}^{+, t}
		}{ \beta },
		\bs{v}_{\Sf{z, 1}}^{+, t}
	)
	(
		\bs{1}_{\tilde{L}}
		\otimes
		\bs{\mu}_{\Sf{z, 0}}^{+, t}
	)
]
,
\end{align*}
with
$
\Lambda
\triangleq
\Delta^{- 1}
$
as derived from \cite{zou2023high}\footnote{
	For symmetric matrices
	$ \bs{A} $,
	$ \bs{B} $,
	and
	$
	\bs{Q}
	\triangleq
	\mathbb{1}_{\tilde{L}}
	\otimes
	\bs{B}
	+
	\Eye_{\tilde{L}}
	\otimes
	(
		\bs{A} - \bs{B}
	)
	$,
	it holds
	$
	\bs{Q}^{-1}
	=
	-
	\mathbb{1}_{\tilde{L}}
	\otimes
	[
		(
			\bs{A} - \bs{B}
		)
		\bs{B}^{-1}
		(
			\bs{A} - \bs{B}
		)
		+
		\tilde{L}
		(
			\bs{A} - \bs{B}
		)
	]^{-1}
	+
	\Eye_{\tilde{L}}
	\otimes
	(
		\bs{A} - \bs{B}
	)^{- 1}
	$.
},
$
\bs{v}_{\Sf{z, 0}}^{-, t}
$,
$
\bs{v}_{\Sf{z}}^{-, t}
$,
$
\bs{\mu}_{\Sf{z}}^{-, t}
$
defined in Lines 7-9 of Algo. \ref{Tab:VASP}, and
$
\bs{v}_{\Sf{z, 1}}^{-, t}
\triangleq
\frac{1}{L}
(
	\bs{v}_{\Sf{z}}^{-, t}
	-
	\bs{v}_{\Sf{z, 0}}^{-, t}
)
$.

For \eqref{Eq:VSP:P_sx}, we compute the projected term in the numerator:
\begin{align*}
f_{\Sf{x}}^{-, t}(\tilde{\bs{x}})
& =
\Normal[
	\tilde{\bs{H}} \tilde{\bs{x}} |
	\bs{1}_{ \tilde{L} }
	\otimes
	\bs{\mu}_{\Sf{z}}^{-, t},
	\Delta(
		\frac{
			\bs{v}_{\Sf{z, 0}}^{-, t}
		}{ \beta },
		\bs{v}_{\Sf{z, 1}}^{-, t}
	)
]
, \\
f_{\Sf{x}}^{-, t}( \tilde{\bs{x}} )
p_{\Sf{x}}^{+, t}( \tilde{\bs{x}} )
& =
\Normal[
	\tilde{\bs{x}} |
	\bs{1}_{\tilde{L}}
	\otimes
	\hat{\bs{\mu}}_{\Sf{x}}^{-, t},
	\hat{\bs{V}}_{\Sf{x}}^{-, t}
]
,
\end{align*}
where
\begin{align*}
\hat{\bs{V}}_{\Sf{x}}^{-, t}
& =
\mathbb{1}_{\tilde{L}} \otimes
[
	\frac{1}{L}(
		\hat{\bs{C}}_{\Sf{x}}^{-, t}
		-
		\hat{\bs{C}}_{\Sf{x, 0}}^{-, t}
	)
]
+
\Eye_{\tilde{L}} \otimes
(
	\frac{1}{ \beta }
	\hat{\bs{C}}_{\Sf{x, 0}}^{-, t}
)
, \\
\bs{1}_{\tilde{L}}
\otimes
\hat{\bs{\mu}}_{\Sf{x}}^{-, t}
& =
\hat{\bs{V}}_{\Sf{x}}^{-, t}
[
	\bs{1}_{\tilde{L}}
	\otimes
	(
		\bs{H}^{\Ts}
		\frac{
			\beta \bs{\mu}_{\Sf{z}}^{-, t}
		}{
			\bs{v}_{\Sf{z}}^{-, t}
		}
		+
		\frac{
			\beta \bs{\mu}_{\Sf{x}}^{+, t}
		}{
			\bs{v}_{\Sf{x}}^{+, t}
		}
	)
]
,
\end{align*}
with the parameters
$
\hat{\bs{C}}_{\Sf{x, 0}}^{-, t}
$,
$
\hat{\bs{C}}_{\Sf{x}}^{-, t}
$,
$
\hat{\bs{\mu}}_{\Sf{x}}^{-, t}
$
defined in Lines 10-12 of Algo. \ref{Tab:VASP}.
Next, we compute the projected term in the numerator:
$
\Proj[
	f_{\Sf{x}}^{-, t}( \tilde{\bs{x}} )
	p_{\Sf{x}}^{+, t}( \tilde{\bs{x}} )
]
=
\Normal[
	\tilde{\bs{x}} |
	\bs{1}_{\tilde{L}} \otimes \hat{\bs{\mu}}_{\Sf{x}}^{-, t},
	\Delta(
		\frac{
			\hat{\bs{v}}_{\Sf{x, 0}}^{-, t}
		}{ \beta },
		\hat{\bs{v}}_{\Sf{x, 1}}^{-, t}
	)
]
$,
where
$
\hat{\bs{v}}_{\Sf{x, 1}}^{-, t}
\triangleq
\frac{1}{L}
(
	\hat{\bs{v}}_{\Sf{x}}^{-, t}
	-
	\hat{\bs{v}}_{\Sf{x, 0}}^{-, t}
)
$,
and
$
\hat{\bs{v}}_{\Sf{x, 0}}^{-, t}
$,
$
\hat{\bs{v}}_{\Sf{x}}^{-, t}
$
are defined in Lines 13-14 of Algo. \ref{Tab:VASP}.
Thus, the fraction in \eqref{Eq:VSP:P_sx} becomes:
$
\Normal[
	\tilde{\bs{x}} |
	\bs{1}_{\tilde{L}} \otimes \bs{\mu}_{\Sf{x}}^{-, t},
	\Delta(
		\frac{
			\bs{v}_{\Sf{x, 0}}^{-, t}
		}{ \beta },
		\bs{v}_{\Sf{x, 1}}^{-, t}
	)
]
$
with parameters
$
\bs{v}_{\Sf{x, 0}}^{-, t}
$,
$
\bs{v}_{\Sf{x, 1}}^{-, t}
$,
$
\bs{v}_{\Sf{x}}^{-, t}
$,
$
\bs{\mu}_{\Sf{x}}^{-, t}
$
given in Lines 15-18 of Algo. \ref{Tab:VASP}.

For \eqref{Eq:VSP:P_px}, we evaluate the numerator similarly to \eqref{Eq:VSP:P_sz}:
$
\Normal[
	\tilde{\bs{x}} |
	\bs{1}_{\tilde{L}}
	\otimes
	\bs{\mu}_{\Sf{x}}^{+, t + 1},
	\Delta(
		\frac{
			\bs{v}_{\Sf{x, 0}}^{+, t + 1}
		}{ \beta },
		\bs{v}_{\Sf{x, 1}}^{+, t + 1}
	)
]
$,
where
\begin{align*}
\hat{\mu}_{\Sf{x}, i}^{+, t + 1}
& \triangleq
\frac{
	\int{\dd m} \,
	\Normal[
		m | \mu_{\Sf{x}, i}^{-, t},
		v_{\Sf{x, 1}, i}^{-, t}
	]
	\EXP[
		\tilde{L} g^{t}_{\Sf{x0}}
	]
	\langle x \rangle
}{
	\EXP[
		g^{t}_{\Sf{x1}}
	]
}
, \\
\hat{v}_{\Sf{x, 1}, i}^{+, t + 1}
& \triangleq
\frac{
	\int{\dd m} \,
	\Normal[
		m | \mu_{\Sf{x}, i}^{-, t},
		v_{\Sf{x, 1}, i}^{-, t}
	]
	\EXP[
		\tilde{L} g^{t}_{\Sf{x0}}
	]
		\langle x \rangle^{2}
}{
	\EXP[
		g^{t}_{\Sf{x1}}
	]
}
-
(
	\hat{\mu}_{\Sf{x}, i}^{+, t + 1}
)^{2}
, \\
\frac{
	\hat{v}_{\Sf{x, 0}, i}^{+, t + 1}
}{ \beta }
& \triangleq
\frac{
	\int{\dd m} \,
	\Normal[
		m | \mu_{\Sf{x}, i}^{-, t},
		v_{\Sf{x, 1}, i}^{-, t}
	]
	\EXP[
		\tilde{L} g^{t}_{\Sf{x0}}
	]
	(
			\langle x^{2} \rangle
			-
			\langle x \rangle^{2}
	)
}{
	\EXP[
		g^{t}_{\Sf{x1}}
	]
}
,
\end{align*}
with parameters
$ \langle a \rangle $,
$ g^{t}_{\Sf{x0}} $,
$ g^{t}_{\Sf{x1}} $
defined in Lines 10-12 of Tab. \ref{Tab:notations}.
The complete fraction is a Gaussian density with:
$
\bs{v}_{\Sf{x, 1}}^{+, t + 1}
\triangleq
\frac{1}{L}
(
	\bs{v}_{\Sf{x}}^{+, t + 1}
	-
	\bs{v}_{\Sf{x, 0}}^{+, t + 1}
)
$,
and parameters
$
\bs{v}_{\Sf{x, 0}}^{+, t + 1}
$,
$
\bs{v}_{\Sf{x}}^{+, t + 1}
$,
$
\bs{\mu}_{\Sf{x}}^{+, t + 1}
$
given in Lines 21-23 of Algo. \ref{Tab:VASP}.

For \eqref{Eq:VSP:P_pz}, we apply the same methodology as in \eqref{Eq:VSP:P_sx} to obtain:
$
\eqref{Eq:VSP:P_pz}
=
\Normal[
	\tilde{\bs{z}} |
	\bs{1}_{\tilde{L}}
	\otimes
	\bs{\mu}_{\Sf{z}}^{+, t + 1},
	\Delta(
		\frac{
			\bs{v}_{\Sf{z, 0}}^{+, t + 1}
		}{ \beta },
		\bs{v}_{\Sf{z, 1}}^{+, t + 1}
	)
]
$
with the parameters defined in Lines 4 and 24-34 of Algo. \ref{Tab:VASP}.

With all surveys explicitly computed, we can now iteratively schedule them to develop a new survey propagation algorithm.

\subsection{The Proposed Algorithm}

We propose the VASP algorithm, detailed in Algo. \ref{Tab:VASP}, for estimating
$ \bs{x}_{0} $.
The VASP algorithm includes VAMP \cite{rangan2019vector} as a special case.
To illustrate this, we analyze the formula for
$ g_{\Sf{x}} $,
which represents the free entropy in VASP and is defined in Line 13 of Tab. \ref{Tab:notations}:
\begin{align}
g_{\Sf{x}}
& =
\frac{1}{L}
\log \int{\dd m} \,
\Normal[
	m | \mu_{\Sf{x}}^{-},
	v_{\Sf{x, 1}}^{-}
]
\EXP[
	L
	f_{\Sf{x0}}(
		m,
		v_{\Sf{x, 0}}^{-}
	)
]
. \label{eq:Gaussian_smoothing2}
\end{align}
By setting
$ v_{\Sf{x, 1}}^{-} \to 0 $,
we find that:
$
\Normal[
	m | \mu_{\Sf{x}}^{-},
	v_{\Sf{x, 1}}^{-}
]
\to
\delta( m - \mu_{\Sf{x}}^{-} )
$,
where
$ m \to \mu_{\Sf{x}}^{-} $
and
$
g_{\Sf{x}}
\to
f_{\Sf{x0}}
$,
showing that the free entropy of VASP reduces to that of VAMP.
The degeneration from VASP to VAMP is similar for
$ g_{\Sf{z}} $
and
$ f_{\Sf{z0}} $
as well.
Notably, VASP maintains the same computational complexity as VAMP, which is
$ O( N^{3} ) $
per iteration, primarily due to matrix inversion as detailed in Lines 10-11 and 24-25 of Algo. \ref{Tab:VASP}.
As noted in VAMP \cite{rangan2019vector}, if the SVD of
$ \bs{H} $
is known, this complexity can be further reduced to
$ O(N^2) $.

Superior to VAMP, the VASP algorithm can accommodate discrete priors (and discrete likelihoods).
To demonstrate this, we first note that for distributions in the exponential family, computing sufficient statistics-specifically, the mean and variance-involves evaluating the first- and second-order derivatives of the distribution's free entropy (i.e., the log partition function) w.r.t. its natural parameter \cite{bishop2006pattern}.
In the context of VAMP \cite[Eq. (16)]{rangan2019vector}, the free entropy is
$
f_{\Sf{x0}}( m, v_{\Sf{x, 0}}^{-} )
$,
and the natural parameter is
$
( m, v_{\Sf{x, 0}}^{-} )
$.
As noted by \cite[Tab. 1]{rangan2011generalized}, the second-order derivative
$
\frac{
	\partial^{2}
}{
	\partial m^{2}
}
f_{\Sf{x0}}( m, v_{\Sf{x, 0}}^{-} )
$
requires that the prior
$ q( x_{0} ) $
be twice differentiable, which limits the algorithm's applicability.
In contrast, VASP avoids this limitation.
It integrates a Gaussian kernel
$
\Normal[
	m | \mu_{\Sf{x}}^{-},
	v_{\Sf{x, 1}}^{-}
]
$
into survey's definition \eqref{eq:Gaussian_smoothing1}.
This Gaussian kernel smoothing maintains the exponential-family nature of the message, resulting in a new free entropy
$ g_{\Sf{x}} $
that is differentiable w.r.t. the natural parameter
$
(
	\mu_{\Sf{x}}^{-},
	v_{\Sf{x, 1}}^{-}
)
$,
regardless of the prior's differentiability.
Notably, handling discrete priors (and likelihoods) is a significant advantage of the VASP algorithm.

Unlike GASP, VASP accounts for correlations between both rows and columns of the matrix
$ \bs{H} $.
For instance, consider the vector
$ \bs{\mu}_{\Sf{z}}^{+} $
defined in Line 34 of Algo. \ref{Tab:VASP}
\begin{align*}
\bs{\mu}_{\Sf{z}}^{+}
& =
\bs{v}_{\Sf{z}}^{+} \odot
(
	\hat{\bs{\mu}}_{\Sf{z}}^{+} \oslash
	\hat{\bs{v}}_{\Sf{z}}^{+}
	-
	\bs{\mu}_{\Sf{z}}^{-} \oslash
	\bs{v}_{\Sf{z}}^{-}
)
, \\
\hat{\bs{\mu}}_{\Sf{z}}^{+}
& \triangleq
\bs{H}
\hat{\bs{C}}_{\Sf{\tilde{x}}}^{+}
[
	\bs{\mu}_{\Sf{x}}^{+} \oslash
	\bs{v}_{\Sf{x}}^{+}
	+
	\bs{H}^{\Ts}
	(
		\bs{\mu}_{\Sf{z}}^{-} \oslash
		\bs{v}_{\Sf{z}}^{-}
	)
]
, \\
\hat{\bs{C}}_{\Sf{\tilde{x}}}^{+}
& \triangleq
[
	\Diag(
		\bs{1} \oslash
		\bs{v}_{\Sf{x}}^{+}
	)
	+
	\bs{H}^{\Ts}
	\Diag(
		\bs{1} \oslash
		\bs{v}_{\Sf{z}}^{-}
	)
	\bs{H}
]^{-1}
.
\end{align*}
Here,
$
\hat{\bs{C}}_{
	\Sf{\tilde{x}}
}^{+}
$
incorporates correlations between any two rows.
In VASP, we approximate:
$
\bs{\mu}_{\Sf{z}}^{+}
\approx
\bs{H} \hat{\bs{C}}_{\Sf{\tilde{x}}}^{+}\bs{\eta}
$.
In contrast, GASP \cite[Eq. (131)]{lucibello2019generalized} approximates:
$
\bs{\mu}_{\Sf{z}}^{+}
\approx
\bs{H}
\bs{\eta}'
$,
which neglects inter-row correlations.
A similar discrepancy is observed between
$ \bs{\mu}_{\Sf{x}}^{-} $
in VASP (defined in Line 17) and its counterpart in GASP \cite[Eq. (137)]{lucibello2019generalized}, where GASP overlooks column correlations in the matrix.

This comparison is analogous to the distinction between VAMP \cite{rangan2019vector} and AMP \cite{donoho2009message}, where VAMP accounted for correlations that AMP did not.
Numerous simulations have shown that VAMP generally outperforms AMP in scenarios involving non-i.i.d. matrices \cite{rangan2019vector,liu2021decentralized}.
In the following simulation section, we will compare VASP with GASP and demonstrate that VASP provides superior performance when the measurement matrix exhibits correlations.

\section{State Evolution of the VASP}
\label{sec:SE}

\subsection{SE and Its Fixed Point Equations}

In this subsection, we propose a set of SE equations to capture the dynamics of VASP.
The SE is based on an assumption derived from two previous studies on mismatched problems \cite{gerbelot2022asymptotic,takahashi2022macroscopic}.
This assumption requires that the parameters of the input and output to the end-point factor nodes in Fig. \ref{Fig:VASP}-namely,
$
\bs{\mu}_{\Sf{x}}^{-, t}
$,
$
\bs{\mu}_{\Sf{z}}^{+, t}
$,
$
\bs{\mu}_{\Sf{x}}^{+, t}
$,
$
\bs{\mu}_{\Sf{z}}^{-, t}
$-all converge to some Gaussian random variables centered around
$ \bs{x}_{0} $
and
$ \bs{z}_{0} $,
up to some rotations.

\begin{assumption}[Empirical Convergence \cite{gerbelot2022asymptotic,takahashi2022macroscopic}]\label{assump:EmpiricalConverge}

At each iteration, the intermediate vectors
$
\bs{\mu}_{\Sf{x}}^{+, t}
$,
$
\bs{\mu}_{\Sf{x}}^{-, t}
$,
$
\bs{\mu}_{\Sf{z}}^{+, t}
$,
$
\bs{\mu}_{\Sf{z}}^{-, t}
$
empirically converge with second-order moment (PL2) towards Gaussian variables as described in LSL
\begin{align}
\lim\limits_{
	M, N \rightarrow \infty
}
\frac{1}{
	\mathtt{v}_{\Sf{x}}^{-, t}
}
\bs{\mu}_{\Sf{x}}^{-, t}
-
\hat{D}_{\Sf{1x}}^{t}
\bs{x}_{0}
& \overset{ \text{PL2} }{=}
\sqrt{
	\hat{F}_{\Sf{1x}}^{t}
}
\xi_{\Sf{1x}}^{t}
, \tag{a1} \\
\lim\limits_{
	M, N \rightarrow \infty
}
\frac{1}{
	\mathtt{v}_{\Sf{x}}^{+, t}
}
\bs{V}^{\Ts}
\bs{\mu}_{\Sf{x}}^{+, t}
-
\hat{D}_{\Sf{2x}}^{t}
\bs{V}^{\Ts} \bs{x}_{0}
& \overset{ \text{PL2} }{=}
\sqrt{
	\hat{F}_{\Sf{2x}}^{t}
}
\xi_{\Sf{2x}}^{t}
, \tag{a2} \\
\lim\limits_{
	M, N \rightarrow \infty
}
\frac{1}{
	\mathtt{v}_{\Sf{z}}^{-, t}
}
\bs{\mu}_{\Sf{z}}^{-, t}
-
\hat{D}_{\Sf{2z}}^{t}
\bs{z}_{0}
& \overset{ \text{PL2} }{=}
\sqrt{
	\hat{F}_{\Sf{2z}}^{t}
}
\xi_{\Sf{2z}}^{t}
, \tag{a3} \\
\lim\limits_{
	M, N \rightarrow \infty
}
\frac{1}{
	\mathtt{v}_{\Sf{z}}^{+, t}
}
\bs{U}^{\Ts}
\bs{\mu}_{\Sf{z}}^{+, t}
-
\hat{D}_{\Sf{1z}}^{t}
\bs{U}^{\Ts} \bs{z}_{0}
& \overset{ \text{PL2} }{=}
\sqrt{
	\hat{F}_{\Sf{1z}}^{t}
}
\xi_{\Sf{1z}}^{t}
, \tag{a4}
\end{align}
where
$ \bs{U} $
and
$ \bs{V} $
are obtained from the SVD of
$
\bs{H}
=
\bs{U} \bs{S} \bs{V}^{\Ts}
$,
and the random variables
$ \{\xi\} $
are i.i.d. standard Gaussian, the coefficients are defined as follows:
$
\hat{D}_{\Sf{1x}}^{t}
=
\Mean[
	\frac{
		\bs{x}_{0}^{\Ts}
		\bs{\mu}_{\Sf{x}}^{-, t}
	}{
		N C_{\Sf{x}}
		\mathtt{v}_{\Sf{x}}^{-, t}
	}
]
$,
$
\hat{F}_{\Sf{1x}}^{t}
=
\Mean[
	\frac{1}{
		N
		(
			\mathtt{v}_{\Sf{x}}^{-, t}
		)^{2}
	}
	\|
		\bs{\mu}_{\Sf{x}}^{-, t}
		-
		\mathtt{v}_{\Sf{x}}^{-, t}
		\hat{D}_{\Sf{1x}}^{t}
		\bs{x}_{0}
	\|_{2}^{2}
]
$,
$
\hat{D}_{\Sf{2x}}^{t}
=
\Mean[
	\frac{
		\bs{x}_{0}^{\Ts}
		\bs{\mu}_{\Sf{x}}^{+, t}
	}{
		N C_{\Sf{x}}
		\mathtt{v}_{\Sf{x}}^{+, t}
	}
]
$,
$
\hat{F}_{\Sf{2x}}^{t}
=
\Mean[
	\frac{1}{
		N
		(
			\mathtt{v}_{\Sf{x}}^{+, t}
		)^{2}
	}
	\|
		\bs{\mu}_{\Sf{x}}^{+, t}
		-
		\mathtt{v}_{\Sf{x}}^{+, t}
		\hat{D}_{\Sf{2x}}^{t}
		\bs{x}_{0}
	\|_{2}^{2}
]
$,
$
\hat{D}_{\Sf{2z}}^{t}
=
\Mean[
	\frac{
		\bs{z}_{0}^{\Ts}
		\bs{\mu}_{\Sf{z}}^{-, t}
	}{
		M C_{\Sf{z}}
		\mathtt{v}_{\Sf{z}}^{-, t}
	}
]
$,
$
\hat{F}_{\Sf{2z}}^{t}
=
\Mean[
	\frac{1}{
		M
		(
			\mathtt{v}_{\Sf{z}}^{-, t}
		)^{2}
	}
	\|
		\bs{\mu}_{\Sf{z}}^{-, t}
		-
		\mathtt{v}_{\Sf{z}}^{-, t}
		\hat{D}_{\Sf{2z}}^{t}
		\bs{z}_{0}
	\|_{2}^{2}
]
$,
$
\hat{D}_{\Sf{1z}}^{t}
=
\Mean[
	\frac{
		\bs{z}_{0}^{\Ts}
		\bs{\mu}_{\Sf{z}}^{+, t}
	}{
		M C_{\Sf{z}}
		\mathtt{v}_{\Sf{z}}^{+, t}
	}
]
$,
$
\hat{F}_{\Sf{1z}}^{t}
=
\Mean[
	\frac{1}{
		M
		(
			\mathtt{v}_{\Sf{z}}^{+, t}
		)^{2}
	}
	\|
		\bs{\mu}_{\Sf{z}}^{+, t}
		-
		\mathtt{v}_{\Sf{z}}^{+, t}
		\hat{D}_{\Sf{1z}}^{t}
		\bs{z}_{0}
	\|_{2}^{2}
]
$,
where
$
C_{\Sf{x}}
=
\Mean[ x_{0}^{2} ]
$,
$
C_{\Sf{z}}
=
\frac{1}{\alpha}
\Mean_{\lambda}[\lambda]
C_{\Sf{x}}
$,
and
$ \lambda $
follows the limiting eigenvalue distribution of
$ \bs{H}^{\Ts} \bs{H} $.
\end{assumption}

\begin{figure}[!t]
\centering
\subfigure[Assumption (a1)]{
\label{Fig:QQplot_SX}
\begin{minipage}[b]{.45\linewidth}
\centering
\includegraphics[scale=0.17]{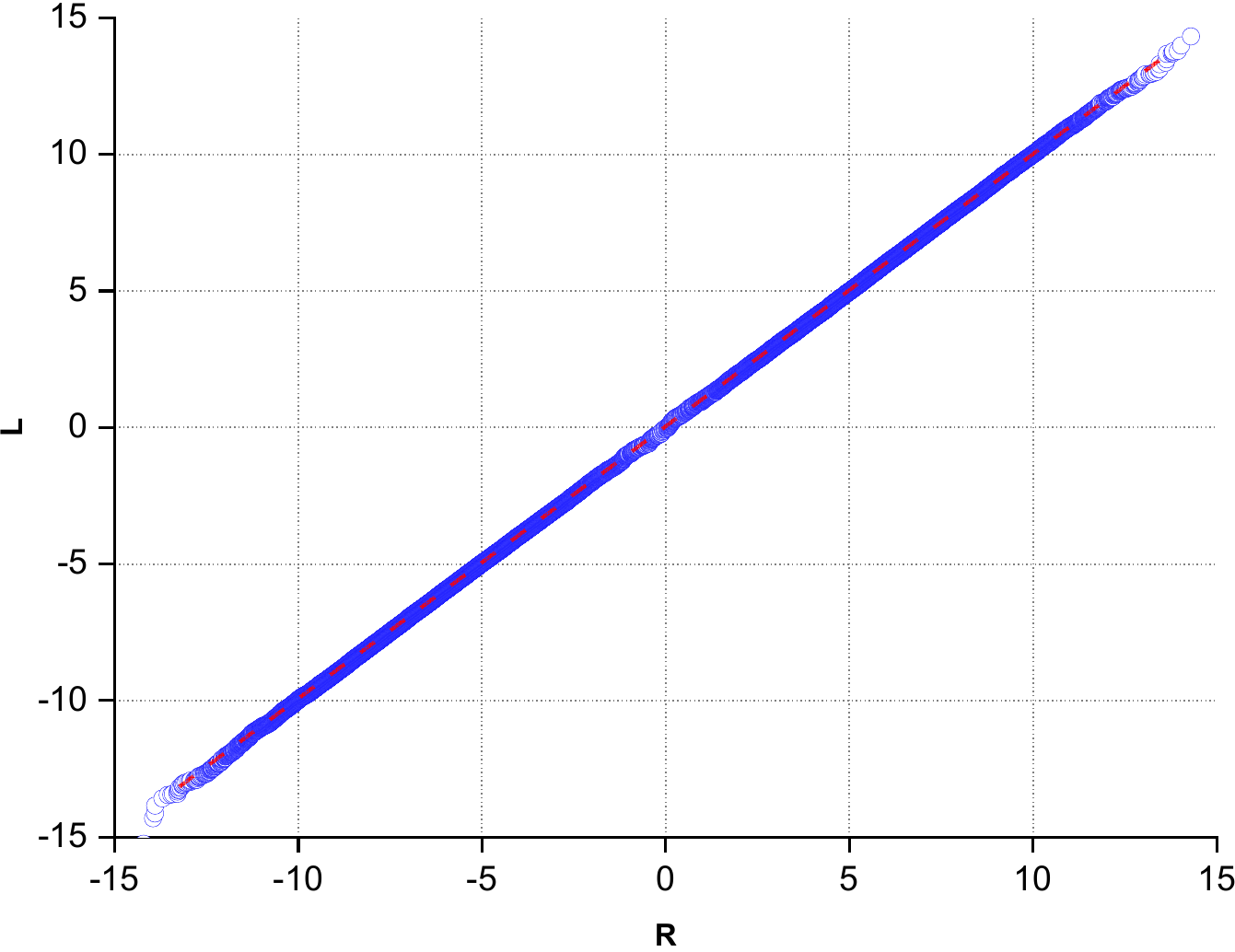}
\end{minipage}
}
\subfigure[Assumption (a2)]{
\label{Fig:QQplot_PX}
\begin{minipage}[b]{.45\linewidth}
\centering
\includegraphics[scale=0.17]{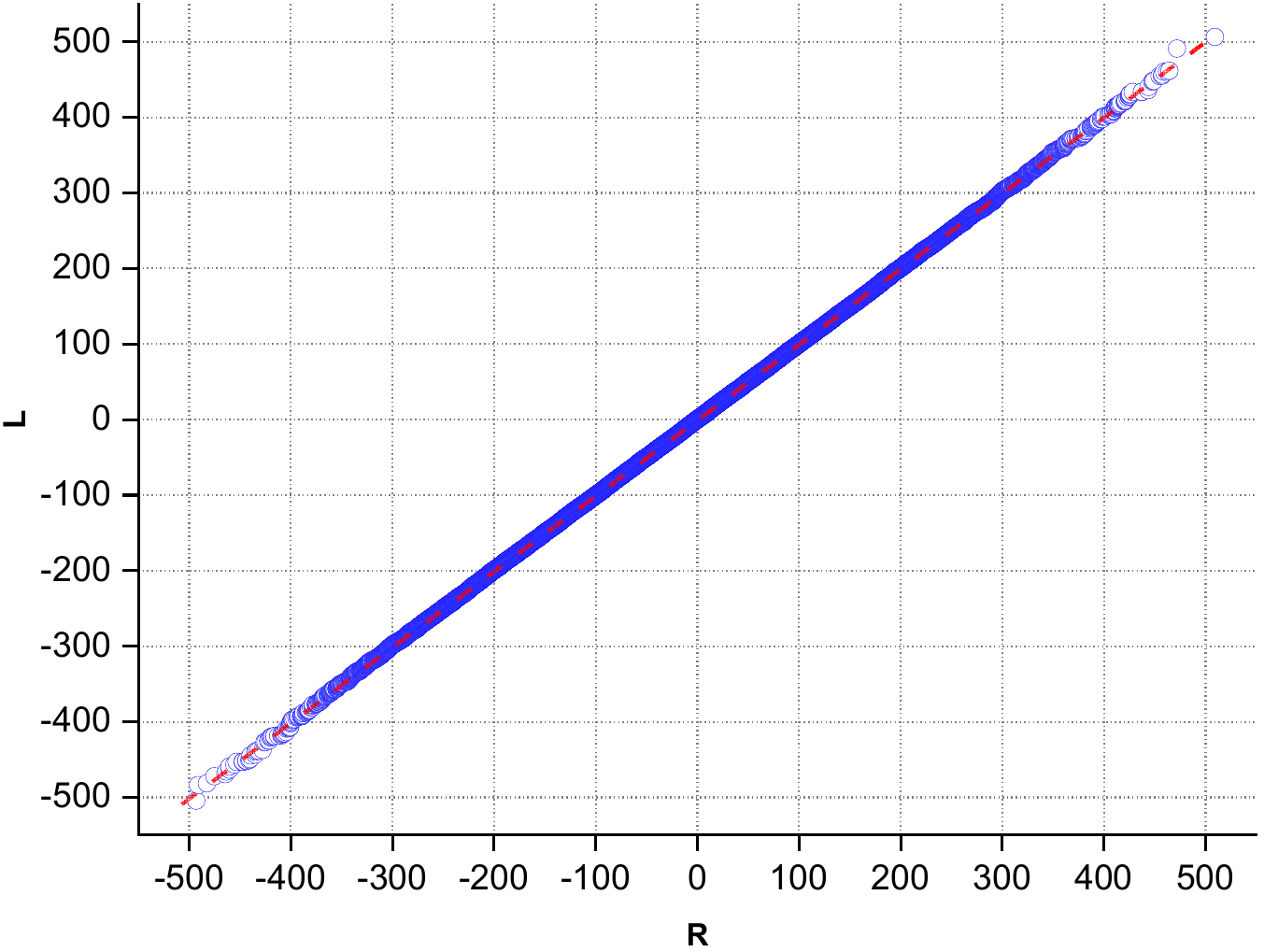}
\end{minipage}
}
\subfigure[Assumption (a3)]{
\label{Fig:QQplot_SZ}
\begin{minipage}[b]{.45\linewidth}
\centering
\includegraphics[scale=0.17]{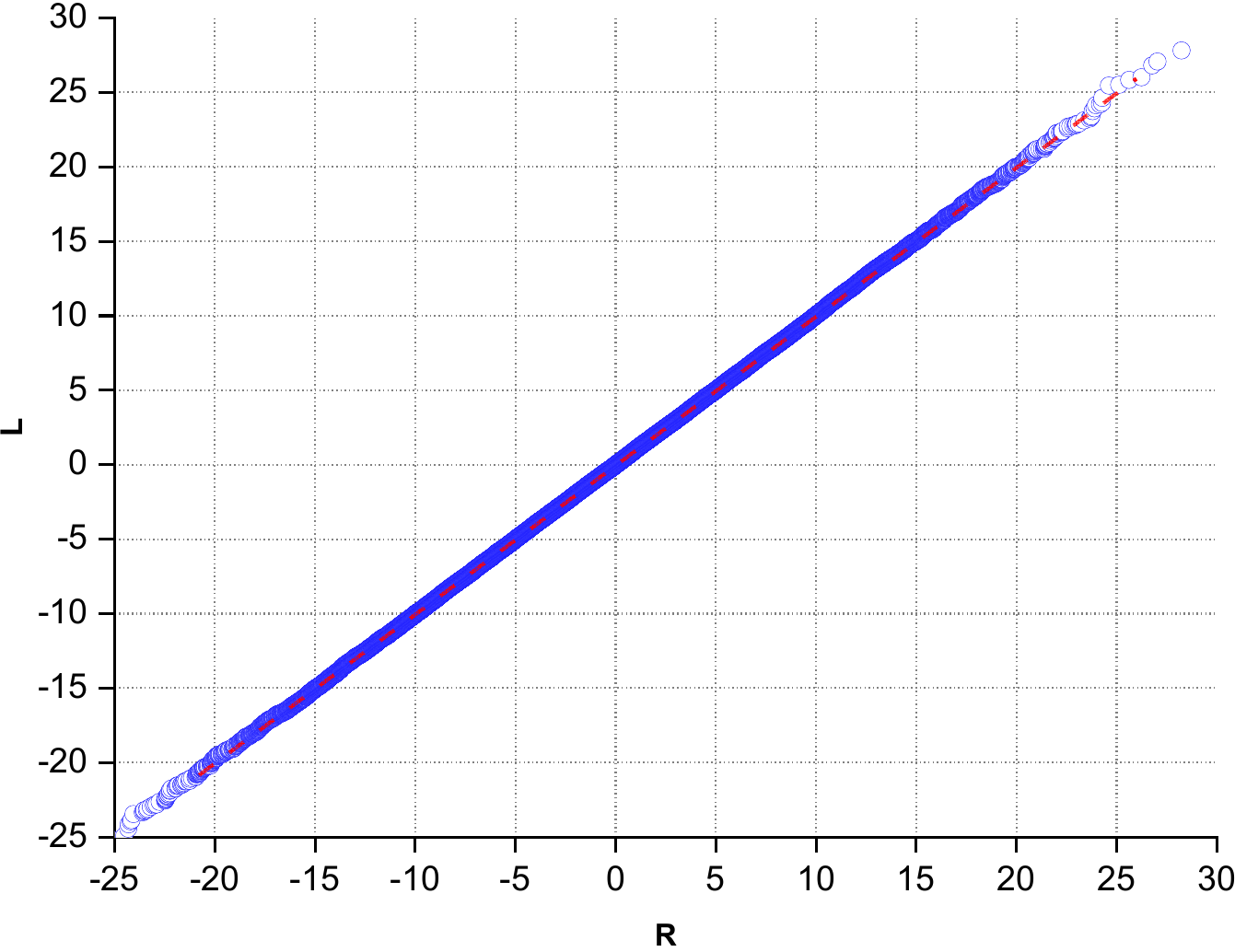}
\end{minipage}
}
\subfigure[Assumption (a4)]{
\label{Fig:QQplot_PZ}
\begin{minipage}[b]{.45\linewidth}
\centering
\includegraphics[scale=0.17]{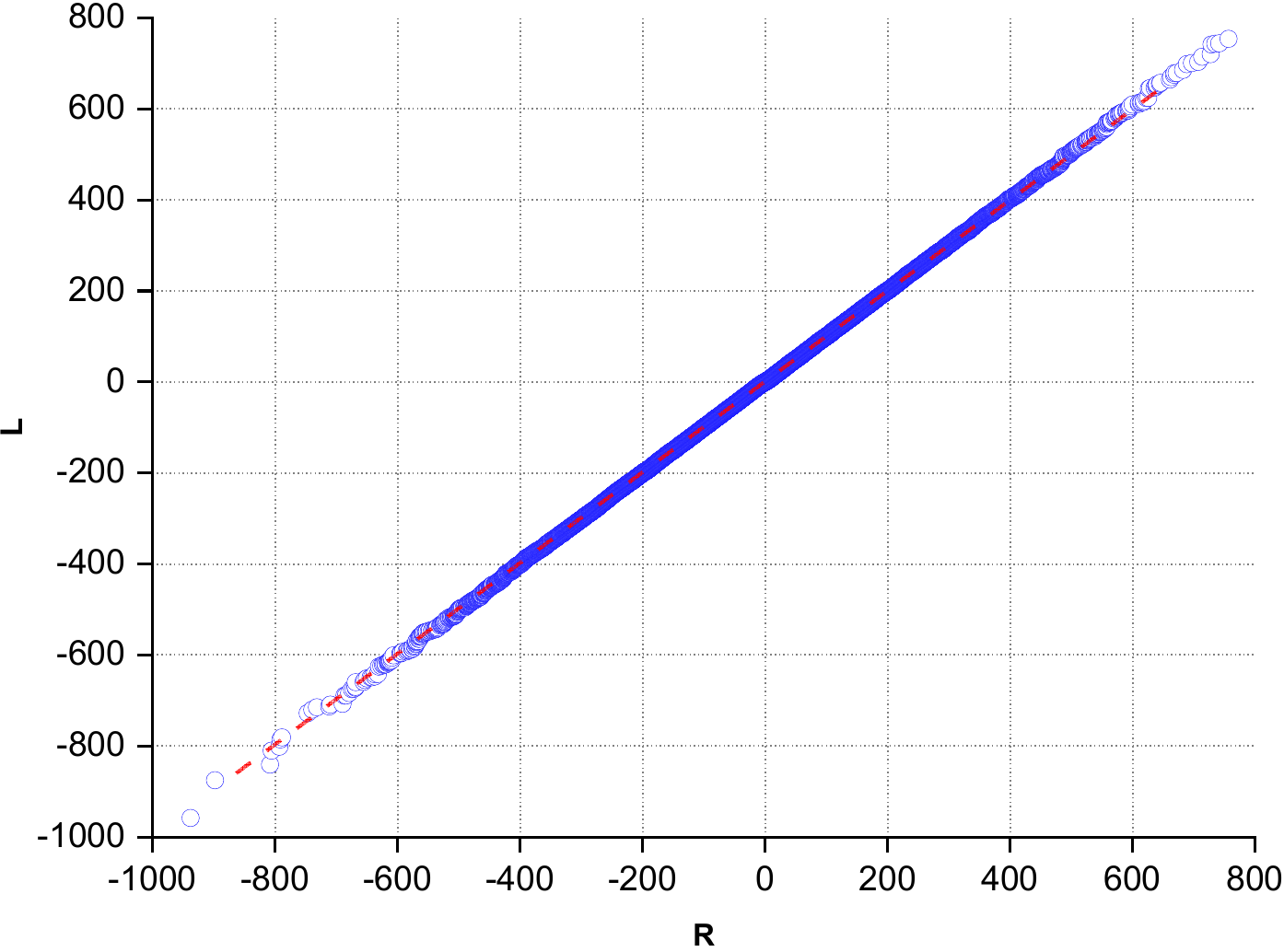}
\end{minipage}
}
\caption{
Validation of Assumption \ref{assump:EmpiricalConverge} (in a QQ-plot, if two distributions compared are similar, their samples will approximately lie on the diagonal line).
}
\label{Fig:QQplot}
\end{figure}

In the case of a matched standard linear model, defined as
$
\bs{y} = \bs{H}\bs{x} + \bs{w}
$,
where both
$ \bs{H} $
and
$ \bs{w} $
are i.i.d. Gaussian.
Assumption (a1) is provably true for the AMP algorithm \cite[Theo. 1]{bayati2011dynamics}.
Under the same conditions, Assumptions (a1) and (a3) are also valid for VAMP \cite[Eqs. (21) and (47)]{rangan2019vector}, with the matrix
$ \bs{H} $
being relaxed to allow right orthogonal invariance.
For a matched GLM, Assumptions (a1)-(a4) have been justified for the ML-VAMP \cite[Eq. (50) in Suppl.]{pandit2020inference}.
In scenarios involving model mismatch, similar assumptions have been explored by \cite{gerbelot2022asymptotic,takahashi2022macroscopic}.
To further validate the above assumptions, we use samples from a running VASP algorithm to generate four quantile-quantile plots (QQ-plots).
As shown in Fig. \ref{Fig:QQplot}, these plots suggest that the empirical convergence assumption holds approximately in the case of interest.

We propose the SE equations in Algo. \ref{Tab:SE} to capture the dynamics of VASP.
Specifically, the per-iteration mean squared error (MSE), defined as
$
\Sf{MSE}^{t}
\triangleq
\frac{1}{C_{\Sf{x}}}
\Mean[
	\frac{1}{N}
	\|
		\hat{\bs{\mu}}_{\Sf{x}}^{+, t}
		-
		\bs{x}_{0}
	\|^{2}
]
$,
is characterized by
$
\Sf{MSE}^{t}
=
\frac{1}{C_{\Sf{x}}}
(
	C_{\Sf{x}}
	+
	F_{\Sf{x}}^{+, t}
	-
	2 D_{\Sf{x}}^{+, t}
)
$,
where the macroscopic parameters are defined as follows:
$
C_{\Sf{x}}
\triangleq
\Mean[ x_{0}^{2} ]
$,
$
D_{\Sf{x}}^{+, t}
\triangleq
\Mean[
	\frac{
		\bs{x}_{0}^{\Ts}
		\hat{\bs{\mu}}_{\Sf{x}}^{+, t}
	}{N}
]
$,
$
F_{\Sf{x}}^{+, t}
\triangleq
\Mean[
	\frac{
		\|
		\hat{\bs{\mu}}_{\Sf{x}}^{+, t}
		\|_{2}^{2}
	}{N}
]
$.
A heuristic derivation of this proposal is provided in the next subsection.
Below, we connect the SE to the 1RSB saddle point equations in Tab. \ref{Tab:Saddle_Point}.

\begin{proposition} \label{Proposition:FixedPoint}
The fixed point equations of the VASP SE align with Tab. \ref{Proposition:Saddle_Point}, which reformulates Kabashima's 1RSB saddle point equations \cite{takahashi2022macroscopic}.
\end{proposition}

\begin{IEEEproof}
Refer to Appendix \ref{Appendix:Fixed_Point} for a sketch of the proof.
It is important to note that while we do not claim absolute rigor for the proposed SE, the provided proof is mathematically rigorous within the context of the proposal.
\end{IEEEproof}

This kind of correspondence is commonly observed in the AMP and GASP literature \cite{takahashi2022macroscopic,gerbelot2022asymptotic,barbier2023compressed,he2017generalized,lucibello2019generalized,antenucci2019approximate}.
As discussed in \cite{takahashi2022macroscopic}, the SE equations are iterations of the variational conditions for the free energy under the RS/RSB ansatz, and their fixed points correspond to the saddle points of the free energy.
It is important to note that the exact number of SE fixed points in a generic AMP or ASP algorithm remains uncertain within the community.
Multiple solutions may exist for the fixed point equations.
In such cases, those solutions that globally minimize the free energy are of particular operational significance and are referred to as (globally) stable solutions \cite{guo2009generic}.
Conversely, solutions that locally minimize the free energy are termed metastable, while those that locally maximize the free energy are called unstable.

The SE-to-replica correspondence suggests that the proposed VASP algorithm asymptotically achieves the ideal estimator \cite{rangan2012asymptotic} with cubic complexity, assuming that the algorithm converges to an SE fixed point and that this fixed point is a globally stable solution of the free energy's 1RSB saddle point equations.
Recall that VAMP serves as the estimator under the RS ansatz \cite{takahashi2022macroscopic}.
Our work extends VAMP to the 1RSB case.

\subsection{A Heuristic Derivation for the SE}

In this subsection, we provide a heuristic derivation for the VASP SE.
This derivation is based on Assumption \ref{assump:EmpiricalConverge} and follows a step-by-step analysis of Algo. \ref{Tab:VASP}.

(S1):
We analyze Lines 5-6 of Algo. \ref{Tab:VASP} to derive the following expressions:
$
D_{\Sf{z}}^{-, t}
=
\Mean[
	\frac{
		\bs{z}_{0}^{\Ts}
		\hat{\bs{\mu}}_{\Sf{z}}^{-, t}
	}{M}
]
$,
$
F_{\Sf{z}}^{-, t}
=
\Mean[
	\frac{
		\|
			\hat{\bs{\mu}}_{\Sf{z}}^{-, t}
		\|_{2}^{2}
	}{M}
]
$,
$
\hat{\mathtt{v}}_{\Sf{z, 1}}^{-, t}
=
\Mean[
	\hat{v}_{\Sf{z, 1}}^{-, t}
]
$,
$
\hat{\mathtt{v}}_{\Sf{z, 0}}^{-, t}
=
\Mean[
	\hat{v}_{\Sf{z, 0}}^{-, t}
]
$,
and
$
\hat{\mathtt{v}}_{\Sf{z}}^{-, t}
=
\hat{\mathtt{v}}_{\Sf{z, 0}}^{-, t}
+
L
\hat{\mathtt{v}}_{\Sf{z, 1}}^{-, t}
$.

(S2):
Substituting Lines 7-9 of Algo. \ref{Tab:VASP} into Assumption (a3), we derive
\begin{align*}
\mathtt{v}_{\Sf{z}}^{-, t}
\hat{D}_{\Sf{2z}}^{t}
& =
\Mean[
	\frac{1}{
		M
		C_{\Sf{z}}
	}
	\bs{z}_{0}^{\Ts}
	\bs{\mu}_{\Sf{z}}^{-, t}
]
=
\frac{
	\mathtt{v}_{\Sf{z}}^{-, t}
	D_{\Sf{z}}^{-, t}
}{
	C_{\Sf{z}}
	\hat{\mathtt{v}}_{\Sf{z}}^{-, t}
}
-
\mathtt{v}_{\Sf{z}}^{-, t}
\hat{D}_{\Sf{1z}}^{t}
, \\
( \mathtt{v}_{\Sf{z}}^{-, t} )^{2}
\hat{F}_{\Sf{2z}}^{t}
& =
\Mean[
	\frac{1}{M}
	\|
		\bs{\mu}_{\Sf{z}}^{-, t}
		-
		\mathtt{v}_{\Sf{z}}^{-, t}
		\hat{D}_{\Sf{2z}}^{t} \bs{z}_{0}
	\|_{2}^{2}
]
\\
& =
\frac{
	(
		\mathtt{v}_{\Sf{z}}^{-, t}
	)^{2}
	F_{\Sf{z}}^{-, t}
}{
	(
		\hat{\mathtt{v}}_{\Sf{z}}^{-, t}
	)^{2}
}
-
\frac{
	(
		\mathtt{v}_{\Sf{z}}^{-, t}
		D_{\Sf{z}}^{-, t}
	)^{2}
}{
	C_{\Sf{z}}
	(
		\hat{\mathtt{v}}_{\Sf{z}}^{-, t}
	)^{2}
}
+
(
	\mathtt{v}_{\Sf{z}}^{-, t}
)^{2}
\hat{F}_{\Sf{1z}}^{t}
-
\\
& \quad
2
\frac{
	(
		\mathtt{v}_{\Sf{z}}^{-, t}
	)^{2}
}{
	\hat{\mathtt{v}}_{\Sf{z}}^{-, t}
}
\sqrt{
	\hat{F}_{\Sf{1z}}^{t}
}
\Mean[
	\frac{1}{M}
	(
		\hat{\bs{\mu}}_{\Sf{z}}^{-, t}
	)^{\Ts}
	\bs{U}
	\bs{\xi}_{\Sf{1z}}^{t}
]
\\
& \overset{ \text{(a)} }{=}
\frac{
	(
		\mathtt{v}_{\Sf{z}}^{-, t}
	)^{2}
	F_{\Sf{z}}^{-, t}
}{
	(
		\hat{\mathtt{v}}_{\Sf{z}}^{-, t}
	)^{2}
}
-
\frac{
	(
		\mathtt{v}_{\Sf{z}}^{-, t}
		D_{\Sf{z}}^{-, t}
	)^{2}
}{
	C_{\Sf{z}}
	(
		\hat{\mathtt{v}}_{\Sf{z}}^{-, t}
	)^{2}
}
-
(
	\mathtt{v}_{\Sf{z}}^{-, t}
)^{2}
\hat{F}_{\Sf{1z}}^{t}
,
\end{align*}
where (a) follows the Stein's lemma\footnote{
	Suppose
	$ x $
	is a normally distributed random variable with mean
	$ m $
	and variance
	$ v $.
	Additionally, let
	$ g( \cdot ) $
	be a function such that the two expectations
	$
	\Mean[ g(x) (x - m) ]
	$
	and
	$
	\Mean[ g'(x) ]
	$
	both exist.
	(The existence of the expectation of any random variable is equivalent to the finiteness of the expectation of its absolute value.)
	Then,
	$
	\Mean[ g(x) (x - m) ]
	=
	v
	\Mean[ g'(x) ]
	$.
}.
Further, we obtain
$
\mathtt{v}_{\Sf{z, 0}}^{-, t}
$,
$
\mathtt{v}_{\Sf{z}}^{-, t}
$,
$
\hat{D}_{\Sf{2z}}^{t}
$,
$
\hat{F}_{\Sf{2z}}^{t}
$,
which correspond to Lines 10-13 of Algo. \ref{Tab:SE}.

(S3):
Substituting Assumptions (a2) and (a3) into Lines 10-12 of Algo. \ref{Tab:VASP}, we obtain
\begin{align*}
\hat{\bs{C}}_{\Sf{x, 0}}^{-, t}
& =
\bs{V}
(
	\frac{1}{
		\mathtt{v}_{\Sf{x, 0}}^{+, t}
	}
	\Eye_{N}
	+
	\frac{1}{
		\mathtt{v}_{\Sf{z, 0}}^{-, t}
	}
	\bs{S}^{\Ts} \bs{S}
)^{-1}
\bs{V}^{\Ts}
, \\
\hat{\bs{C}}_{\Sf{x}}^{-, t}
& =
\bs{V}
(
	\frac{1}{
		\mathtt{v}_{\Sf{x}}^{+, t}
	}
	\Eye_{N}
	+
	\frac{1}{
		\mathtt{v}_{\Sf{z}}^{-, t}
	}
	\bs{S}^{\Ts} \bs{S}
)^{-1}
\bs{V}^{\Ts}
, \\
\hat{\bs{\mu}}_{\Sf{x}}^{-, t}
& =
\bs{V}
(
	\frac{1}{
		\mathtt{v}_{\Sf{x}}^{+, t}
	}
	\Eye_{N}
	+
	\frac{1}{
		\mathtt{v}_{\Sf{z}}^{-, t}
	}
	\bs{S}^{\Ts} \bs{S}
)^{-1}
[
	(
		\hat{D}_{\Sf{2x}}^{t}
		+
		\hat{D}_{\Sf{2z}}^{t}
		\bs{S}^{\Ts} \bs{S}
	)
	\bs{V}^{\Ts}
	\bs{x}_{0}
	+
\\
& \quad
	\sqrt{
		\hat{F}_{\Sf{2x}}^{t}
	}
	\bs{\xi}_{\Sf{2x}}^{t}
	+
	\sqrt{
		\hat{F}_{\Sf{2z}}^{t}
	}
	\bs{S}^{\Ts} \bs{U}^{\Ts}
	\bs{\xi}_{\Sf{2z}}^{t}
]
,
\end{align*}
and
$
D_{\Sf{x}}^{-, t}
=
\Mean[
	\frac{
		\bs{x}_{0}^{\Ts}
		\hat{\bs{\mu}}_{\Sf{x}}^{-, t}
	}{N}
]
$,
$
F_{\Sf{x}}^{-, t}
=
\Mean[
	\frac{
		\|
			\hat{\bs{\mu}}_{\Sf{x}}^{-, t}
		\|_{2}^{2}
	}{N}
]
$,
$
\hat{\mathtt{v}}_{\Sf{x, 0}}^{-, t}
=
\Mean[
	\frac{
		\Trace(
			\hat{\bs{C}}_{\Sf{x, 0}}^{-, t}
		)
	}{N}
]
$,
$
\hat{\mathtt{v}}_{\Sf{x}}^{-, t}
=
\Mean[
	\frac{
		\Trace(
			\hat{\bs{C}}_{\Sf{x}}^{-, t}
		)
	}{N}
]
$.

(S4):
Combining Assumption (a1) with Lines 15-18 of Algo. \ref{Tab:VASP} yields
\begin{align*}
\mathtt{v}_{\Sf{x}}^{-, t}
\hat{D}_{\Sf{1x}}^{t}
& =
\Mean[
	\frac{1}{
		N
		C_{\Sf{x}}
	}
	\bs{x}_{0}^{\Ts}
	\bs{\mu}_{\Sf{x}}^{-, t}
]
=
\frac{
	\mathtt{v}_{\Sf{x}}^{-, t}
	D_{\Sf{x}}^{-, t}
}{
	C_{\Sf{x}}
	\hat{\mathtt{v}}_{\Sf{x}}^{-, t}
}
-
\mathtt{v}_{\Sf{x}}^{-, t}
\hat{D}_{\Sf{2x}}^{t}
, \\
(
	\mathtt{v}_{\Sf{x}}^{-, t}
)^{2}
\hat{F}_{\Sf{1x}}^{t}
& =
\Mean[
	\frac{1}{N}
	\|
		\bs{\mu}_{\Sf{x}}^{-, t}
		-
		\mathtt{v}_{\Sf{x}}^{-, t}
		\hat{D}_{\Sf{1x}}^{t}
		\bs{x}_{0}
	\|_{2}^{2}
]
\\
& =
\frac{
	(
		\mathtt{v}_{\Sf{x}}^{-, t}
	)^{2}
	F_{\Sf{x}}^{-, t}
}{
	(
		\hat{\mathtt{v}}_{\Sf{x}}^{-, t}
	)^{2}
}
-
\frac{
	(
		\mathtt{v}_{\Sf{x}}^{-, t}
		D_{\Sf{x}}^{-, t}
	)^{2}
}{
	C_{\Sf{x}}
	(
		\hat{\mathtt{v}}_{\Sf{x}}^{-, t}
	)^{2}
}
-
(
	\mathtt{v}_{\Sf{x}}^{-, t}
)^{2}
\hat{F}_{\Sf{2x}}^{t}
,
\end{align*}
along with
$
\hat{D}_{\Sf{1x}}^{t}
$,
$
\hat{F}_{\Sf{1x}}^{t}
$,
$
\mathtt{v}_{\Sf{x, 0}}^{-, t}
$,
$
\mathtt{v}_{\Sf{x}}^{-, t}
$
as defined in Algo. \ref{Tab:SE}.

(S5):
We analyze Lines 19-20 of Algo. \ref{Tab:VASP} and obtain
$
D_{\Sf{x}}^{+, t + 1}
=
\Mean[
	\frac{
		\bs{x}_{0}^{\Ts}
		\hat{\bs{\mu}}_{\Sf{x}}^{+, t + 1}
	}{N}
]
$,
$
F_{\Sf{x}}^{+, t + 1}
=
\Mean[
	\frac{
		\|
			\hat{\bs{\mu}}_{\Sf{x}}^{+, t + 1}
		\|_{2}^{2}
	}{N}
]
$,
$
\hat{\mathtt{v}}_{\Sf{x, 1}}^{+, t + 1}
=
\Mean[
	\hat{v}_{\Sf{x, 1}}^{+, t + 1}
]
$,
$
\hat{\mathtt{v}}_{\Sf{x, 0}}^{+, t + 1}
=
\Mean[
	\hat{v}_{\Sf{x, 0}}^{+, t + 1}
]
$,
$
\hat{\mathtt{v}}_{\Sf{x}}^{+, t + 1}
=
\hat{\mathtt{v}}_{\Sf{x, 0}}^{+, t + 1}
+
L
\hat{\mathtt{v}}_{\Sf{x, 1}}^{+, t + 1}
$.

(S6):
Similarly to (S4), substituting Lines 21-23 of Algo. \ref{Tab:VASP} into Assumption (a2) yields
$
\hat{D}_{\Sf{2x}}^{t + 1}
$
and
$
\hat{F}_{\Sf{2x}}^{t + 1}
$.

(S7):
Similar to (S3), we rewrite Lines 27-29 of Algo. \ref{Tab:VASP} as
\begin{align*}
&
\hat{\bs{C}}_{\Sf{z, 0}}^{+, t + 1}
=
\bs{U} \bs{S}
(
	\frac{1}{
		\mathtt{v}_{\Sf{x, 0}}^{+, t + 1}
	}
	\Eye_{N}
	+
	\frac{1}{
		\mathtt{v}_{\Sf{z, 0}}^{-, t}
	}
	\bs{S}^{\Ts} \bs{S}
)^{-1}
\bs{S}^{\Ts} \bs{U}^{\Ts}
, \\
&
\hat{\bs{C}}_{\Sf{z}}^{+, t + 1}
=
\bs{U} \bs{S}
(
	\frac{1}{
		\mathtt{v}_{\Sf{x}}^{+, t + 1}
	}
	\Eye_{N}
	+
	\frac{1}{
		\mathtt{v}_{\Sf{z}}^{-, t} }
	\bs{S}^{\Ts} \bs{S}
)^{-1}
\bs{S}^{\Ts} \bs{U}^{\Ts}
, \\
&
\hat{\bs{\mu}}_{\Sf{z}}^{+, t + 1}
=
\bs{U} \bs{S}
(
	\frac{1}{
		\mathtt{v}_{\Sf{x}}^{+, t + 1}
	}
	\Eye_{N}
	+
	\frac{1}{
		\mathtt{v}_{\Sf{z}}^{-, t} }
	\bs{S}^{\Ts} \bs{S}
)^{-1}
\times
\\
& \quad
[
	(
		\hat{D}_{\Sf{2x}}^{t + 1}
		+
		\hat{D}_{\Sf{2z}}^{t}
		\bs{S}^{\Ts} \bs{S}
	)
	\bs{V}^{\Ts}
	\bs{x}_{0}
	+
	\sqrt{
		\hat{F}_{\Sf{2x}}^{t + 1}
	}
	\bs{\xi}_{\Sf{2x}}^{t + 1}
	+
	\sqrt{
		\hat{F}_{\Sf{2z}}^{t}
	}
	\bs{S}^{\Ts} \bs{U}^{\Ts}
	\bs{\xi}_{\Sf{2z}}^{t}
]
,
\end{align*}
and further get
$
D_{\Sf{z}}^{+, t + 1}
=
\Mean[
	\frac{
		\bs{z}_{0}^{\Ts}
		\hat{\bs{\mu}}_{\Sf{z}}^{+, t + 1}
	}{M}
]
$,
$
F_{\Sf{z}}^{+, t + 1}
=
\Mean[
	\frac{
		\|
			\hat{\bs{\mu}}_{\Sf{z}}^{+, t + 1}
		\|_{2}^{2}
	}{M}
]
$,
$
\hat{\mathtt{v}}_{\Sf{z, 0}}^{+, t + 1}
=
\Mean[
	\frac{
		\Trace(
			\hat{\bs{C}}_{\Sf{z, 0}}^{+, t + 1}
		)
	}{M}
]
$,
$
\hat{\mathtt{v}}_{\Sf{z}}^{+, t + 1}
=
\Mean[
	\frac{
		\Trace(
			\hat{\bs{C}}_{\Sf{z}}^{+, t + 1}
		)
	}{M}
]
$.

(S8):
Similar to (S2), substituting Lines 32-34 of Algo. \ref{Tab:VASP} into Assumption (a4) yields
$
\hat{D}_{\Sf{1z}}^{t + 1}
$
and
$
\hat{F}_{\Sf{1z}}^{t + 1}
$.

\section{Simulation Results}\label{sec:simulation}

We perform Monte Carlo simulations to assess the performance of VASP and its SE.
We focus on symbol detection in multiple-input multiple-output (MIMO) systems
\begin{align*}
\bs{y}
& =
\bs{z}_{0} + \bs{w}
, \quad \quad \,
\bs{z}_{0}
\triangleq
\bs{H} \bs{x}_{0}
, \quad
\bs{H}
\triangleq
\bs{H}_{\Sf{w}}
\bs{R}^{\frac{1}{2}}
, \\
p( y | z_{0} )
& =
\Normal[ y | z_{0}, v_{\Sf{T}} ]
, \quad
p( x_{0} )
\propto
\EXP\left[
-
\frac{
	(
		| x_{0} | - 1
	)^{2}
}{
	2 c
}
\right]
, \\
q( y | z_{0} )
& =
\Normal[ y | z_{0}, v_{\Sf{F}} ]
, \quad
q(x_{0})
=
\frac{
	\delta( x_{0} + 1 )
	+
	\delta( x_{0} - 1 )
}{2}
,
\end{align*}
where the measurement matrix
$ \bs{H}_{\Sf{w}} $
is randomly drawn from a white Gaussian ensemble with i.i.d., zero-mean, and
${1}/{N}$-variance elements.
The deterministic correlation matrix
$ \bs{R} $
is constructed using the Kronecker correlation model \cite{paulraj2003introduction}, with the
$ ( i, j ) $-th element given by
$
R_{ i, j }
=
\rho^{ |i - j| }
$
where
$
\rho \in [ 0, 1 )
$
is the correlation factor.
When
$ \rho = 0 $,
the correlated model reduces to the standard i.i.d. model considered by \cite{lucibello2019generalized}.
For the postulated prior
$ q( x_{0} ) $,
we use standard BPSK modulation.
To introduce model mismatch, the ground-truth prior
$ p( x_{0} ) $
is allowed to differ slightly;
however, as
$ c \to 0 $,
the model mismatch disappears, making the ground truth identical to the postulated prior.
By default, the simulation parameters are set as follows, unless otherwise specified\footnote{
	The Julia code for the simulations in this paper is available at \url{https://github.com/PeiKaLunCi/VASP\_JL}.
}:
$ T = 30 $,
$ N = 1000 $,
$ \alpha = 2 $,
$ L = 4 $,
$ c = 0 $,
$ v_{\Sf{T}} = v_{\Sf{F}} = 0.1 $,
$ \rho = 0 $.
For evaluating estimation accuracy, we use the MSE metric:
$
\Sf{MSE}
=
\|
	\hat{\bs{x}}
	-
	\bs{x}_{0}
\|^{2}
/
\|
	\bs{x}_{0}
\|^{2}
$.
We compare four algorithms in our study:
\begin{itemize}

\item
VASP:
Detailed in Algo. \ref{Tab:VASP}, this algorithm uses the postulated distributions
$ q( x_{0} ) $
and
$ q( y | z_{0} ) $
as inputs;

\item
VAMP:
Described in \cite[Algo. 3]{rangan2019vector}, VAMP also takes the postulated distributions
$ q( x_{0} ) $
and
$ q( y | z_{0} ) $
as inputs;

\item
GASP:
Outlined in \cite[Algo. 1]{lucibello2019generalized}, GASP similarly uses the postulated distributions
$ q( x_{0} ) $
and
$ q( y | z_{0} ) $
as inputs;

\item
Bayes-optimal:
This is obtained by solving
$
\hat{\bs{x}}
=
\ArgMax_{ \bs{x}_{0} }
p( \bs{x}_{0} )
p( \bs{y} | \bs{H} \bs{x}_{0} )
$.
However, this approach is generally NP-hard if the support of
$ p( \bs{x}_{0} ) $
is discrete \cite{rangan2012asymptotic}, and thus cannot be simulated in such cases.
When
$ p( \bs{x}_{0} ) $
is continuous, the Bayes-optimal estimator can be approximated using an effective algorithm such as VAMP, which takes the ground-truth distributions
$ p( \bs{x}_{0} ) $
and
$ p( \bs{y} | \bs{z}_{0} ) $
as inputs \cite{takahashi2022macroscopic}.

\end{itemize}

\begin{figure}[!t]
\centering
\subfigure[i.i.d. matrix ($\rho=0$)]{
\label{Fig:NMM_Iter_iid_MSE_0001}
\begin{minipage}[b]{.45\linewidth}
\centering
\includegraphics[scale=0.17]{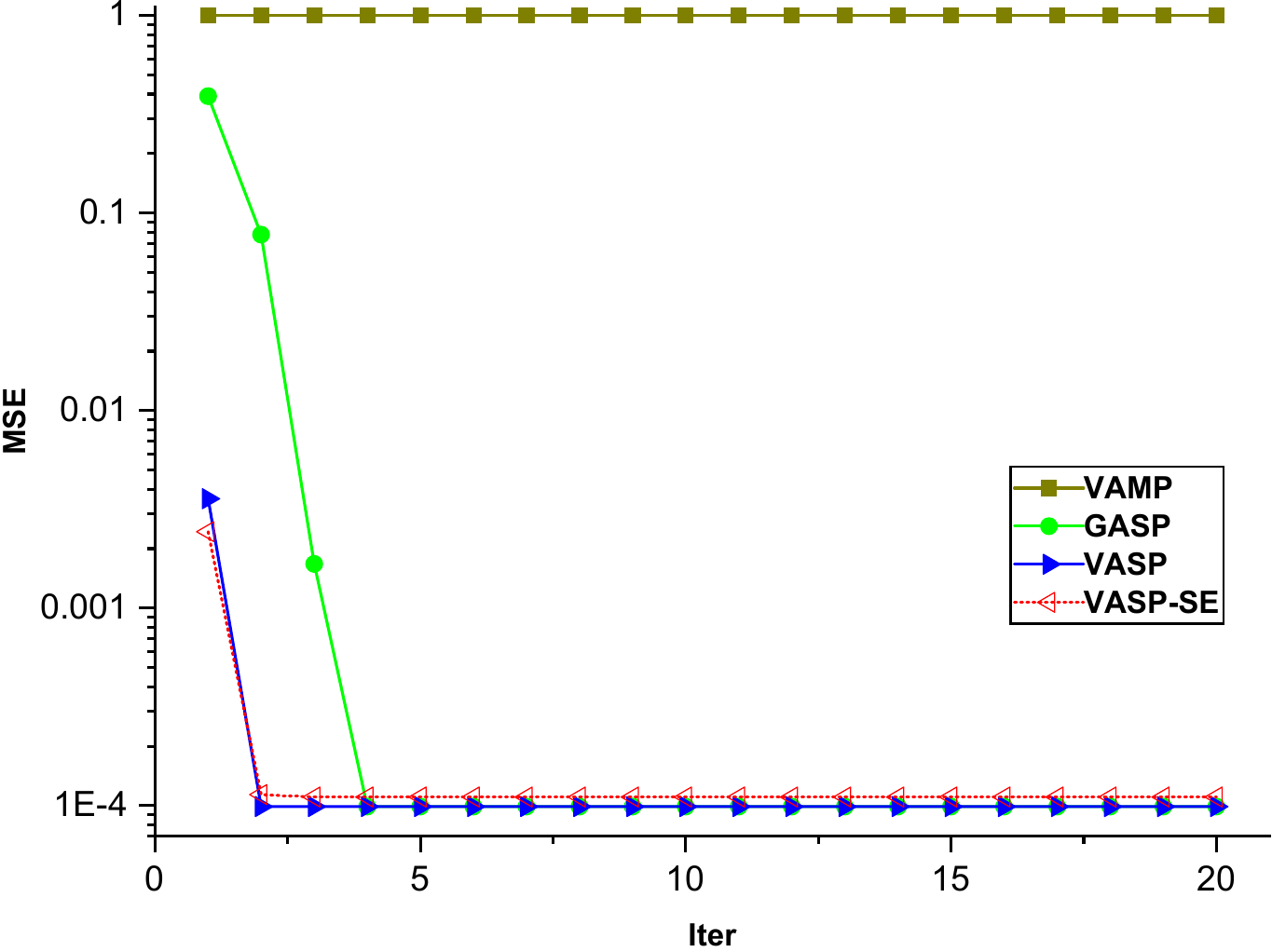}
\end{minipage}
}
\subfigure[correlated matrix ($\rho=0.4$)]{
\label{Fig:NMM_Iter_Correlated_MSE_0001}
\begin{minipage}[b]{.45\linewidth}
\centering
\includegraphics[scale=0.17]{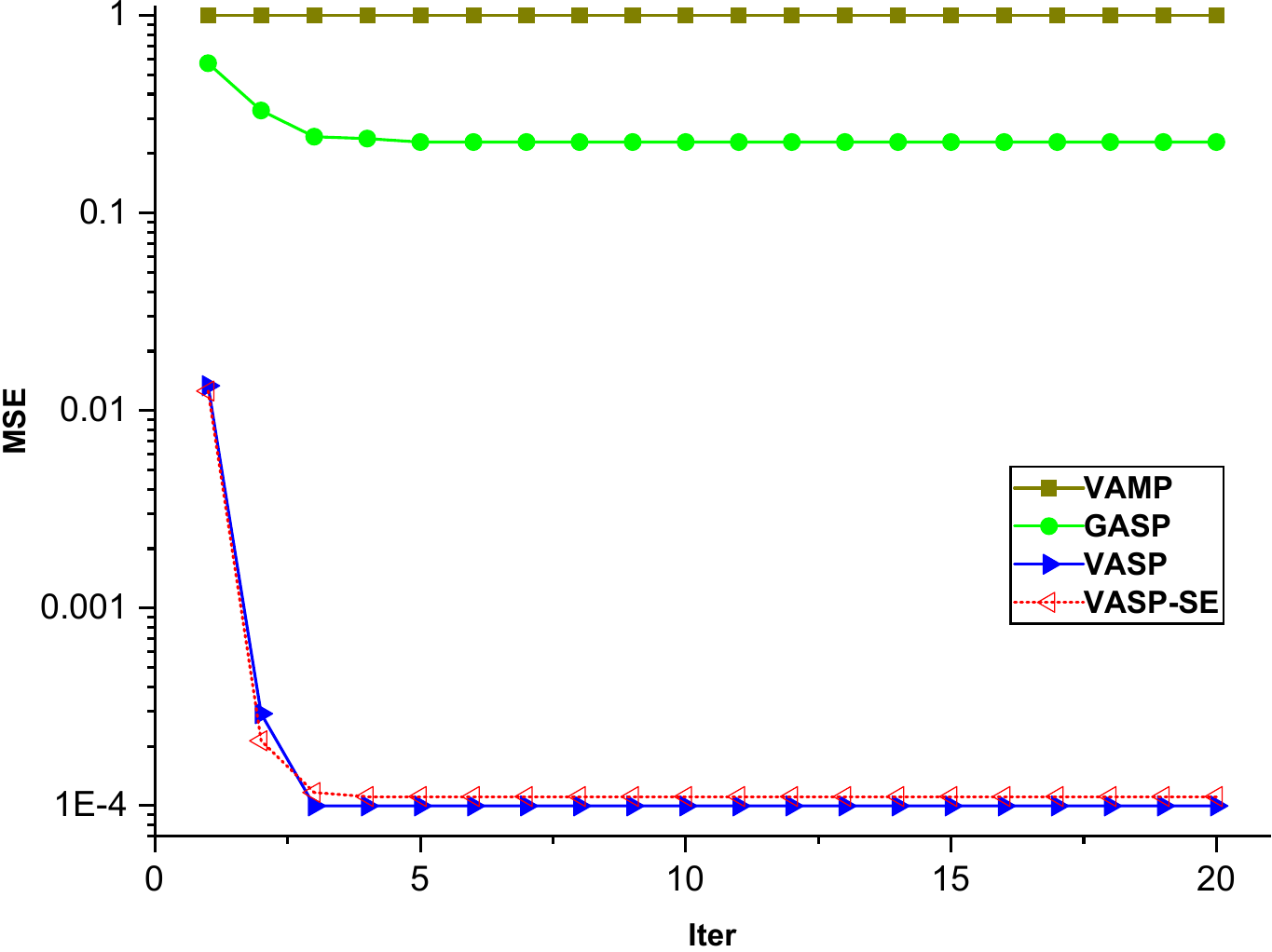}
\end{minipage}
}
\caption{
Comparison in model-matched cases ($ c = 0 $)
}
\label{Fig:NMM_Iter_MSE_c_0}
\end{figure}

\textbf{Matched model:}
Fig. \ref{Fig:NMM_Iter_MSE_c_0} presents the simulation results in the default model-matched setting.
Here,
$ c = 0 $,
meaning both the true and postulated priors are discrete, specifically following a BPSK constellation.
The VAMP algorithm \cite{rangan2019vector} requires a twice differentiable prior, making it inapplicable in this case.
For simplicity, we set its normalized MSE as a constant value of
$ 1 $.
Additionally, the Bayes-optimal estimator \cite{rangan2012asymptotic} cannot be simulated due to its NP-hard complexity.
Thus, we simulate GASP and VASP.
VASP proves resilient to increasing matrix correlation, significantly outperforming GASP, particularly in medium-correlation scenarios.
This advantage stems from VASP's use of vector surveys, which treat the measurement matrix
$ \bs{H} $
as a whole, as previously discussed.

\begin{figure}[!t]
\centering
\subfigure[i.i.d. matrix ($\rho=0$)]{
\label{Fig:NMM_Iter_iid_MSE}
\begin{minipage}[b]{.45\linewidth}
\centering
\includegraphics[scale=0.17]{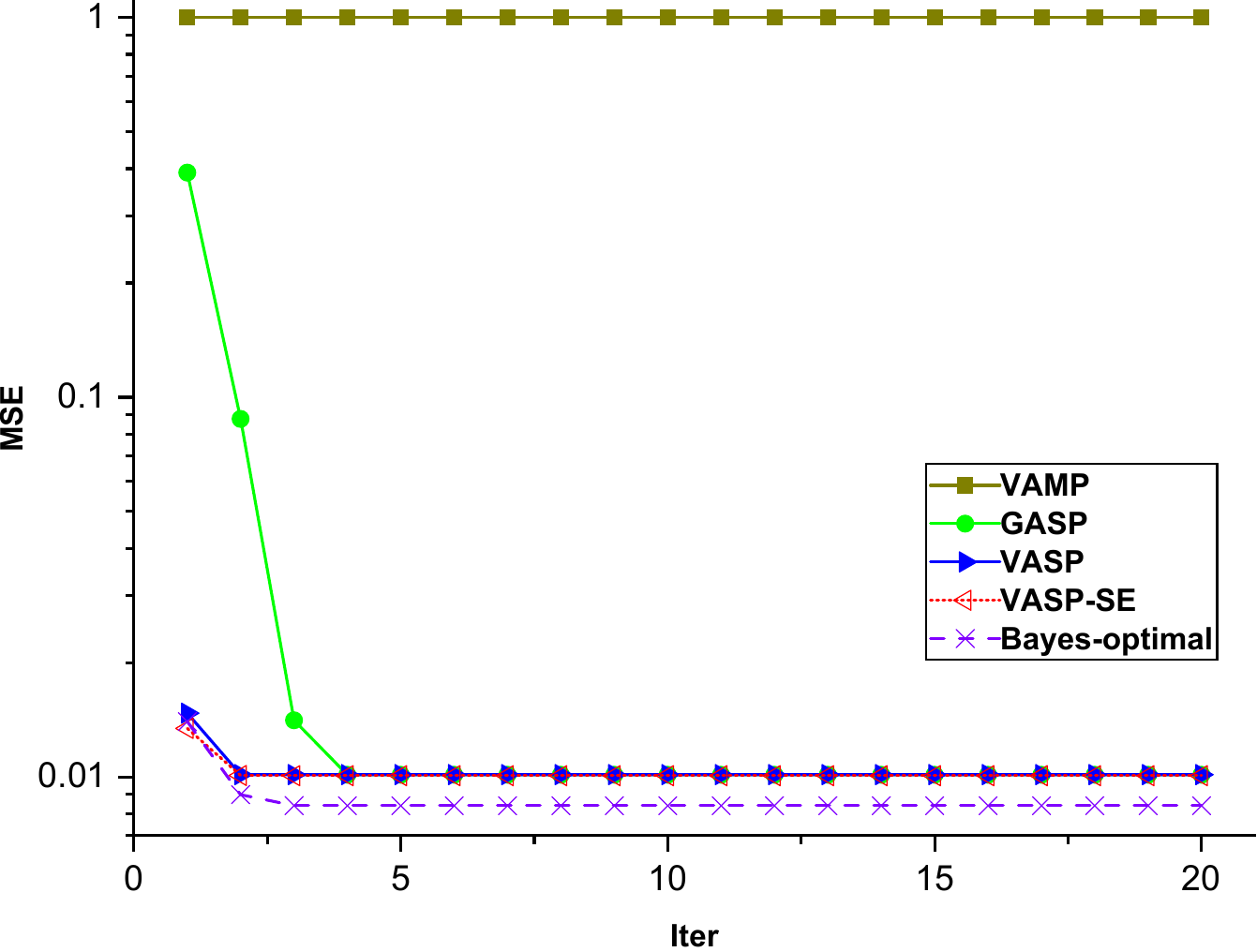}
\end{minipage}
}
\subfigure[correlated matrix ($\rho=0.4$)]{
\label{Fig:NMM_Iter_Correlated_MSE}
\begin{minipage}[b]{.45\linewidth}
\centering
\includegraphics[scale=0.17]{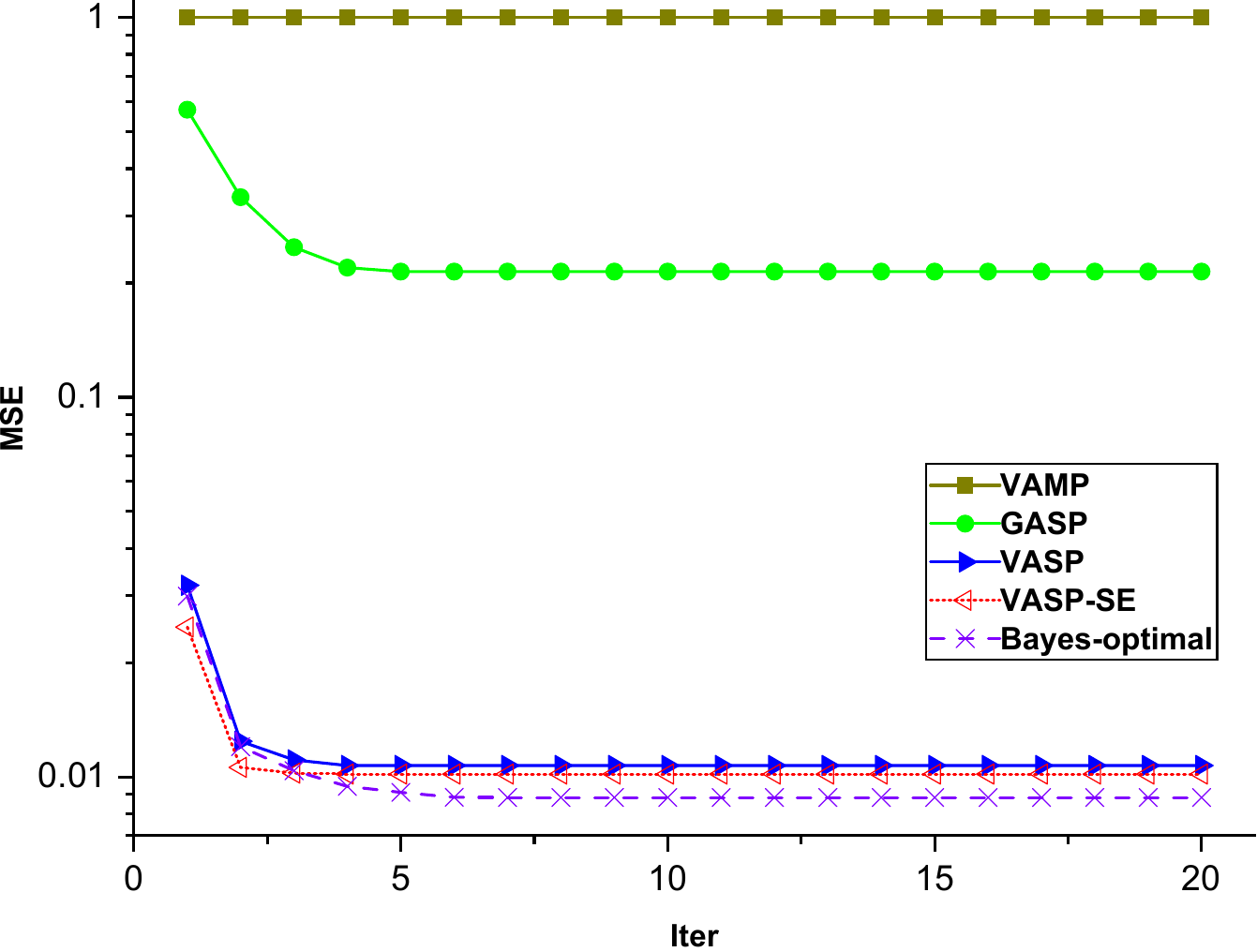}
\end{minipage}
}
\caption{
Comparison in model-mismatched cases ($ c = 0.01 $)
}
\label{Fig:NMM_Iter_MSE_c_0_01}
\end{figure}

\textbf{Mismatched model:}
Fig. \ref{Fig:NMM_Iter_MSE_c_0_01} shows the results for a model-mismatch scenario where
$ c = 0.01 $
and
$ p(x_0) \neq q(x_0) $.
In this case, VASP outperforms both VAMP and GASP, achieving an MSE that is close to that of the Bayes-optimal estimator.
Although the shift from
$ c = 0 $
(matched model) to
$ c = 0.01 $
(mismatched model) might seem minor, the change from a discrete-support prior to a continuous-support one is significant.
This is reflected in the increase in the MSE of the Bayes-optimal estimator, which rises from below
$ 10^{ - 4 } $
(as seen in Figure \ref{Fig:NMM_Iter_MSE_c_0}) to around
$ 10^{ - 2 } $
(in Figure \ref{Fig:NMM_Iter_MSE_c_0_01}), representing a
$100$-fold increase.

\begin{figure}[!t]
\centering
\subfigure[Model-matched ($c = 0$)]{
\label{Fig:MD_Correlated_cT1_0e_5}
\begin{minipage}[b]{.45\linewidth}
\centering
\includegraphics[scale=0.17]{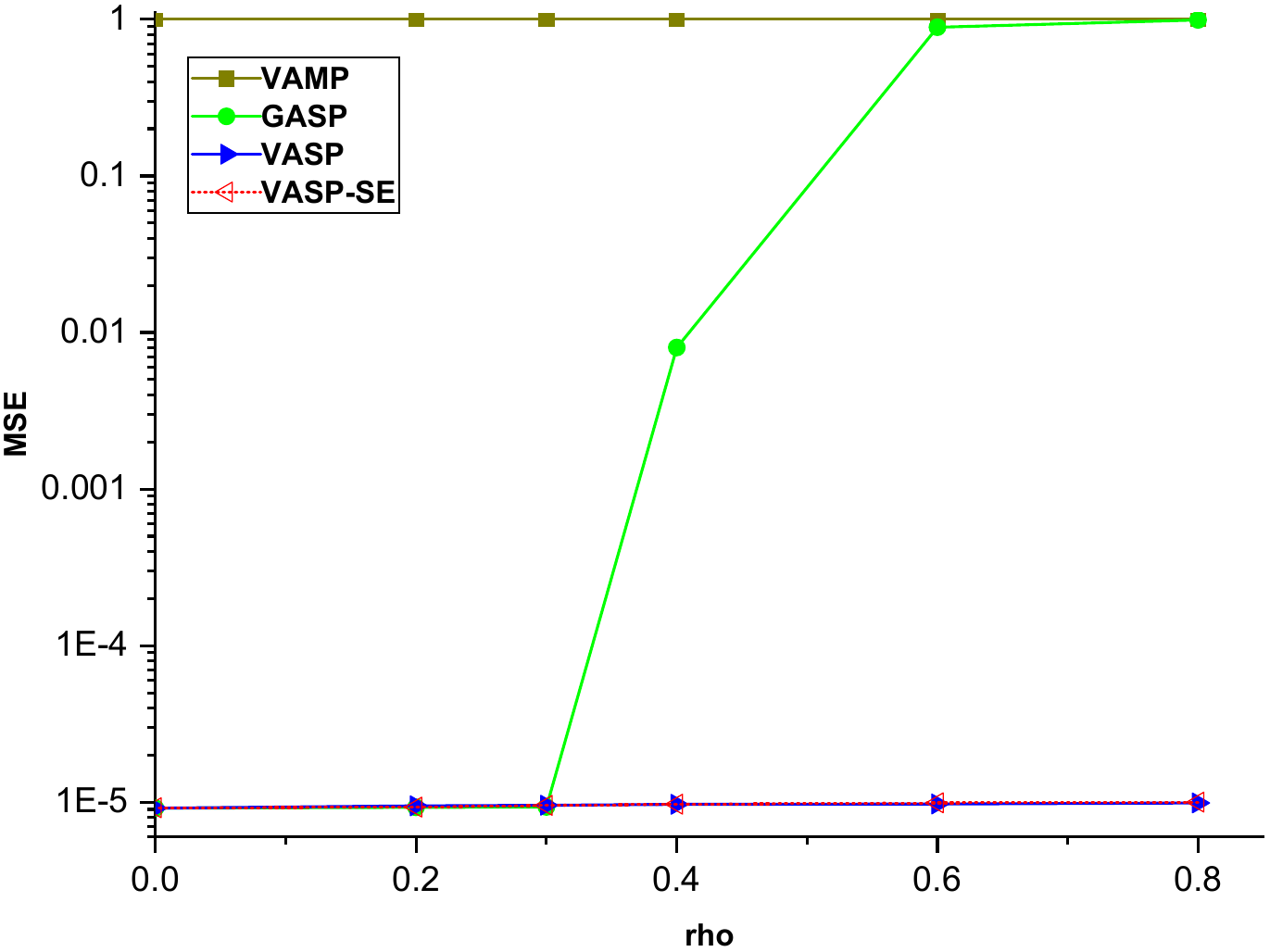}
\end{minipage}
}
\subfigure[Model-mismatched ($c = 0.001$)]{
\label{Fig:MD_Correlated_cT0_001}
\begin{minipage}[b]{.45\linewidth}
\centering
\includegraphics[scale=0.17]{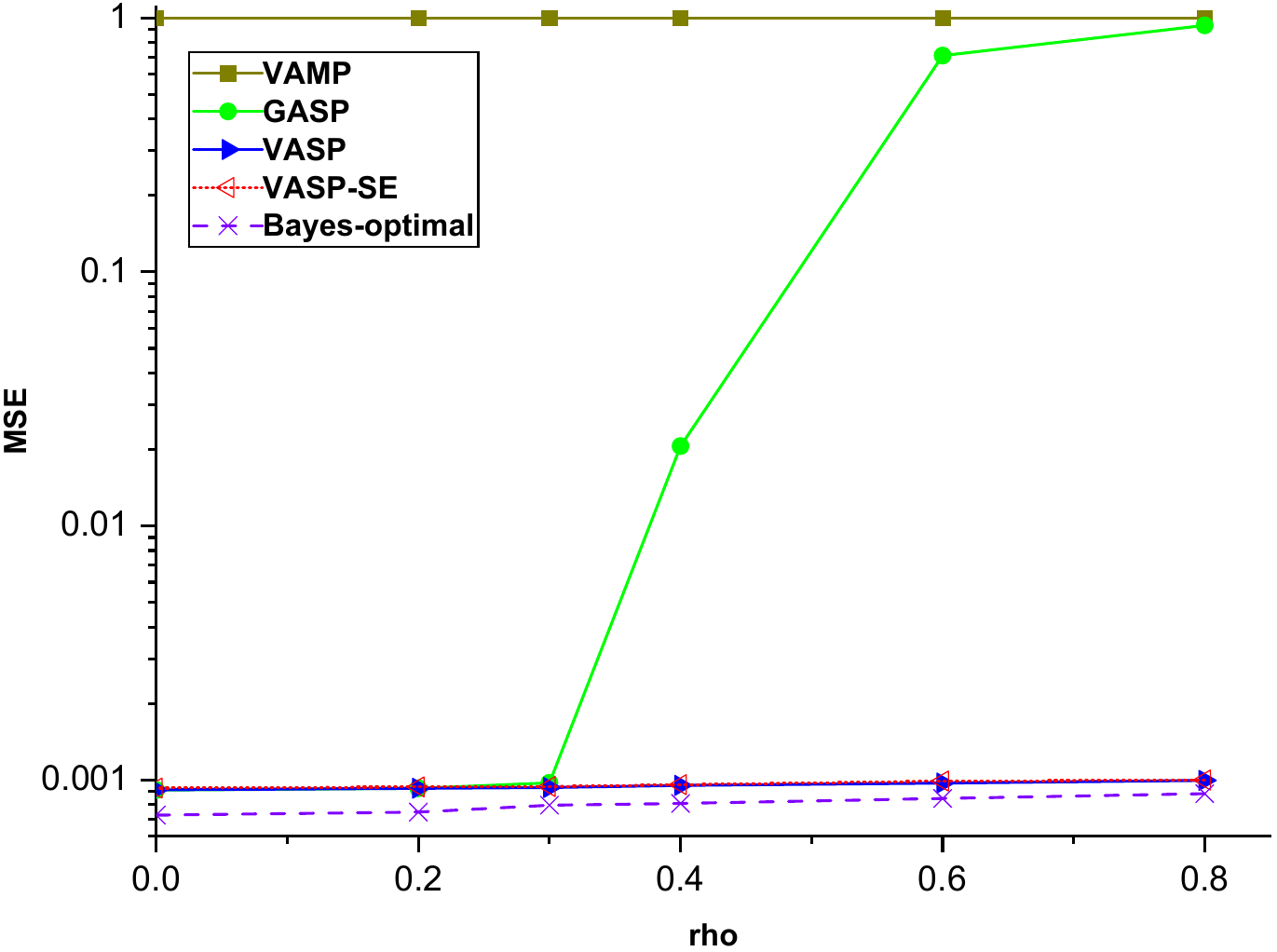}
\end{minipage}
}
\caption{
Influence of matrix correlation
($
v_{\Sf{T}} = v_{\Sf{F}} = 0.01
$)
}
\label{Fig:MD_Iter_ct_vt_vf_0_01}
\end{figure}

\textbf{Matrix correlation:}
Fig. \ref{Fig:MD_Iter_ct_vt_vf_0_01} examines the impact of correlation in both the matched and mismatched models.
The correlation factor
$ \rho $
is varied from
$ 0 $
(uncorrelated) to
$ 0.8 $
(highly correlated).
It can be observed that the MSE of VASP remains relatively stable, while the MSE of GASP significantly worsens as the correlation increases, especially in the medium to high range (e.g.,
$ \rho > 0.3 $).

\begin{figure}[!t]
\centering
\subfigure[i.i.d. matrix ($\rho=0$)]{
\label{Fig:PR_Iter_IID_ct}
\begin{minipage}[b]{.45\linewidth}
\centering
\includegraphics[scale=0.17]{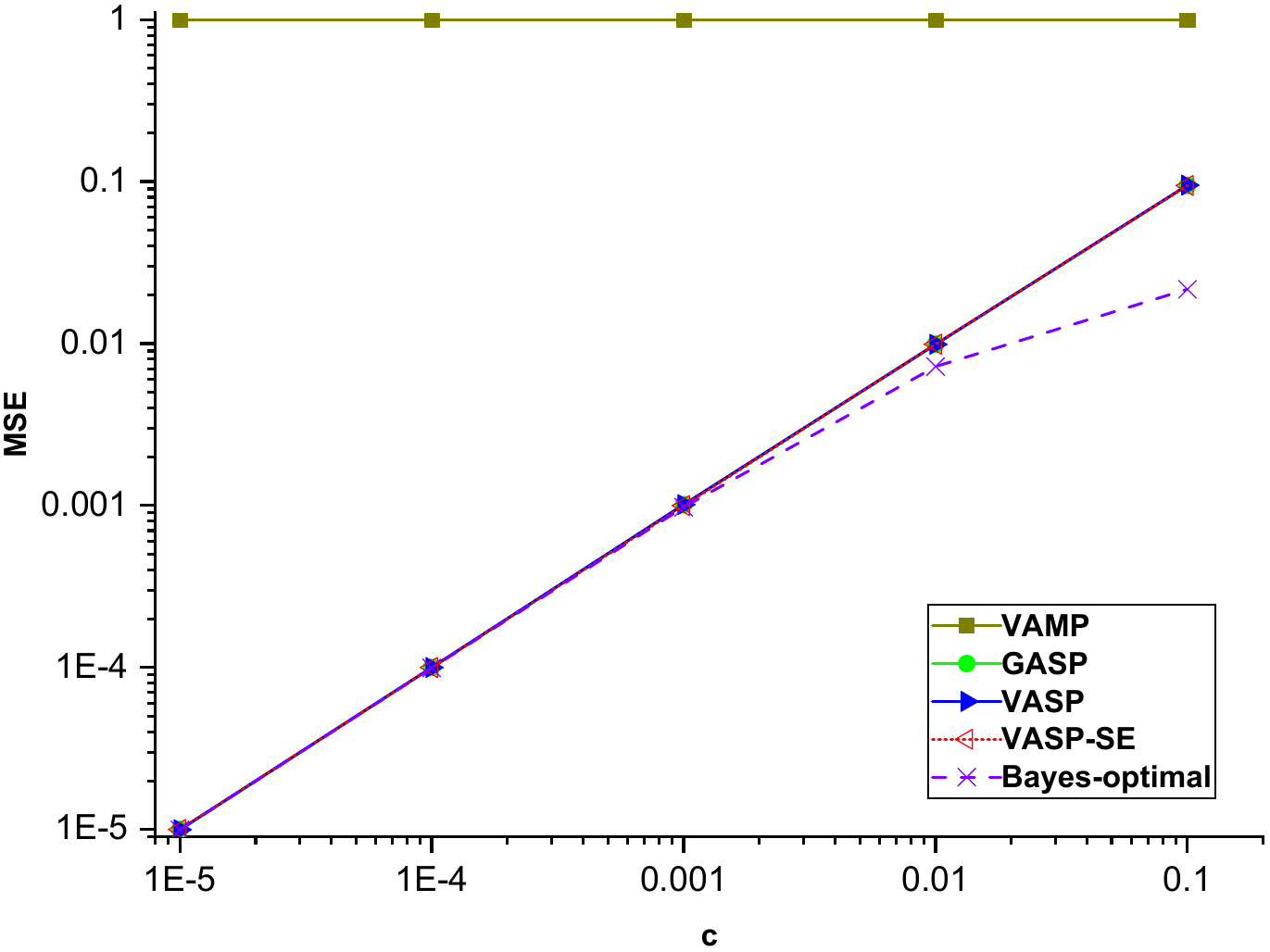}
\end{minipage}
}
\subfigure[correlated matrix ($\rho=0.4$)]{
\label{Fig:PR_Iter_Correlated_ct}
\begin{minipage}[b]{.45\linewidth}
\centering
\includegraphics[scale=0.17]{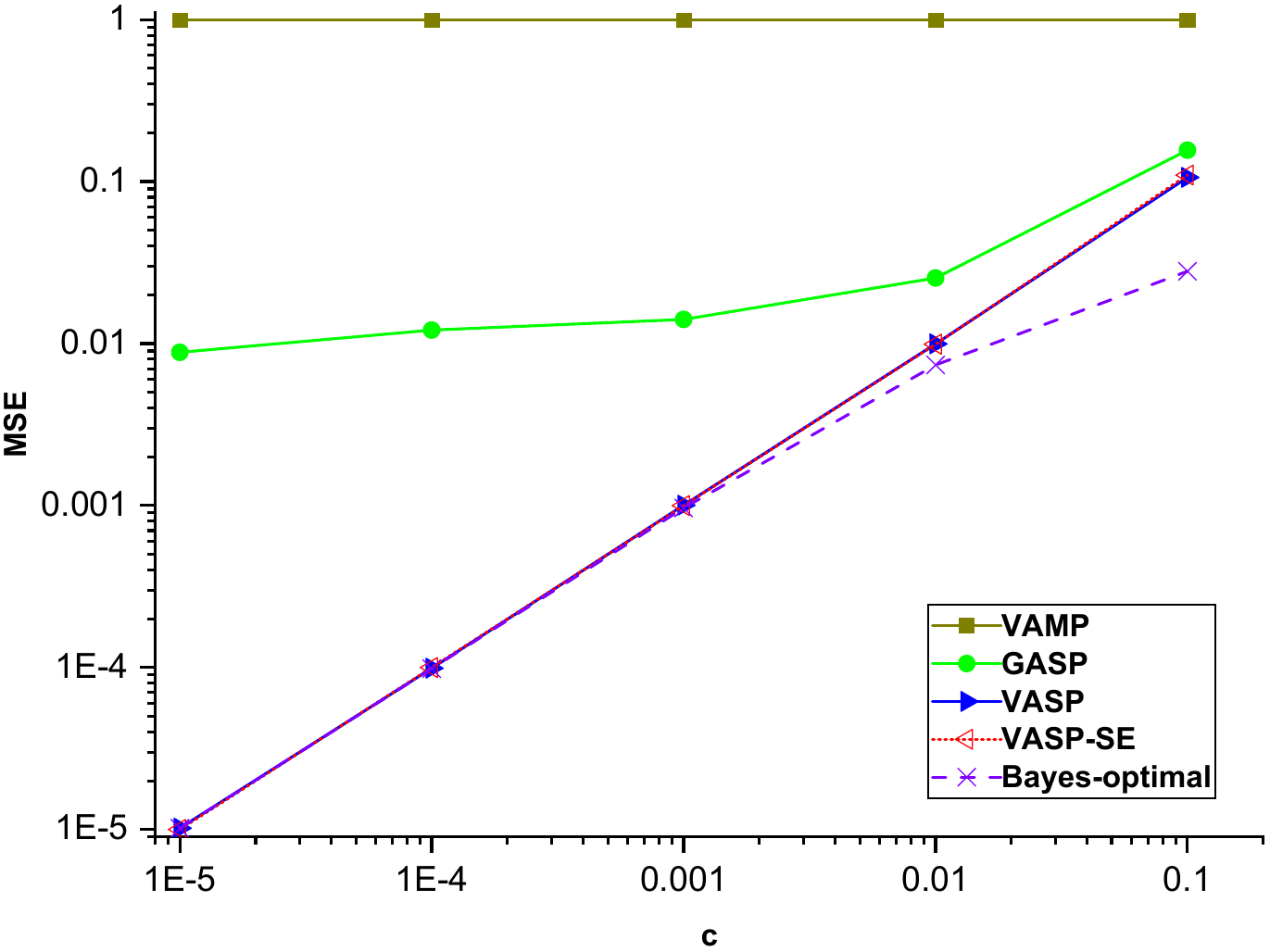}
\end{minipage}
}
\caption{
Influence of prior-only mismatch
($
\alpha = 4
$)
}
\label{Fig:MD_Iter_ct_vt_vf_a_4_c}
\end{figure}

\begin{figure}[!t]
\centering
\subfigure[i.i.d. matrix ($\rho=0$)]{
\label{Fig:PR_Iter_IID_vt}
\begin{minipage}[b]{.45\linewidth}
\centering
\includegraphics[scale=0.17]{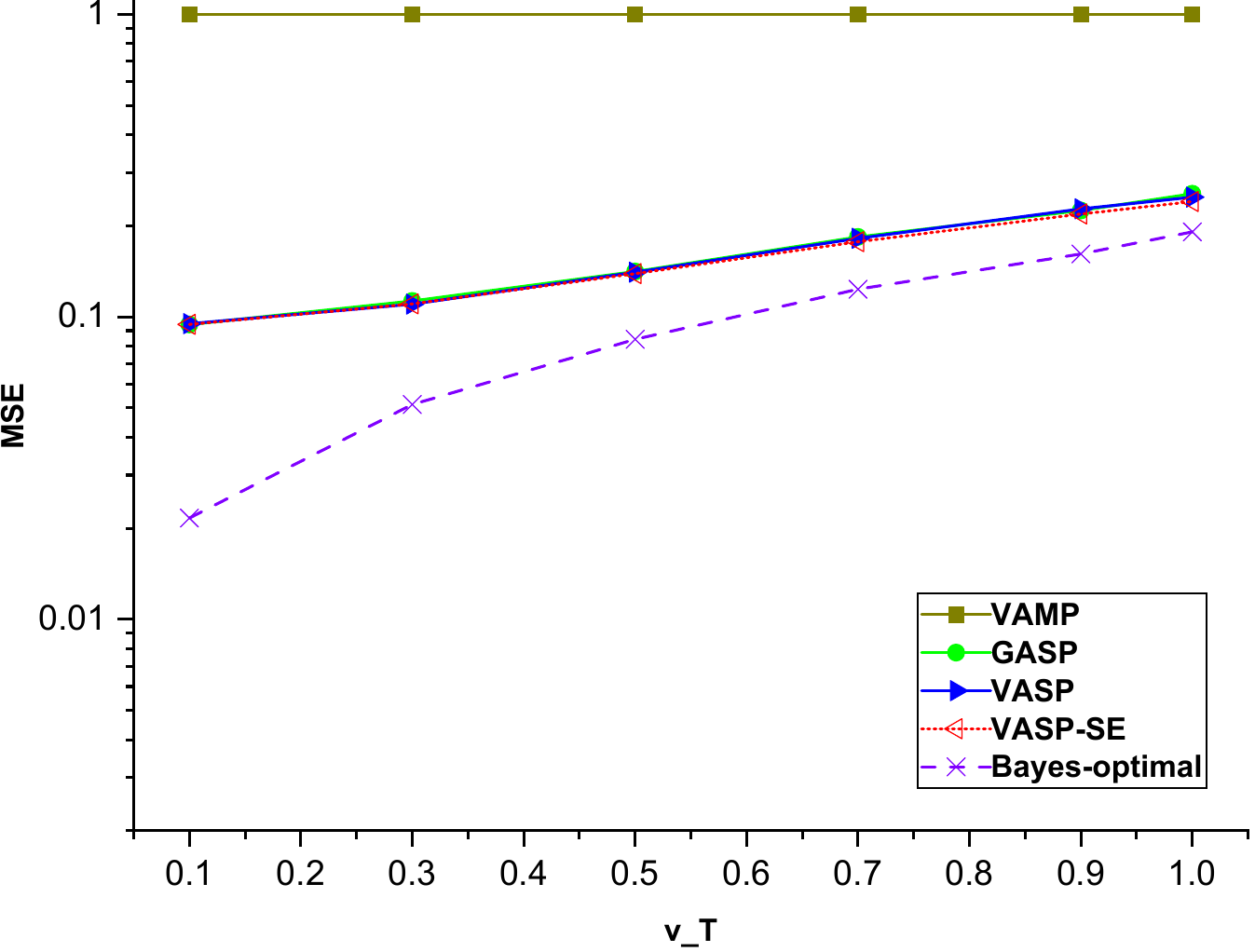}
\end{minipage}
}
\subfigure[correlated matrix ($\rho=0.4$)]{
\label{Fig:PR_Iter_Correlated_vt}
\begin{minipage}[b]{.45\linewidth}
\centering
\includegraphics[scale=0.17]{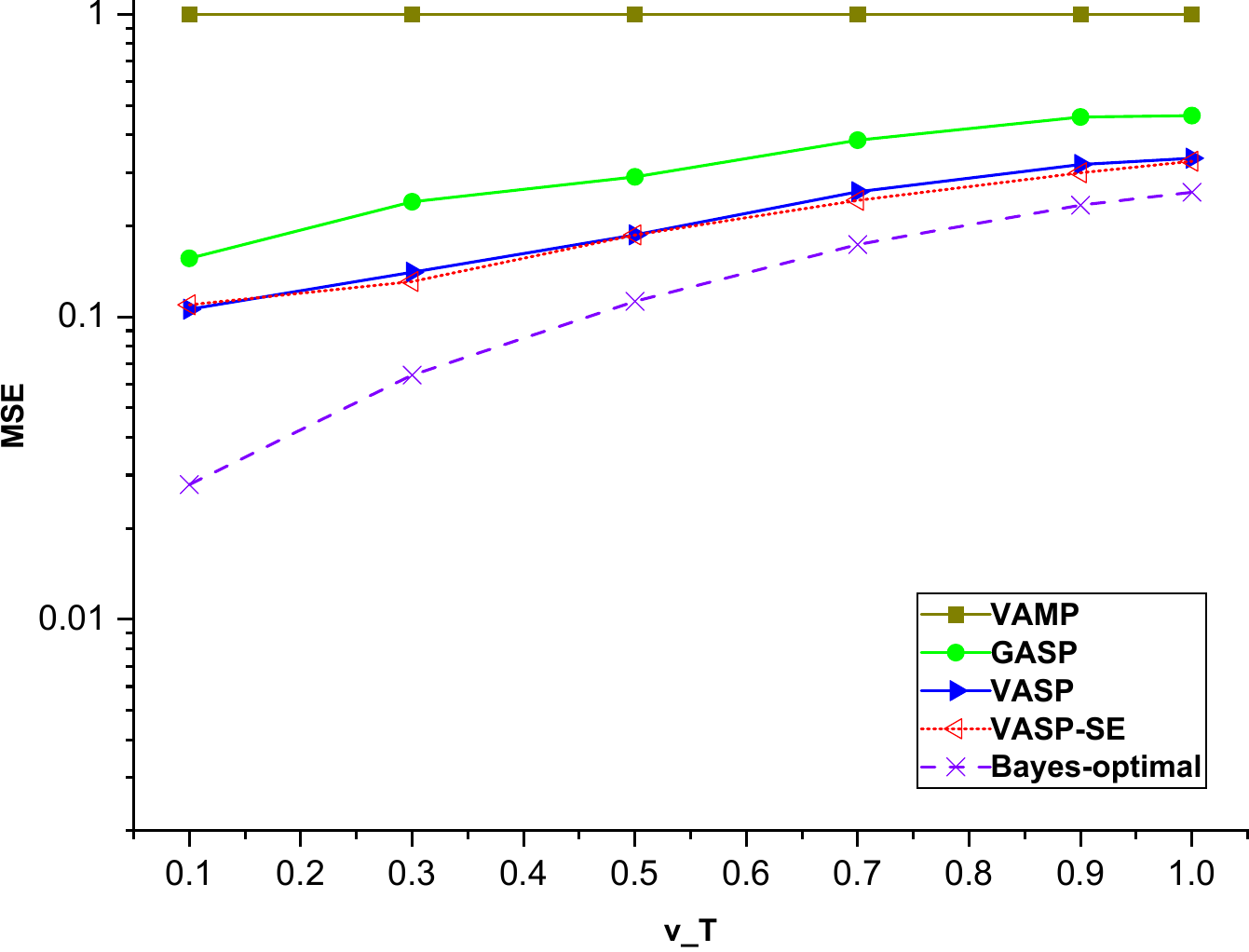}
\end{minipage}
}
\caption{
Influence of prior and likelihood mismatch
($\alpha = 4$, $c = 0.1$)
}
\label{Fig:MD_Iter_ct_vt_vf_a_4_vT}
\end{figure}

\textbf{Mismatched parameters:}
Figs. \ref{Fig:MD_Iter_ct_vt_vf_a_4_c} and \ref{Fig:MD_Iter_ct_vt_vf_a_4_vT} illustrate the effect of parameter mismatches in the model, considering a dimension ratio of
$ \alpha = 4 $.
In Fig. \ref{Fig:MD_Iter_ct_vt_vf_a_4_c}, we explore the impact of prior mismatch by varying the prior parameter
$ c $
between
$ 10^{ - 5 } $
and
$ 0.1 $,
while keeping the likelihood parameters matched.
We observe that VASP converges to the Bayes-optimal estimator as long as the mismatch in parameter
$ c $
is relatively small, i.e.,
$ c \leq 10^{ - 3 } $.
In Fig. \ref{Fig:MD_Iter_ct_vt_vf_a_4_vT}, we examine the case where mismatches exist in both the prior and likelihood.
Here, we use a mismatched prior parameter of
$ c = 0.1 $
(compared to the matched case where
$ c = 0 $)
and vary the likelihood parameter
$ v_{\Sf{T}} $
between
$ 0.1 $
and
$ 1 $
(matched value:
$ v_{\Sf{T}} = 0.1 $).
We observe that although VASP still outperforms GASP, the performance gap between VASP and the Bayes-optimal estimator widens compared to the prior-only mismatch scenario.

\section{Conclusion}\label{sec:Conlcusion}
Originally designed to solve high-dimensional linear inverse problems, AMP has a significant impact on signal processing and statistical inference.
Over the past two decades, it has been adapted into various forms to suit different data types and models.
Recently, two AMP variants, VAMP \cite{rangan2019vector} and GASP \cite{lucibello2019generalized}, have proven effective even in the model-mismatched setting, where the assumed generative model differs from the true model \cite{takahashi2022macroscopic,lucibello2019generalized}.
However, we found that the MAP versions of both VAMP and GASP have certain limitations.
Specifically, VAMP cannot handle non-differentiable priors or likelihoods, and GASP requires the measurement matrix to have i.i.d. elements.
To address these limitations, this paper introduces a new algorithm, VASP, which leverages survey propagation to resolve the differentiability issue and utilizes vector-form messages to exploit inter-element correlations in the measurement matrix.
Simulations demonstrated that VASP significantly outperforms VAMP and GASP in terms of estimation accuracy.
Moreover, the SE derived heuristically for VASP closely matches the per-iteration MSE of the algorithm.
Interestingly, the fixed point equations of this SE are identical to the saddle point equations of the free energy, as computed by \cite{takahashi2022macroscopic} under the 1RSB ansatz.
This correspondence between SE and the 1RSB analysis suggests that if the 1RSB ansatz is correct and the algorithm converges to the SE fixed point, the proposed VASP implements the estimator in \eqref{Eq:Postulated_Posterior_Estimator} with only cubic computational complexity.

Regarding statistical inference in the model-mismatched setting, we have the following observations:
(a) While algorithms like VAMP \cite{takahashi2022macroscopic}, GASP \cite{lucibello2019generalized}, and VASP can achieve accurate estimates in certain cases, many fundamental questions about model mismatch remain unanswered.
For instance, it is still unclear what degree of model mismatch will cause the PPE \eqref{Eq:Postulated_Posterior_Estimator} to exhibit an RSB structure in the extremum conditions of its free energy \cite{takahashi2022macroscopic}, and what step of RSB is sufficient.
(b) With a mismatched model, the PPE is generally inferior to the Bayes-optimal estimator, which assumes the correct prior and likelihood.
Message passing algorithms such as VAMP, GASP, and VASP aim to approximate the PPE with feasible complexity, so it is natural that their accuracy falls short of the Bayes-optimal estimator.
(c) If model mismatch exceeds a certain threshold, the global configuration space of the estimation problem \eqref{Eq:Postulated_Posterior_Estimator} may separate into many distant clusters.
In such cases, BP might not hold globally, although it may converge within a specific cluster.
SP, on the other hand, considers all clusters and assigns each a probability proportional to the number of configurations within it, summing up the weighted beliefs from each cluster \cite{braunstein2005survey}.
This is especially important in dealing with generation-inference mismatches, where the inference model might otherwise get trapped in local minima or misleading configurations.

\appendices

\section{Fixed Point Equations of  SE}\label{Appendix:Fixed_Point}

In the following, we analyze the fixed point equations of the SE.
To do this, we first remove the iteration indices
$ t $
and
$ t + 1 $
from Algo. \ref{Tab:SE}, since once a fixed point is reached, the results remain unchanged.
We then simplify the expressions as follows:
\begin{itemize}

\item
Introduce auxiliary parameters
$
\hat{\chi}_{\Sf{1x}}
\triangleq
\frac{1}{
	v_{\Sf{x, 0}}^{-}
}
$,
$
\hat{H}_{\Sf{1x}}
\triangleq
-
\frac{1}{L}
(
	\frac{1}{
		v_{\Sf{x}}^{-}
	}
	-
	\frac{1}{
		v_{\Sf{x, 0}}^{-}
	}
)
$,
$
\hat{\chi}_{\Sf{2x}}
\triangleq
\frac{1}{
	v_{\Sf{x, 0}}^{+}
}
$,
$
\hat{H}_{\Sf{2x}}
\triangleq
-
\frac{1}{L}
(
	\frac{1}{
		v_{\Sf{x}}^{+}
	}
	-
	\frac{1}{
		v_{\Sf{x, 0}}^{+}
	}
)
$,
$
\hat{\chi}_{\Sf{1z}}
\triangleq
\frac{1}{
	v_{\Sf{z, 0}}^{+}
}
$,
$
\hat{H}_{\Sf{1z}}
\triangleq
-
\frac{1}{L}
(
	\frac{1}{
		v_{\Sf{z}}^{+}
	}
	-
	\frac{1}{
		v_{\Sf{z, 0}}^{+}
	}
)
$,
$
\hat{\chi}_{\Sf{2z}}
\triangleq
\frac{1}{
	v_{\Sf{z, 0}}^{-}
}
$,
$
\hat{H}_{\Sf{2z}}
\triangleq
-
\frac{1}{L}
(
	\frac{1}{
		v_{\Sf{z}}^{-}
	}
	-
	\frac{1}{
		v_{\Sf{z, 0}}^{-}
	}
)
$;

\item
Merge Lines 10-13 of Algo. \ref{Tab:SE} into Lines 36-39, and obtain
$
D_{\Sf{z}}^{-}
=
D_{\Sf{z}}^{+}
$,
$
F_{\Sf{z}}^{-}
=
F_{\Sf{z}}^{+}
$,
$
\hat{\mathtt{v}}_{\Sf{z, 0}}^{-}
=
\hat{\mathtt{v}}_{\Sf{z, 0}}^{+}
$,
$
\hat{\mathtt{v}}_{\Sf{z}}^{-}
=
\hat{\mathtt{v}}_{\Sf{z}}^{+}
$;

\item
Define
$
D_{\Sf{z}}
\triangleq
D_{\Sf{z}}^{-}
$,
$
F_{\Sf{z}}
\triangleq
F_{\Sf{z}}^{-}
$,
$
\chi_{\Sf{z}}
\triangleq
\hat{\mathtt{v}}_{\Sf{z, 0}}^{-}
$,
$
H_{\Sf{z}}
\triangleq
\frac{1}{L}
(
	\hat{\mathtt{v}}_{\Sf{z}}^{-}
	-
	\hat{\mathtt{v}}_{\Sf{z, 0}}^{-}
)
$;

\item
Simplify Lines 4-13 and 32-39 of Algo. \ref{Tab:SE} into Lines 11-14, 19-22, and 27-30 of Tab. \ref{Proposition:Saddle_Point};

\item
Merge Lines 18-22 of Algo. \ref{Tab:SE} into Lines 28-31 of the same algorithm;

\item
define
$
D_{\Sf{x}}
$,
$
F_{\Sf{x}}
$,
$
\chi_{\Sf{x}}
$,
$
H_{\Sf{x}}
$
similarly;

\item
Simplify Lines 14-31 of Algo. \ref{Tab:SE} into Lines 7-10, 15-18, and 23-26 of Tab. \ref{Proposition:Saddle_Point}.

\end{itemize}

\section{A Useful Lemma}\label{Appendix:Summary}

\begin{lemma}\label{lemma:Decom_Formula}
A matrix
$
\bs{Q}
=
\mathcal{Q}^{
	\tau + 1
}_{
	\tilde{L}
}(
	C, D, F, \chi, H
)
$
in the 1RSB form can be decomposed as below \cite{shinzato2008perceptron}
\begin{align*}
\bs{Q}
& =
\bs{F} \bs{D}_{\Sf{q}} \bs{F}^{\Cts}
, \quad
\bs{F}
\triangleq
\left[
\begin{array}{ cccc }
	1
	&
	\\
	&
	\frac{1}{
		\sqrt{
			\frac{\tau}{\tilde{L}}
		}
	}
	\bs{F}_{
		\frac{\tau}{\tilde{L}}
	}
	\otimes
	\frac{1}{
		\sqrt{\tilde{L}}
	}
	\bs{F}_{\tilde{L}}
	\\
\end{array}
\right]
, \\
\bs{D}_{ \Sf{q} }^{ (i, j) }
& \triangleq
\begin{cases}
	C
	&
	i = j = 0
	\\
	\sqrt{\tau} D
	&
	( i = 0, \, j = 1 )
	\,
	\text{or}
	\,
	( i = 1, \, j = 0 )
	\\
	\bs{D}_{\Sf{a}}^{
		(i - 1, i - 1)
	}
	&
	(
		i \geq 1, j \geq 1, i = j
	)
	\\
	0
	&
	\text{others}
	\\
\end{cases}
,
\end{align*}
where
$
\bs{D}_{\Sf{a}}
\triangleq
\Diag[
	\chi \bs{1}_{\tau}
	+
	\tilde{L} H
	\bs{1}_{
		\frac{\tau}{\tilde{L}}
	}
	\otimes \bs{e}_{\tilde{L}}
	+
	\tau F \bs{e}_{\tau}
]
$
and
$
\bs{F}_{\tilde{L}}
$
represents an
$
\tilde{L} \times \tilde{L}
$
discrete Fourier transform matrix with
$(m, n)$-th
element given by
$
\frac{1}{
	\sqrt{ \tilde{L} }
}
\EXP[
	-
	2 \pi
	\imagUnit
	\frac{
		( m - 1 ) ( n - 1 )
	}{ \tilde{L} }
]
$.
\end{lemma}

\begin{table}[!t]
\centering
\caption{The reformulated 1RSB saddle point equations}
\label{Tab:Saddle_Point}
\begin{tabular}{ c }
\toprule[1pt]
\begin{minipage}[b]{.85\linewidth}
\centering
\includegraphics[scale=0.85]{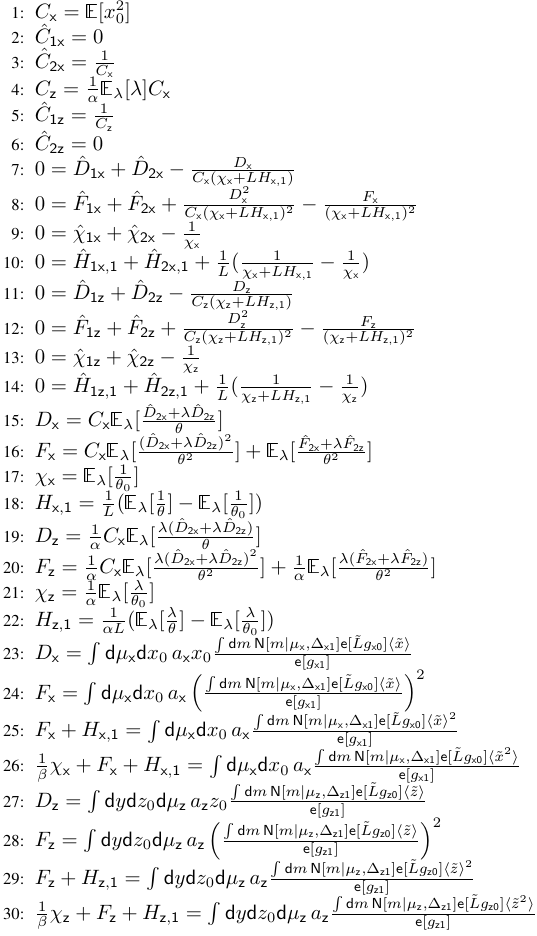}
\end{minipage}
\\ \bottomrule[1pt]
\end{tabular}
\end{table}

\begin{algorithm}[!t]
\caption{The VASP Algorithm}
\label{Tab:VASP}
\footnotesize
\DontPrintSemicolon
\SetAlgoLined
\TB{Input:}
$
\bs{y}
$,
$
\bs{H}
$,
$
q( \bs{x}_{0} )
$,
$
q( \bs{y} | \bs{z}_{0} )
$,
$
L
$
\;
\TB{Initialize:}
$
\bs{\mu}_{\Sf{z}}^{+, 1}
$,
$
\bs{v}_{\Sf{z, 0}}^{+, 1}
$,
$
\bs{v}_{\Sf{z}}^{+, 1}
$,
$
\bs{\mu}_{\Sf{x}}^{+, 1}
$,
$
\bs{v}_{\Sf{x, 0}}^{+, 1}
$,
$
\bs{v}_{\Sf{x}}^{+, 1}
$
\;
\For{
	$ t \leftarrow 1 $ \KwTo $T$
}{
	$
	\bs{v}_{\Sf{z, 1}}^{+, t}
	=
	\frac{1}{L}
	(
		\bs{v}_{\Sf{z}}^{+, t}
		-
		\bs{v}_{\Sf{z, 0}}^{+, t}
	)
	$
	\;
	$
	(
		\hat{\bs{\mu}}_{\Sf{z}}^{-, t},
		\hat{\bs{v}}_{\Sf{z, 0}}^{-, t},
		\hat{\bs{v}}_{\Sf{z, 1}}^{-, t}
	)
	=
	\Mean\Var[
		\bs{z} |
		\bs{\mu}_{\Sf{z}}^{+, t},
		\bs{v}_{\Sf{z, 0}}^{+, t},
		\bs{v}_{\Sf{z, 1}}^{+, t}, L
	]
	$
	\;
	
	$
	\hat{\bs{v}}_{\Sf{z}}^{-, t}
	=
	\hat{\bs{v}}_{\Sf{z, 0}}^{-, t}
	+
	L \hat{\bs{v}}_{\Sf{z, 1}}^{-, t}
	$
	\;
	$
	\bs{v}_{\Sf{z, 0}}^{-, t}
	=
	\bs{1} \oslash
	(
		\bs{1} \oslash
		\hat{\bs{v}}_{\Sf{z, 0}}^{-, t}
		-
		\bs{1} \oslash
		\bs{v}_{\Sf{z, 0}}^{+, t}
	)
	$
	\;
	$
	\bs{v}_{\Sf{z}}^{-, t}
	=
	\bs{1} \oslash
	(
		\bs{1} \oslash
		\hat{\bs{v}}_{\Sf{z}}^{-, t}
		-
		\bs{1} \oslash
		\bs{v}_{\Sf{z}}^{+, t}
	)
	$
	\;
	$
	\bs{\mu}_{\Sf{z}}^{-, t}
	=
	\bs{v}_{\Sf{z}}^{-, t} \odot
	(
		\hat{\bs{\mu}}_{\Sf{z}}^{-, t} \oslash
		\hat{\bs{v}}_{\Sf{z}}^{-, t}
		-
		\bs{\mu}_{\Sf{z}}^{+, t} \oslash
		\bs{v}_{\Sf{z}}^{+, t}
	)
	$
	\;
	$
	\hat{\bs{C}}_{\Sf{x, 0}}^{-, t}
	=
	[
		\Diag(
			\bs{1} \oslash
			\bs{v}_{\Sf{x, 0}}^{+, t}
		)
		+
		\bs{H}^{\Ts}
		\Diag(
			\bs{1} \oslash
			\bs{v}_{\Sf{z, 0}}^{-, t}
		)
		\bs{H}
	]^{-1}
	$
	\;
	$
	\hat{\bs{C}}_{\Sf{x}}^{-, t}
	=
	[
		\Diag(
			\bs{1} \oslash
			\bs{v}_{\Sf{x}}^{+, t}
		)
		+
		\bs{H}^{\Ts}
		\Diag(
			\bs{1} \oslash
			\bs{v}_{\Sf{z}}^{-, t}
		)
		\bs{H}
	]^{-1}
	$
	\;
	$
	\hat{\bs{\mu}}_{\Sf{x}}^{-, t}
	=
	\hat{\bs{C}}_{\Sf{x}}^{-, t}
	[
		\bs{\mu}_{\Sf{x}}^{+, t} \oslash \bs{v}_{\Sf{x}}^{+, t}
		+ \bs{H}^{\Ts}
		(
			\bs{\mu}_{\Sf{z}}^{-, t} \oslash
			\bs{v}_{\Sf{z}}^{-, t}
		)
	]
	$
	\;
	$
	\hat{\bs{v}}_{\Sf{x, 0}}^{-, t}
	=
	\Sf{d}(
		\hat{\bs{C}}_{\Sf{x, 0}}^{-, t}
	)
	$
	\;
	$
	\hat{\bs{v}}_{\Sf{x}}^{-, t}
	=
	\Sf{d}(
		\hat{\bs{C}}_{\Sf{x}}^{-, t}
	)
	$
	\;
	$
	\bs{v}_{\Sf{x, 0}}^{-, t}
	=
	\bs{1} \oslash
	(
		\bs{1} \oslash
		\hat{\bs{v}}_{\Sf{x, 0}}^{-, t}
		-
		\bs{1} \oslash
		\bs{v}_{\Sf{x, 0}}^{+, t}
	)
	$
	\;
	$
	\bs{v}_{\Sf{x}}^{-, t}
	=
	\bs{1} \oslash
	(
		\bs{1} \oslash
		\hat{\bs{v}}_{\Sf{x}}^{-, t}
		-
		\bs{1} \oslash
		\bs{v}_{\Sf{x}}^{+, t}
	)
	$
	\;
	$
	\bs{\mu}_{\Sf{x}}^{-, t}
	=
	\bs{v}_{\Sf{x}}^{-, t} \odot
	(
		\hat{\bs{\mu}}_{\Sf{x}}^{-, t} \oslash
		\hat{\bs{v}}_{\Sf{x}}^{-, t}
		-
		\bs{\mu}_{\Sf{x}}^{+, t} \oslash
		\bs{v}_{\Sf{x}}^{+, t}
	)
	$
	\;
	$
	\bs{v}_{\Sf{x, 1}}^{-, t}
	=
	\frac{1}{L}
	(
		\bs{v}_{\Sf{x}}^{-, t}
		-
		\bs{v}_{\Sf{x, 0}}^{-, t}
	)
	$
	\;
	$
	(
		\hat{\bs{\mu}}_{\Sf{x}}^{+, t + 1}, \hat{\bs{v}}_{\Sf{x, 0}}^{+, t + 1},
		\hat{\bs{v}}_{\Sf{x, 1}}^{+, t + 1}
	)
	=
	\Mean\Var[
		\bs{x} |
		\bs{\mu}_{\Sf{x}}^{-, t},
		\bs{v}_{\Sf{x, 0}}^{-, t},
		\bs{v}_{\Sf{x, 1}}^{-, t}, L
	]
	$
	\;
	$
	\hat{\bs{v}}_{\Sf{x}}^{+, t + 1}
	=
	\hat{\bs{v}}_{\Sf{x, 0}}^{+, t + 1}
	+
	L \hat{\bs{v}}_{\Sf{x, 1}}^{+, t + 1}
	$
	\;
	$
	\bs{v}_{\Sf{x, 0}}^{+, t + 1}
	=
	\bs{1} \oslash
	(
		\bs{1} \oslash
		\hat{\bs{v}}_{\Sf{x, 0}}^{+, t + 1}
		-
		\bs{1} \oslash
		\bs{v}_{\Sf{x, 0}}^{-, t}
	)
	$
	\;
	$
	\bs{v}_{\Sf{x}}^{+, t + 1}
	=
	\bs{1} \oslash
	(
		\bs{1} \oslash
		\hat{\bs{v}}_{\Sf{x}}^{+, t + 1}
		-
		\bs{1} \oslash
		\bs{v}_{\Sf{x}}^{-, t}
	)
	$
	\;
	$
	\bs{\mu}_{\Sf{x}}^{+, t + 1}
	=
	\bs{v}_{\Sf{x}}^{+, t + 1}
	\odot
	(
		\hat{\bs{\mu}}_{\Sf{x}}^{+, t + 1} \oslash
		\hat{\bs{v}}_{\Sf{x}}^{+, t + 1}
		-
		\bs{\mu}_{\Sf{x}}^{-, t} \oslash
		\bs{v}_{\Sf{x}}^{-, t}
	)
	$
	\;
	$
	\hat{\bs{C}}_{\Sf{\tilde{x}, 0}}^{+, t + 1}
	=
	[
		\Diag(
			\bs{1} \oslash
			\bs{v}_{\Sf{x, 0}}^{+, t + 1}
		)
		+
		\bs{H}^{\Ts}
		\Diag(
			\bs{1} \oslash
			\bs{v}_{\Sf{z, 0}}^{-, t}
		)
		\bs{H}
	]^{-1}
	$
	\;
	$
	\hat{\bs{C}}_{\Sf{\tilde{x}}}^{+, t + 1}
	=
	[
		\Diag(
			\bs{1} \oslash
			\bs{v}_{\Sf{x}}^{+, t + 1}
		)
		+
		\bs{H}^{\Ts}
		\Diag(
			\bs{1} \oslash
			\bs{v}_{\Sf{z}}^{-, t}
		)
		\bs{H}
	]^{-1}
	$
	\;
	$
	\hat{\bs{\mu}}_{\Sf{\tilde{x}}}^{+, t + 1}
	=
	\hat{\bs{C}}_{\Sf{\tilde{x}}}^{+, t + 1}
	[
		\bs{\mu}_{\Sf{x}}^{+, t + 1} \oslash
		\bs{v}_{\Sf{x}}^{+, t + 1}
		+
		\bs{H}^{\Ts}
		(
			\bs{\mu}_{\Sf{z}}^{-, t} \oslash
			\bs{v}_{\Sf{z}}^{-, t}
		)
	]
	$
	\;
	$
	\hat{\bs{C}}_{\Sf{z, 0}}^{+, t + 1}
	=
	\bs{H}
	\hat{\bs{C}}_{\Sf{\tilde{x}, 0}}^{+, t + 1}
	\bs{H}^{\Ts}
	$
	\;
	$
	\hat{\bs{C}}_{\Sf{z}}^{+, t + 1}
	=
	\bs{H}
	\hat{\bs{C}}_{\Sf{\tilde{x}}}^{+, t + 1}
	\bs{H}^{\Ts}
	$
	\;
	$
	\hat{\bs{\mu}}_{\Sf{z}}^{+, t + 1}
	=
	\bs{H}
	\hat{\bs{\mu}}_{\Sf{\tilde{x}}}^{+, t + 1}
	$
	\;
	$
	\hat{\bs{v}}_{\Sf{z, 0}}^{+, t + 1}
	=
	\Sf{d}(
		\hat{\bs{C}}_{\Sf{z, 0}}^{+, t + 1}
	)
	$
	\;
	$
	\hat{\bs{v}}_{\Sf{z}}^{+, t + 1}
	=
	\Sf{d}(
		\hat{\bs{C}}_{\Sf{z}}^{+, t + 1}
	)
	$
	\;
	$
	\bs{v}_{\Sf{z, 0}}^{+, t + 1}
	=
	\bs{1} \oslash
	(
		\bs{1} \oslash
		\hat{\bs{v}}_{\Sf{z, 0}}^{+, t + 1}
		-
		\bs{1} \oslash
		\bs{v}_{\Sf{z, 0}}^{-, t}
	)
	$
	\;
	$
	\bs{v}_{\Sf{z}}^{+, t + 1}
	=
	\bs{1} \oslash
	(
		\bs{1} \oslash
		\hat{\bs{v}}_{\Sf{z}}^{+, t + 1}
		-
		\bs{1} \oslash
		\bs{v}_{\Sf{z}}^{-, t}
	)
	$
	\;
	$
	\bs{\mu}_{\Sf{z}}^{+, t + 1}
	=
	\bs{v}_{\Sf{z}}^{+, t + 1} \odot
	(
		\hat{\bs{\mu}}_{\Sf{z}}^{+, t + 1} \oslash
		\hat{\bs{v}}_{\Sf{z}}^{+, t + 1}
		-
		\bs{\mu}_{\Sf{z}}^{-, t} \oslash
		\bs{v}_{\Sf{z}}^{-, t}
	)
	$
	\;
}
\TB{Output:}
$
\hat{\bs{x}}=\hat{\bs{\mu}}_{\Sf{x}}^{+}
$
\end{algorithm}

\begin{algorithm}[!t]
\caption{State Evolution of VASP}
\label{Tab:SE}
\scriptsize
\DontPrintSemicolon
\SetAlgoLined
\TB{Input:}
$
p( \bs{x}_{0} )
$,
$
p( \bs{y} | \bs{z}_{0} )
$,
$
q( \bs{x}_{0} )
$,
$
q( \bs{y} | \bs{z}_{0} )
$,
$
L
$
\;
\TB{Initialize:}
$
\hat{D}_{\Sf{1z}}^{1}
$,
$
\hat{F}_{\Sf{1z}}^{1}
$,
$
\mathtt{v}_{\Sf{z, 0}}^{+, 1}
$,
$
\mathtt{v}_{\Sf{z}}^{+, 1}
$,
$
\hat{D}_{\Sf{2x}}^{1}
$,
$
\hat{F}_{\Sf{2x}}^{1}
$,
$
\mathtt{v}_{\Sf{x, 0}}^{+, 1}
$,
$
\mathtt{v}_{\Sf{x}}^{+, 1}
$
\;
\For{
	$ t \leftarrow 1 $ \KwTo $T$
}{
	$
	\mathtt{v}_{\Sf{z, 1}}^{+, t}
	=
	\frac{1}{L}
	(
		\mathtt{v}_{\Sf{z}}^{+, t}
		-
		\mathtt{v}_{\Sf{z, 0}}^{+, t}
	)
	$
	\;
	$
	D_{\Sf{z}}^{-, t}
	=
	\int{
		\dd y \dd z_{0} \dd \mu_{\Sf{z}}
	} \,
	a^{t}_{\Sf{z}} z_{0}
	\frac{
		\int{\dd m} \,
		\Normal[
			m | \mu_{\Sf{z}},
			\mathtt{v}_{\Sf{z, 1}}^{+, t}
		]
		\EXP[
			\tilde{L} g^{t}_{\Sf{z0}}
		]
		\langle \tilde{z} \rangle
	}{
		\EXP[
			g^{t}_{\Sf{z1}}
		]
	}
	$
	\;
	$
	F_{\Sf{z}}^{-, t}
	=
	\int{
		\dd y \dd z_{0} \dd \mu_{\Sf{z}}
	} \,
	a^{t}_{\Sf{z}}
	[
		\frac{
			\int{\dd m} \,
			\Normal[
				m | \mu_{\Sf{z}},
				\mathtt{v}_{\Sf{z, 1}}^{+, t}
			]
			\EXP[
				\tilde{L} g^{t}_{\Sf{z0}}
			]
			\langle \tilde{z} \rangle
		}{
			\EXP[
				g^{t}_{\Sf{z1}}
			]
		}
	]^{2}
	$
	\;
	$
	\hat{\mathtt{v}}_{\Sf{z, 1}}^{-, t}
	=
	\int{
		\dd y \dd z_{0} \dd \mu_{\Sf{z}}
	} \,
	a^{t}_{\Sf{z}}
	\frac{
		\int{\dd m} \,
		\Normal[
			m | \mu_{\Sf{z}},
			\mathtt{v}_{\Sf{z, 1}}^{+, t}
		]
		\EXP[
			\tilde{L} g^{t}_{\Sf{z0}}
		]
		\langle \tilde{z} \rangle^{2}
	}{
		\EXP[
			g^{t}_{\Sf{z1}}
		]
	}
	-
	F_{\Sf{z}}^{-, t}
	$
	\;
	$
	\frac{1}{ \beta }
	\hat{\mathtt{v}}_{\Sf{z, 0}}^{-, t}
	=
	\int{
		\dd y \dd z_{0} \dd \mu_{\Sf{z}}
	} \,
	a^{t}_{\Sf{z}}
	\frac{
		\int{\dd m} \,
		\Normal[
			m | \mu_{\Sf{z}},
			\mathtt{v}_{\Sf{z, 1}}^{+, t}
		]
		\EXP[
			\tilde{L} g^{t}_{\Sf{z0}}
		]
		(
			\langle \tilde{z}^{2} \rangle
			-
			\langle \tilde{z} \rangle^{2}
		)
	}{
		\EXP[
			g^{t}_{\Sf{z1}}
		]
	}
	$
	\;
	$
	\hat{\mathtt{v}}_{\Sf{z}}^{-, t}
	=
	\hat{\mathtt{v}}_{\Sf{z, 0}}^{-, t}
	+
	L \hat{\mathtt{v}}_{\Sf{z, 1}}^{-, t}
	$
	\;
	$
	\mathtt{v}_{\Sf{z, 0}}^{-, t}
	=
	(
		\frac{1}{
			\hat{\mathtt{v}}_{\Sf{z, 0}}^{-, t}
		}
		-
		\frac{1}{
			\mathtt{v}_{\Sf{z, 0}}^{+, t}
		}
	)^{-1}
	$
	\;
	$
	\mathtt{v}_{\Sf{z}}^{-, t}
	=
	(
		\frac{1}{
			\hat{\mathtt{v}}_{\Sf{z}}^{-, t}
		}
		-
		\frac{1}{
			\mathtt{v}_{\Sf{z}}^{+, t}
		}
	)^{-1}
	$
	\;
	$
	\hat{D}_{\Sf{2z}}^{t}
	=
	\frac{
		D_{\Sf{z}}^{-, t}
	}{
		C_{\Sf{z}}
		\hat{\mathtt{v}}_{\Sf{z}}^{-, t}
	}
	-
	\hat{D}_{\Sf{1z}}^{t}
	$
	\;
	$
	\hat{F}_{\Sf{2z}}^{t}
	=
	\frac{
		F_{\Sf{z}}^{-, t}
	}{
		(
			\hat{\mathtt{v}}_{\Sf{z}}^{-, t}
		)^{2}
	}
	-
	\frac{
		(
			D_{\Sf{z}}^{-, t}
		)^{2}
	}{
		C_{\Sf{z}}
		(
			\hat{\mathtt{v}}_{\Sf{z}}^{-, t}
		)^{2}
	}
	-
	\hat{F}_{\Sf{1z}}^{t}
	$
	\;
	$
	\hat{\mathtt{v}}_{\Sf{x, 0}}^{-, t}
	=
	\Mean_{\lambda}[
		(
			\frac{1}{
				\mathtt{v}_{\Sf{x, 0}}^{+, t}
			}
			+
			\frac{\lambda}{
				\mathtt{v}_{\Sf{z, 0}}^{-, t}
			}
		)^{-1}
	]
	$
	\;
	$
	\hat{\mathtt{v}}_{\Sf{x}}^{-, t}
	=
	\Mean_{\lambda}[
		(
			\frac{1}{
				\mathtt{v}_{\Sf{x}}^{+, t}
			}
			+
			\frac{\lambda}{
				\mathtt{v}_{\Sf{z}}^{-, t}
			}
		)^{-1}
	]
	$
	\;
	$
	D_{\Sf{x}}^{-, t}
	=
	C_{\Sf{x}}
	\Mean_{\lambda}[
		\frac{
			\hat{D}_{\Sf{2x}}^{t}
			+
			\hat{D}_{\Sf{2z}}^{t} \lambda
		}{
			\frac{1}{
				\mathtt{v}_{\Sf{x}}^{+, t}
			}
			+
			\frac{\lambda}{
				\mathtt{v}_{\Sf{z}}^{-, t}
			}
		}
	]
	$
	\;
	$
	F_{\Sf{x}}^{-, t}
	=
	C_{\Sf{x}}
	\Mean_{\lambda}[
		\frac{
			(
				\hat{D}_{\Sf{2x}}^{t}
				+
				\hat{D}_{\Sf{2z}}^{t} \lambda
			)^{2}
		}{
			(
				\frac{1}{
					\mathtt{v}_{\Sf{x}}^{+, t}
				}
				+
				\frac{\lambda}{
					\mathtt{v}_{\Sf{z}}^{-, t}
				}
			)^{2}
		}
	]
	+
	\Mean_{\lambda}[
		\frac{
			\hat{F}_{\Sf{2x}}^{t}
			+
			\hat{F}_{\Sf{2z}}^{t} \lambda
		}{
			(
				\frac{1}{
					\mathtt{v}_{\Sf{x}}^{+, t}
				}
				+
				\frac{\lambda}{
					\mathtt{v}_{\Sf{z}}^{-, t}
				}
			)^{2}
		}
	]
	$
	\;
	$
	\mathtt{v}_{\Sf{x, 0}}^{-, t}
	=
	(
		\frac{1}{
			\hat{\mathtt{v}}_{\Sf{x, 0}}^{-, t}
		}
		-
		\frac{1}{
			\mathtt{v}_{\Sf{x, 0}}^{+, t}
		}
	)^{-1}
	$
	\;
	$
	\mathtt{v}_{\Sf{x}}^{-, t}
	=
	(
		\frac{1}{
			\hat{\mathtt{v}}_{\Sf{x}}^{-, t}
		}
		-
		\frac{1}{
			\mathtt{v}_{\Sf{x}}^{+, t}
		}
	)^{-1}
	$
	\;
	$
	\hat{D}_{\Sf{1x}}^{t}
	=
	\frac{
		D_{\Sf{x}}^{-, t}
	}{
		C_{\Sf{x}}
		\hat{\mathtt{v}}_{\Sf{x}}^{-, t}
	}
	-
	\hat{D}_{\Sf{2x}}^{t}
	$
	\;
	$
	\hat{F}_{\Sf{1x}}^{t}
	=
	\frac{
		F_{\Sf{x}}^{-, t}
	}{
		(
			\hat{\mathtt{v}}_{\Sf{x}}^{-, t}
		)^{2}
	}
	-
	\frac{
		(
			D_{\Sf{x}}^{-, t}
		)^{2}
	}{
		C_{\Sf{x}}
		(
			\hat{\mathtt{v}}_{\Sf{x}}^{-, t}
		)^{2}
	}
	-
	\hat{F}_{\Sf{2x}}^{t}
	$
	\;
	$
	\mathtt{v}_{\Sf{x, 1}}^{-, t}
	=
	\frac{1}{L}
	(
		\mathtt{v}_{\Sf{x}}^{-, t}
		-
		\mathtt{v}_{\Sf{x, 0}}^{-, t}
	)
	$
	\;
	$
	D_{\Sf{x}}^{+, t + 1}
	=
	\int{
		\dd \mu_{\Sf{x}} \dd x_{0}
	} \,
	a^{t}_{\Sf{x}} x_{0}
	\frac{
		\int{\dd m} \,
		\Normal[
			m | \mu_{\Sf{x}},
			\mathtt{v}_{\Sf{x, 1}}^{-, t}
		]
		\EXP[
			\tilde{L} g^{t}_{\Sf{x0}}
		]
		\langle \tilde{x} \rangle
	}{
		\EXP[
			g^{t}_{\Sf{x1}}
		]
	}
	$
	\;
	$
	F_{\Sf{x}}^{+, t + 1}
	=
	\int{
		\dd \mu_{\Sf{x}} \dd x_{0}
	} \,
	a^{t}_{\Sf{x}}
	[
		\frac{
			\int{\dd m} \,
			\Normal[
				m | \mu_{\Sf{x}},
				\mathtt{v}_{\Sf{x, 1}}^{-, t}
			]
			\EXP[
				\tilde{L} g^{t}_{\Sf{x0}}
			]
			\langle \tilde{x} \rangle
		}{
			\EXP[
				g^{t}_{\Sf{x1}}
			]
		}
	]^{2}
	$
	\;
	$
	\hat{\mathtt{v}}_{\Sf{x, 1}}^{+, t + 1}
	=
	\int{
		\dd \mu_{\Sf{x}} \dd x_{0}
	} \,
	a^{t}_{\Sf{x}}
	\frac{
		\int{\dd m} \,
		\Normal[
			m | \mu_{\Sf{x}},
			\mathtt{v}_{\Sf{x, 1}}^{-, t}
		]
		\EXP[
			\tilde{L} g^{t}_{\Sf{x0}}
		]
		\langle \tilde{x} \rangle^{2}
	}{
		\EXP[
			g^{t}_{\Sf{x1}}
		]
	}
	-
	F_{\Sf{x}}^{+, t + 1}
	$
	\;
	$
	\frac{1}{ \beta }
	\hat{\mathtt{v}}_{\Sf{x, 0}}^{+, t + 1}
	=
	\int{
		\dd \mu_{\Sf{x}} \dd x_{0}
	} \,
	a^{t}_{\Sf{x}}
	\frac{
		\int{\dd m} \,
		\Normal[
			m | \mu_{\Sf{x}},
			\mathtt{v}_{\Sf{x, 1}}^{-, t}
		]
		\EXP[
			\tilde{L} g^{t}_{\Sf{x0}}
		]
		(
			\langle \tilde{x}^{2} \rangle
			-
			\langle \tilde{x} \rangle^{2}
		)
	}{
		\EXP[
			g^{t}_{\Sf{x1}}
		]
	}
	$
	\;
	$
	\hat{\mathtt{v}}_{\Sf{x}}^{+, t + 1}
	=
	\hat{\mathtt{v}}_{\Sf{x, 0}}^{+, t + 1}
	+
	L \hat{\mathtt{v}}_{\Sf{x, 1}}^{+, t + 1}
	$
	\;
	$
	\mathtt{v}_{\Sf{x, 0}}^{+, t + 1}
	=
	(
		\frac{1}{
			\hat{\mathtt{v}}_{\Sf{x, 0}}^{+, t + 1}
		}
		-
		\frac{1}{
			\mathtt{v}_{\Sf{x, 0}}^{-, t}
		}
	)^{-1}
	$
	\;
	$
	\mathtt{v}_{\Sf{x}}^{+, t + 1}
	=
	(
		\frac{1}{
			\hat{\mathtt{v}}_{\Sf{x}}^{+, t + 1}
		}
		-
		\frac{1}{
			\mathtt{v}_{\Sf{x}}^{-, t}
		}
	)^{-1}
	$
	\;
	$
	\hat{D}_{\Sf{2x}}^{t + 1}
	=
	\frac{
		D_{\Sf{x}}^{+, t + 1}
	}{
		C_{\Sf{x}}
		\hat{\mathtt{v}}_{\Sf{x}}^{+, t + 1}
	}
	-
	\hat{D}_{\Sf{1x}}^{t}
	$
	\;
	$
	\hat{F}_{\Sf{2x}}^{t + 1}
	=
	\frac{
		F_{\Sf{x}}^{+, t + 1}
	}{
		(
			\hat{\mathtt{v}}_{\Sf{x}}^{+, t + 1}
		)^{2}
	}
	-
	\frac{
		(
			D_{\Sf{x}}^{+, t + 1}
		)^{2}
	}{
		C_{\Sf{x}}
		(
			\hat{\mathtt{v}}_{\Sf{x}}^{+, t + 1}
		)^{2}
	}
	-
	\hat{F}_{\Sf{1x}}^{t}
	$
	\;
	$
	\hat{\mathtt{v}}_{\Sf{z, 0}}^{+, t + 1}
	=
	\frac{1}{\alpha}
	\Mean_{\lambda}[
		\frac{\lambda}{
			(
				\frac{1}{
					\mathtt{v}_{\Sf{x, 0}}^{+, t + 1}
				}
				+
				\frac{\lambda}{
					\mathtt{v}_{\Sf{z, 0}}^{-, t}
				}
			)
		}
	]
	$
	\;
	$
	\hat{\mathtt{v}}_{\Sf{z}}^{+, t + 1}
	=
	\frac{1}{\alpha}
	\Mean_{\lambda}[
		\frac{\lambda}{
			(
				\frac{1}{
					\mathtt{v}_{\Sf{x}}^{+, t + 1}
				}
				+
				\frac{\lambda}{
					\mathtt{v}_{\Sf{z}}^{-, t}
				}
			)
		}
	]
	$
	\;
	$
	D_{\Sf{z}}^{+, t + 1}
	=
	\frac{
		C_{\Sf{x}}
	}{\alpha}
	\Mean_{\lambda}[
		\frac{
			\lambda
			(
				\hat{D}_{\Sf{2x}}^{t + 1}
				+
				\hat{D}_{\Sf{2z}}^{t} \lambda
			)
		}{
			\frac{1}{
				\mathtt{v}_{\Sf{x}}^{+, t + 1}
			}
			+
			\frac{\lambda}{
				\mathtt{v}_{\Sf{z}}^{-, t}
			}
		}
	]
	$
	\;
	$
	F_{\Sf{z}}^{+, t + 1}
	=
	\frac{
		C_{\Sf{x}}
	}{\alpha}
	\Mean_{\lambda}[
		\frac{
			\lambda (
				\hat{D}_{\Sf{2x}}^{t + 1}
				+
				\hat{D}_{\Sf{2z}}^{t} \lambda
			)^{2}
		}{
			(
				\frac{1}{
					\mathtt{v}_{\Sf{x}}^{+, t + 1}
				}
				+
				\frac{\lambda}{
					\mathtt{v}_{\Sf{z}}^{-, t}
				}
			)^{2}
		}
	]
	+
	\frac{1}{\alpha}
	\Mean_{\lambda}[
		\frac{
			\lambda
			(
				\hat{F}_{\Sf{2x}}^{t + 1}
				+
				\hat{F}_{\Sf{2z}}^{t} \lambda
			)
		}{
			(
				\frac{1}{
					\mathtt{v}_{\Sf{x}}^{+, t + 1}
				}
				+ \frac{\lambda}{
					\mathtt{v}_{\Sf{z}}^{-, t}
				}
			)^{2}
		}
	]
	$
	\;
	$
	\mathtt{v}_{\Sf{z, 0}}^{+, t + 1}
	=
	(
		\frac{1}{
			\hat{\mathtt{v}}_{\Sf{z, 0}}^{+, t + 1}
		}
		-
		\frac{1}{
			\mathtt{v}_{\Sf{z, 0}}^{-, t}
		}
	)^{-1}
	$
	\;
	$
	\mathtt{v}_{\Sf{z}}^{+, t + 1}
	=
	(
		\frac{1}{
			\hat{\mathtt{v}}_{\Sf{z}}^{+, t + 1}
		}
		-
		\frac{1}{
			\mathtt{v}_{\Sf{z}}^{-, t}
		}
	)^{-1}
	$
	\;
	$
	\hat{D}_{\Sf{1z}}^{t + 1}
	=
	\frac{
		D_{\Sf{z}}^{+, t + 1}
	}{
		C_{\Sf{z}}
		\hat{\mathtt{v}}_{\Sf{z}}^{+, t + 1}
	}
	-
	\hat{D}_{\Sf{2z}}^{t}
	$
	\;
	$
	\hat{F}_{\Sf{1z}}^{t + 1}
	=
	\frac{
		F_{\Sf{z}}^{+, t + 1}
	}{
		(
			\hat{\mathtt{v}}_{\Sf{z}}^{+, t + 1}
		)^{2}
	}
	-
	\frac{
		(
			D_{\Sf{z}}^{+, t + 1}
		)^{2}
	}{
		C_{\Sf{z}}
		(
			\hat{\mathtt{v}}_{\Sf{z}}^{+, t + 1}
		)^{2}
	}
	-
	\hat{F}_{\Sf{2z}}^{t}
	$
	\;
}
\TB{Output:}
$
D_{\Sf{x}}^{+, T + 1}
$,
$
F_{\Sf{x}}^{+, T + 1}
$,
$
\hat{\mathtt{v}}_{\Sf{x, 0}}^{+, T + 1}
$,
$
\hat{\mathtt{v}}_{\Sf{x, 1}}^{+, T + 1}
$
\end{algorithm}

\begin{table*}[!t]
\footnotesize
\centering
\caption{Useful Notations}
\label{Tab:notations}
\begin{tabular}{ l l l }
\toprule[1pt]
\TB{Line}
&
\TB{
Parameters associated with
$\bs{z}_{0}$
}
&
\TB{
Parameters associated with
$\bs{x}_{0}$
}
\\ \midrule[0.5pt]
\rownumber
&
$
m_{\Sf{z}}
\triangleq
\frac{
	\imagUnit \sqrt{
		2 \eta_{\Sf{z}}
		(
			-
			\hat{F}_{\Sf{1z}}
		)
	}
	\mu_{\Sf{z}}
}{
	\hat{\chi}_{\Sf{1z}}
	-
	L \hat{H}_{\Sf{1z, 1}}
}
$
&
$
m_{\Sf{x}}
\triangleq
\frac{
	\imagUnit \sqrt{
		2 \eta_{\Sf{x}}
		(
			-
			\hat{F}_{\Sf{1x}}
		)
	}
	\mu_{\Sf{x}}
}{
	\hat{\chi}_{\Sf{1x}}
	-
	L \hat{H}_{\Sf{1x, 1}}
}
$
\\
\rownumber
&
$
\Delta_{\Sf{z0}}
\triangleq
\frac{1}{
	\hat{\chi}_{\Sf{1z}}
}
$
&
$
\Delta_{\Sf{x0}}
\triangleq
\frac{1}{
	\hat{\chi}_{\Sf{1x}}
}
$
\\
\rownumber
&
$
\Delta_{\Sf{z1}}
\triangleq
\frac{1}{ L }
(
	\frac{1}{
		\hat{\chi}_{\Sf{1z}}
		-
		L \hat{H}_{\Sf{1z, 1}}
	}
	-
	\frac{1}{
		\hat{\chi}_{\Sf{1z}}
	}
)
$
&
$
\Delta_{\Sf{x1}}
\triangleq
\frac{1}{ L }
(
	\frac{1}{
		\hat{\chi}_{\Sf{1x}}
		-
		L \hat{H}_{\Sf{1x, 1}}
	}
	-
	\frac{1}{
		\hat{\chi}_{\Sf{1x}}
	}
)
$
\\
\rownumber
&
$
b_{\Sf{z}}
\triangleq
\EXP\{
	- \frac{1}{2}
	[
		\hat{C}_{\Sf{1z}}
		-
		\frac{
			(
				- \hat{D}_{\Sf{1z}}
			)^{2}
		}{
			- \hat{F}_{\Sf{1z}}
		}
	]
	z_{0}^{2}
	+
	\imagUnit \sqrt{2 \eta_{\Sf{z}}}
	\frac{
		- \hat{D}_{\Sf{1z}}
	}{
		\sqrt{
			- \hat{F}_{\Sf{1z}}
		}
	}
	z_{0} \mu_{\Sf{z}}
\}
$
&
$
b_{\Sf{x}}
\triangleq
\EXP\{
	- \frac{1}{2}
	[
		\hat{C}_{\Sf{1x}}
		-
		\frac{
			(
				- \hat{D}_{\Sf{1x}}
			)^{2}
		}{
			- \hat{F}_{\Sf{1x}}
		}
	]
	x_{0}^{2}
	+
	\imagUnit \sqrt{2 \eta_{\Sf{x}}}
	\frac{
		- \hat{D}_{\Sf{1x}}
	}{
		\sqrt{
			- \hat{F}_{\Sf{1x}}
		}
	}
	x_{0} \mu_{\Sf{x}}
\}
$
\\
\rownumber
&
$
\langle a \rangle
\triangleq
\frac{
	\int{\dd \tilde{z}} \,
	q^{\beta}(y | \tilde{z})
	\Normal[
		\tilde{z} | m,
		\frac{1}{\beta}
		\Delta_{\Sf{z0}}
	]
	a
}{
	\int{\dd \tilde{z}} \,
	q^{\beta}(y | \tilde{z})
	\Normal[
		\tilde{z} | m,
		\frac{1}{\beta}
		\Delta_{\Sf{z0}}	
	]
}
$
&
$
\langle a \rangle
\triangleq
\frac{
	\int{
		\dd \tilde{x}
	} \,
	\Normal[
		\tilde{x}
		| m,
		\frac{1}{\beta}
		\Delta_{\Sf{x0}}
	]
	q^{\beta}(
		\tilde{x}
	)
	a
}{
	\int{
		\dd \tilde{x}
	} \,
	\Normal[
		\tilde{x}
		| m,
		\frac{1}{\beta}
		\Delta_{\Sf{x0}}
	]
	q^{\beta}(
		\tilde{x}
	)
}
$
\\
\rownumber
&
$
g_{\Sf{z0}}
\triangleq
\log
\int{\dd \tilde{z}} \,
q^{\beta}(y | \tilde{z})
\Normal[
	\tilde{z} | m,
	\frac{1}{\beta}
	\Delta_{\Sf{z0}}
]
$
&
$
g_{\Sf{x0}}
\triangleq
\log
\int{
	\dd \tilde{x}
} \,
\Normal[
	\tilde{x}
	| m,
	\frac{1}{\beta}
	\Delta_{\Sf{x0}}
]
q^{\beta}(
	\tilde{x}
)
$
\\
\rownumber
&
$
g_{\Sf{z1}}
\triangleq
\log
\int{\dd m} \,
\Normal[
	m | m_{\Sf{z}},
	\Delta_{\Sf{z1}}
]
\EXP[
	\tilde{L} g_{\Sf{z0}}
]
$
&
$
g_{\Sf{x1}}
\triangleq
\log
\int{\dd m} \,
\Normal[
	m | m_{\Sf{x}},
	\Delta_{\Sf{x1}}
]
\EXP[
	\tilde{L} g_{\Sf{x0}}
]
$
\\
\rownumber
&
$
\eta_{\Sf{z}}
=
\frac{1}{
	2 \hat{F}_{\Sf{1z}}
}
(
	\hat{\chi}_{\Sf{1z}}
	-
	L \hat{H}_{\Sf{1z, 1}}
)^{2}
$
&
$
\eta_{\Sf{x}}
=
\frac{1}{
	2 \hat{F}_{\Sf{1x}}
}
(
	\hat{\chi}_{\Sf{1x}}
	-
	L \hat{H}_{\Sf{1x, 1}}
)^{2}
$
\\
\multirow{2}*{\rownumber}
&
$
a_{\Sf{z}}
\triangleq
p(y | z_{0})
\Normal\left[
	z_{0}
	\left|
		\frac{
			(
				- \hat{D}_{\Sf{1z}}
			)
			(
				\hat{\chi}_{\Sf{1z}}
				-
				L \hat{H}_{\Sf{1z, 1}}
			)
		}{
			( - \hat{F}_{\Sf{1z}} )
			[
				\hat{C}_{\Sf{1z}}
				-
				\frac{
					(
						- \hat{D}_{\Sf{1z}}
					)^{2}
				}{
					- \hat{F}_{\Sf{1z}}
				}
			]
		}
		\mu_{\Sf{z}},
		\frac{1}
		{
			\hat{C}_{\Sf{1z}}
			-
			\frac{
				(
					- \hat{D}_{\Sf{1z}}
				)^{2}
			}{
				- \hat{F}_{\Sf{1z}}
			}
		}
	\right.
\right]
\times
$
&
\multirow{2}*{
	$
	a_{\Sf{x}}
	\triangleq
	\Normal[
		\mu_{\Sf{x}} |
		\frac{
			\hat{D}_{\Sf{1x}}
		}{
			\hat{\chi}_{\Sf{1x}}
			-
			L \hat{H}_{\Sf{1x, 1}}
		}
		x_{0},
		\frac{
			\hat{F}_{\Sf{1x}}
		}{
			(
				\hat{\chi}_{\Sf{1x}}
				- L \hat{H}_{\Sf{1x, 1}}
			)^{2}
		}]
	p(x_{0})
	$
}
\\
&
$
\quad \quad \quad
\Normal\left[
	\mu_{\Sf{z}}
	\left|
		0, \frac{
			\hat{F}_{\Sf{1z}}
			[
				\hat{C}_{\Sf{1z}}
				-
				\frac{
					(
						- \hat{D}_{\Sf{1z}}
					)^{2}
				}{
					- \hat{F}_{\Sf{1z}}
				}
			]
		}{
			\hat{C}_{\Sf{1z}} (
				\hat{\chi}_{\Sf{1z}}
				-
				L \hat{H}_{\Sf{1z, 1}}
			)^{2}
		}
	\right.
\right]
$
&
\\ \midrule[0.5pt]
\rownumber
&
$
\langle a \rangle
\triangleq
\frac{
	\int{ \dd \tilde{z} } \,
	q^{ \beta }(y | \tilde{z})
	\Normal[
		\tilde{z} | m,
		\frac{1}{ \beta }
		v_{\Sf{z, 0}}^{+, t}
	]
	a
}{
	\int{ \dd \tilde{z} } \,
	q^{ \beta }(y | \tilde{z})
	\Normal[
		\tilde{z} | m,
		\frac{1}{ \beta }
		v_{\Sf{z, 0}}^{+, t}
	]
}
$
&
$
\langle a \rangle
\triangleq
\frac{
	\int{ \dd \tilde{x} } \,
	\Normal[
		\tilde{x} | m,
		\frac{1}{ \beta }
		v_{\Sf{x, 0}}^{-, t}
	]
	q^{ \beta }(\tilde{x})
	a
}{
	\int{ \dd \tilde{x} } \,
	\Normal[
		\tilde{x} | m,
		\frac{1}{ \beta }
		v_{\Sf{x, 0}}^{-, t}
	]
	q^{ \beta }(\tilde{x})
}
$
\\
\rownumber
&
$
g^{t}_{\Sf{z0}}
\triangleq
\log
\int{ \dd \tilde{z} } \,
q^{ \beta }(y | \tilde{z})
\Normal[
	\tilde{z} | m,
	\frac{1}{ \beta }
	v_{\Sf{z, 0}}^{+, t}
]
$
&
$
g^{t}_{\Sf{x0}}
\triangleq
\log
\int{ \dd \tilde{x} } \,
\Normal[
	\tilde{x} | m,
	\frac{1}{ \beta }
	v_{\Sf{x, 0}}^{-, t}
]
q^{ \beta }(\tilde{x})
$
\\
\rownumber
&
$
g^{t}_{\Sf{z1}}
\triangleq
\log \int{\dd m} \,
\Normal[
	m | \mu_{\Sf{z}}^{+, t},
	v_{\Sf{z, 1}}^{+, t}
]
\EXP[
	\tilde{L} g^{t}_{\Sf{z0}}
]
$
&
$
g^{t}_{\Sf{x1}}
\triangleq
\log \int{\dd m} \,
\Normal[
	m | \mu_{\Sf{x}}^{-, t},
	v_{\Sf{x, 1}}^{-, t}
]
\EXP[
	\tilde{L} g^{t}_{\Sf{x0}}
]
$
\\ \midrule[0.5pt]
\rownumber
&
$
g^{t}_{\Sf{z}}
\triangleq
\frac{1}{L}
g^{t}_{\Sf{z1}}
$
&
$
g^{t}_{\Sf{x}}
\triangleq
\frac{1}{L}
g^{t}_{\Sf{x1}}
$
\\
\rownumber
&
$
\Gamma_{\Sf{z0}}^{t}
\triangleq
\frac{2}{L - 1}
(
	\frac{
		\partial g^{t}_{\Sf{z}}
	}{
		\partial v_{\Sf{z, 1}}^{+, t}
	}
	-
	L \frac{
		\partial g^{t}_{\Sf{z}}
	}{
		\partial v_{\Sf{z, 0}}^{+, t}
	}
)
$
&
$
\Gamma_{\Sf{x0}}^{t}
\triangleq
\frac{2}{L - 1}
(
	\frac{
		\partial g^{t}_{\Sf{x}}
	}{
		\partial v_{\Sf{x, 1}}^{-, t}
	}
	-
	L \frac{
		\partial g^{t}_{\Sf{x}}
	}{
		\partial v_{\Sf{x, 0}}^{-, t}
	}
)
$
\\
\rownumber
&
$
\Gamma_{\Sf{z}}^{t}
\triangleq
-
2 \frac{
	\partial g^{t}_{\Sf{z}}
}{
	\partial v_{\Sf{z, 1}}^{+, t}
}
+
L (
	\frac{
		\partial g^{t}_{\Sf{z}}
	}{
		\partial \mu_{\Sf{z}}^{+, t}
	}
)^{2}
$
&
$
\Gamma_{\Sf{x}}^{t}
\triangleq
-
2 \frac{
	\partial g^{t}_{\Sf{x}}
}{
	\partial v_{\Sf{x, 1}}^{-, t}
}
+
L (
	\frac{
		\partial g^{t}_{\Sf{x}}
	}{
		\partial \mu_{\Sf{x}}^{-, t}
	}
)^{2}
$
\\
\rownumber
&
$
\hat{\mu}_{\Sf{z}}^{-, t}
\triangleq
v_{\Sf{z}}^{+, t}
\frac{
	\partial g^{t}_{\Sf{z}}
}{
	\partial \mu_{\Sf{z}}^{+, t}
}
+
\mu_{\Sf{z}}^{+, t}
$
&
$
\hat{\mu}_{\Sf{x}}^{+, t + 1}
\triangleq
v_{\Sf{x}}^{-, t}
\frac{
	\partial g^{t}_{\Sf{x}}
}{
	\partial \mu_{\Sf{x}}^{-, t}
}
+
\mu_{\Sf{x}}^{-, t}
$
\\
\rownumber
&
$
\hat{v}_{\Sf{z, 0}}^{-, t}
\triangleq
v_{\Sf{z, 0}}^{+, t}
-
(
	v_{\Sf{z, 0}}^{+, t}
)^{2}
\Gamma_{\Sf{z0}}^{t}
$
&
$
\hat{v}_{\Sf{x, 0}}^{+, t + 1}
\triangleq
v_{\Sf{x, 0}}^{-, t}
-
(
	v_{\Sf{x, 0}}^{-, t}
)^{2}
\Gamma_{\Sf{x0}}^{t}
$
\\
\rownumber
&
$
\hat{v}_{\Sf{z}}^{-, t}
\triangleq
v_{\Sf{z}}^{+, t}
-
(
	v_{\Sf{z}}^{+, t}
)^{2}
\Gamma_{\Sf{z}}^{t}
$
&
$
\hat{v}_{\Sf{x}}^{+, t + 1}
\triangleq
v_{\Sf{x}}^{-, t}
-
(
	v_{\Sf{x}}^{-, t}
)^{2}
\Gamma_{\Sf{x}}^{t}
$
\\ \midrule[0.5pt]
\rownumber
&
$
f^{t}_{\Sf{z0}}(
	y, m,
	v_{\Sf{z, 0}}^{+, t}
)
\triangleq
\max\limits_{ \tilde{z} }
[
	\log q( y | \tilde{z} )
	-
	\frac{1}{
		2 v_{\Sf{z, 0}}^{+, t}
	}
	( \tilde{z} - m )^{2}
]
$
&
$
f^{t}_{\Sf{x0}}(
	m,
	v_{\Sf{x, 0}}^{-, t}
)
\triangleq
\max\limits_{ \tilde{x} }
[
	\log q( \tilde{x} )
	-
	\frac{1}{
		2 v_{\Sf{x, 0}}^{-, t}
	}
	( \tilde{x} - m )^{2}
]
$
\\
\rownumber
&
$
g^{t}_{\Sf{z1}}
=
\log \int{\dd m} \,
\Normal[
	m | \mu_{\Sf{z}}^{+, t},
	v_{\Sf{z, 1}}^{+, t}
]
\EXP[
	L
	f^{t}_{\Sf{z0}}(
		y, m,
		v_{\Sf{z, 0}}^{+, t}
	)
]
$
&
$
g^{t}_{\Sf{x1}}
=
\log \int{\dd m} \,
\Normal[
	m | \mu_{\Sf{x}}^{-, t},
	v_{\Sf{x, 1}}^{-, t}
]
\EXP[
	L
	f^{t}_{\Sf{x0}}(
		m,
		v_{\Sf{x, 0}}^{-, t}
	)
]
$
\\
\rownumber
&
$
\Gamma_{\Sf{z0}}^{t}
=
-
2
(
	\frac{
		\partial g^{t}_{\Sf{z}}
	}{
		\partial v_{\Sf{z, 1}}^{+, t}
	}
	-
	L \frac{
		\partial g^{t}_{\Sf{z}}
	}{
		\partial v_{\Sf{z, 0}}^{+, t}
	}
)
$
&
$
\Gamma_{\Sf{x0}}^{t}
=
-
2
(
	\frac{
		\partial g^{t}_{\Sf{x}}
	}{
		\partial v_{\Sf{x, 1}}^{-, t}
	}
	- L \frac{
		\partial g^{t}_{\Sf{x}}
	}{
		\partial v_{\Sf{x, 0}}^{-, t}
	}
)
$
\\ \midrule[0.5pt]
\multirow{2}*{\rownumber}
&
$
a^{t}_{\Sf{z}}
\triangleq
p(y | z_{0})
\Normal\left[
	z_{0}
	\left|
		\frac{
			(
				- \hat{D}_{\Sf{1z}}^{t}
			)
			(
				\hat{\chi}_{\Sf{1z}}^{t}
				-
				L \hat{H}_{\Sf{1z, 1}}^{t}
			)
		}{
			( - \hat{F}_{\Sf{1z}}^{t} )
			[
				\hat{C}_{\Sf{1z}}
				-
				\frac{
					(
						- \hat{D}_{\Sf{1z}}^{t}
					)^{2}
				}{
					- \hat{F}_{\Sf{1z}}^{t}
				}
			]
		}
		\mu_{\Sf{z}},
		\frac{1}
		{
			\hat{C}_{\Sf{1z}}
			-
			\frac{
				(
					- \hat{D}_{\Sf{1z}}^{t}
				)^{2}
			}{
				- \hat{F}_{\Sf{1z}}^{t}
			}
		}
	\right.
\right]
\times
$
&
\multirow{2}*{
	$
	a^{t}_{\Sf{x}}
	\triangleq
	\Normal[
		\mu_{\Sf{x}} |
		\frac{
			\hat{D}_{\Sf{1x}}^{t}
		}{
			\hat{\chi}_{\Sf{1x}}^{t}
			-
			L \hat{H}_{\Sf{1x, 1}}^{t}
		}
		x_{0},
		\frac{
			\hat{F}_{\Sf{1x}}^{t}
		}{
			(
				\hat{\chi}_{\Sf{1x}}^{t}
				-
				L \hat{H}_{\Sf{1x, 1}}^{t}
			)^{2}
		}
	]
	p(x_{0})
	$
}
\\
&
$
\quad \quad \quad
\Normal\left[
	\mu_{\Sf{z}}
	\left|
		0, \frac{
			\hat{F}_{\Sf{1z}}^{t}
			[
				\hat{C}_{\Sf{1z}}
				-
				\frac{
					(
						- \hat{D}_{\Sf{1z}}^{t}
					)^{2}
				}{
					- \hat{F}_{\Sf{1z}}^{t}
				}
			]
		}{
			\hat{C}_{\Sf{1z}} (
				\hat{\chi}_{\Sf{1z}}^{t}
				- L \hat{H}_{\Sf{1z, 1}}^{t}
			)^{2}
		}
	\right.
\right]
$
&
\\ \bottomrule[1pt]
\end{tabular}
\end{table*}

\section*{Acknowledgment}

The authors would like to thank Prof. Carlo Lucibello for many helpful discussions and his sharing of the GASP code.

\ifCLASSOPTIONcaptionsoff
	\newpage
\fi

\bibliographystyle{IEEEtran}
\normalem
\bibliography{IEEEabrv,./CQ_Original,./CQ_bib}

\begin{thebibliography}{10}
\providecommand{\url}[1]{#1}
\csname url@samestyle\endcsname
\providecommand{\newblock}{\relax}
\providecommand{\bibinfo}[2]{#2}
\providecommand{\BIBentrySTDinterwordspacing}{\spaceskip=0pt\relax}
\providecommand{\BIBentryALTinterwordstretchfactor}{4}
\providecommand{\BIBentryALTinterwordspacing}{\spaceskip=\fontdimen2\font plus
\BIBentryALTinterwordstretchfactor\fontdimen3\font minus
  \fontdimen4\font\relax}
\providecommand{\BIBforeignlanguage}[2]{{%
\expandafter\ifx\csname l@#1\endcsname\relax
\typeout{** WARNING: IEEEtran.bst: No hyphenation pattern has been}%
\typeout{** loaded for the language `#1'. Using the pattern for}%
\typeout{** the default language instead.}%
\else
\language=\csname l@#1\endcsname
\fi
#2}}
\providecommand{\BIBdecl}{\relax}
\BIBdecl

\bibitem{kabashima2003cdma}
Y.~Kabashima, ``A {CDMA} multiuser detection algorithm on the basis of belief
  propagation,'' \emph{J. Phys. A. Math. Gen.}, vol.~36, no.~43, p. 11111,
  2003.

\bibitem{donoho2009message}
D.~L. Donoho, A.~Maleki, and A.~Montanari, ``Message-passing algorithms for
  compressed sensing,'' \emph{Proc. Nat. Acad. Sci.}, vol. 106, no.~45, pp.
  18\,914--18\,919, 2009.

\bibitem{bayati2011dynamics}
M.~Bayati and A.~Montanari, ``The dynamics of message passing on dense graphs,
  with applications to compressed sensing,'' \emph{{IEEE} Trans. Inf. Theory},
  vol.~57, no.~2, pp. 764--785, 2011.

\bibitem{javanmard2013state}
A.~Javanmard and A.~Montanari, ``State evolution for general approximate
  message passing algorithms, with applications to spatial coupling,''
  \emph{Information and Inference: A Journal of the IMA}, vol.~2, no.~2, pp.
  115--144, 2013.

\bibitem{rangan2011generalized}
S.~Rangan, ``Generalized approximate message passing for estimation with random
  linear mixing,'' in \emph{Proc. {IEEE} Int. Symp. Inf. Theory.}\hskip 1em
  plus 0.5em minus 0.4em\relax IEEE, 2011, pp. 2168--2172.

\bibitem{rangan2019vector}
S.~Rangan, P.~Schniter, and A.~K. Fletcher, ``Vector approximate message
  passing,'' \emph{{IEEE} Trans. Inf. Theory}, vol.~65, no.~10, pp. 6664--6684,
  2019.

\bibitem{schniter2016vector}
P.~Schniter, S.~Rangan, and A.~K. Fletcher, ``Vector approximate message
  passing for the generalized linear model,'' in \emph{Proc. Asilomar Conf.
  Signals, Syst., Comput.}\hskip 1em plus 0.5em minus 0.4em\relax IEEE, 2016,
  pp. 1525--1529.

\bibitem{fletcher2018inference}
A.~K. Fletcher, S.~Rangan, and P.~Schniter, ``Inference in deep networks in
  high dimensions,'' in \emph{{IEEE} Int. Symp. on Inf. Theory.}\hskip 1em plus
  0.5em minus 0.4em\relax IEEE, 2018, pp. 1884--1888.

\bibitem{he2018bayesian}
H.~He, C.-K. Wen, and S.~Jin, ``Bayesian optimal data detector for hybrid
  mmwave {MIMO}-{OFDM} systems with low-resolution {ADCs},'' \emph{{IEEE} J.
  Sel. Topics Signal Process.}, vol.~12, no.~3, pp. 469--483, 2018.

\bibitem{ma2017orthogonal}
J.~Ma and L.~Ping, ``Orthogonal amp,'' \emph{IEEE Access}, vol.~5, pp.
  2020--2033, 2017.

\bibitem{antenucci2019approximate}
F.~Antenucci, F.~Krzakala, P.~Urbani, and L.~Zdeborov{\'a}, ``Approximate
  survey propagation for statistical inference,'' \emph{J. Stat. Mech., Theory
  Exp.}, vol. 2019, no.~2, p. 023401, 2019.

\bibitem{lucibello2019generalized}
C.~Lucibello, L.~Saglietti, and Y.~Lu, ``Generalized approximate survey
  propagation for high-dimensional estimation,'' in \emph{Int. Conf. Mach.
  Learn.}\hskip 1em plus 0.5em minus 0.4em\relax PMLR, 2019, pp. 4173--4182.

\bibitem{takahashi2022macroscopic}
T.~Takahashi and Y.~Kabashima, ``Macroscopic analysis of vector approximate
  message passing in a model-mismatched setting,'' \emph{{IEEE} Trans. Inf.
  Theory}, vol.~68, no.~8, pp. 5579--5600, 2022.

\bibitem{rangan2012asymptotic}
S.~Rangan, A.~K. Fletcher, and V.~K. Goyal, ``Asymptotic analysis of {MAP}
  estimation via the replica method and applications to compressed sensing,''
  \emph{{IEEE} Trans. Inf. Theory}, vol.~58, no.~3, pp. 1902--1923, 2012.

\bibitem{mezard1987spin}
M.~M{\'e}zard, G.~Parisi, and M.~A. Virasoro, \emph{Spin glass theory and
  beyond: {An} {Introduction} to the {Replica} {Method} and {Its}
  {Applications}}.\hskip 1em plus 0.5em minus 0.4em\relax World Sci. Publishing
  Comp., 1987, vol.~9.

\bibitem{braunstein2005survey}
A.~Braunstein, M.~M{\'e}zard, and R.~Zecchina, ``Survey propagation: {An}
  algorithm for satisfiability,'' \emph{Random Structures \& Algorithms},
  vol.~27, no.~2, pp. 201--226, 2005.

\bibitem{mezard2001bethe}
M.~M{\'e}zard and G.~Parisi, ``The bethe lattice spin glass revisited,''
  \emph{Eur. Phys. J. B.}, vol.~20, pp. 217--233, 2001.

\bibitem{mezard2003cavity}
------, ``The cavity method at zero temperature,'' \emph{J. Statist. Phys.},
  vol. 111, pp. 1--34, 2003.

\bibitem{pearl1988probabilistic}
J.~Pearl, \emph{Probabilistic reasoning in intelligent systems: networks of
  plausible inference}.\hskip 1em plus 0.5em minus 0.4em\relax Morgan kaufmann,
  1988.

\bibitem{bromiley2003products}
P.~Bromiley, ``Products and convolutions of {Gaussian} probability density
  functions,'' \emph{Tina-Vision Memo (Internal Report)}, vol.~3, no.~4, 2003.

\bibitem{mezard2009information}
M.~Mezard and A.~Montanari, \emph{Information, physics, and computation}.\hskip
  1em plus 0.5em minus 0.4em\relax Oxford University Press, 2009.

\bibitem{zou2021multi}
Q.~Zou, H.~Zhang, and H.~Yang, ``Multi-layer bilinear generalized approximate
  message passing,'' \emph{{IEEE} Trans. Signal Process.}, vol.~69, pp.
  4529--4543, 2021.

\bibitem{zou2023high}
Q.~Zou and H.~Zhang, ``High-{Dimensional} {Multiple}-{Measurement}-{Vector}
  {Problem}: {Mutual} {Information} and {Message} {Passing},'' \emph{{IEEE}
  Trans. Signal Process.}, 2023.

\bibitem{hubbard1959calculation}
J.~Hubbard, ``Calculation of partition functions,'' \emph{Phys. Rev. Lett.},
  vol.~3, no.~2, p.~77, 1959.

\bibitem{stratonovich1957method}
R.~Stratonovich, ``On a method of calculating quantum distribution functions,''
  in \emph{Soviet Physics Doklady}, vol.~2, 1957, p. 416.

\bibitem{tulino2013support}
A.~M. Tulino, G.~Caire, S.~Verd{\'u}, and S.~Shamai, ``Support recovery with
  sparsely sampled free random matrices,'' \emph{{IEEE} Trans. Inf. Theory},
  vol.~59, no.~7, pp. 4243--4271, 2013.

\bibitem{bishop2006pattern}
C.~M. Bishop and N.~M. Nasrabadi, \emph{Pattern recognition and machine
  learning}.\hskip 1em plus 0.5em minus 0.4em\relax Springer, 2006, vol.~4,
  no.~4.

\bibitem{gradenigo2020solving}
G.~Gradenigo, M.~C. Angelini, L.~Leuzzi, and F.~Ricci-Tersenghi, ``Solving the
  spherical p-spin model with the cavity method: equivalence with the replica
  results,'' \emph{J. Stat. Mech., Theory Exp.}, vol. 2020, no.~11, p. 113302,
  2020.

\bibitem{zou2018concise}
Q.~Zou, H.~Zhang, C.-K. Wen, S.~Jin, and R.~Yu, ``Concise derivation for
  generalized approximate message passing using expectation propagation,''
  \emph{{IEEE} Signal Process. Lett.}, vol.~25, no.~12, pp. 1835--1839, 2018.

\bibitem{pandit2020inference}
P.~Pandit, M.~Sahraee-Ardakan, S.~Rangan, P.~Schniter, and A.~K. Fletcher,
  ``Inference with deep generative priors in high dimensions,'' \emph{{IEEE} J.
  Sel. Areas. Inf. Theory.}, vol.~1, no.~1, pp. 336--347, 2020.

\bibitem{opper2005expectation}
M.~Opper, O.~Winther, and M.~J. Jordan, ``Expectation consistent approximate
  inference.'' \emph{J. Mach. Learn. Res.}, vol.~6, no.~12, 2005.

\bibitem{minka2001family}
T.~P. Minka, ``A family of algorithms for approximate {Bayesian} inference,''
  Ph.D. dissertation, Massachusetts Institute of Technology, 2001.

\bibitem{liu2021decentralized}
S.~Liu, H.~Zhang, and Q.~Zou, ``Decentralized channel estimation for the uplink
  of grant-free massive machine-type communications,'' \emph{{IEEE} Trans.
  Commun.}, vol.~70, no.~2, pp. 967--979, 2021.

\bibitem{gerbelot2022asymptotic}
C.~Gerbelot, A.~Abbara, and F.~Krzakala, ``Asymptotic errors for
  teacher-student convex generalized linear models (or: {How} to prove
  {Kabashima’s} replica formula),'' \emph{{IEEE} Trans. Inf. Theory},
  vol.~69, no.~3, pp. 1824--1852, 2022.

\bibitem{barbier2023compressed}
D.~Barbier, C.~Lucibello, L.~Saglietti, F.~Krzakala, and L.~Zdeborova,
  ``Compressed sensing with l0-norm: statistical physics analysis and
  algorithms for signal recovery,'' \emph{arXiv preprint arXiv:2304.12127},
  2023.

\bibitem{he2017generalized}
H.~He, C.-K. Wen, and S.~Jin, ``Generalized expectation consistent signal
  recovery for nonlinear measurements,'' in \emph{{IEEE} Int. Symp. on Inf.
  Theory.}\hskip 1em plus 0.5em minus 0.4em\relax IEEE, 2017, pp. 2333--2337.

\bibitem{guo2009generic}
D.~Guo, T.~Tanaka, and M.~Honig, ``Generic multiuser detection and statistical
  physics,'' \emph{Advances in Multiuser Detection}, vol.~99, p. 251, 2009.

\bibitem{paulraj2003introduction}
A.~Paulraj, R.~Nabar, and D.~Gore, \emph{Introduction to space-time wireless
  communications}.\hskip 1em plus 0.5em minus 0.4em\relax Cambridge university
  press, 2003.

\bibitem{shinzato2008perceptron}
T.~Shinzato and Y.~Kabashima, ``Perceptron capacity revisited: classification
  ability for correlated patterns,'' \emph{J. Phys. A. Math. Theor.}, vol.~41,
  no.~32, p. 324013, 2008.

\end{thebibliography}

\end{document}